\documentclass[prb,aps,twocolumn,showpacs,floats,floatfix]{revtex4}
\usepackage{graphicx}  
\usepackage{dcolumn}   
\usepackage{amsmath}
\usepackage{amssymb}
\usepackage{subfigure}
\usepackage{color}

\usepackage{bm}
\usepackage{natbib}

\begin{document}
\bibliographystyle{apsrev}
\title{%
Ideal, Defective, and Gold--Promoted Rutile $\mathbf{TiO_{2}(110)}$ Surfaces: 
Structures, Energies, Dynamics, and Thermodynamics from PBE+U}
\author{Matteo Farnesi Camellone}
\email{matteo.farnesi@theochem.rub.de}
\affiliation{Lehrstuhl f\"ur Theoretische Chemie, Ruhr--Universit\"at Bochum,
44780 Bochum, Germany}
\author{Piotr M. Kowalski}         
\altaffiliation[Present address: ]{Helmholtz Centre Potsdam, Telegrafenberg,
  14473 Potsdam, Germany}     
\affiliation{Lehrstuhl f\"ur Theoretische Chemie, Ruhr--Universit\"at Bochum,                            
44780 Bochum, Germany}
\author{Dominik Marx}
\affiliation{Lehrstuhl f\"ur Theoretische Chemie, Ruhr--Universit\"at Bochum,
44780 Bochum, Germany}

\date{\today}
\begin{abstract}
Extensive first principles calculations are carried out to investigate 
gold-promoted $\mathrm{TiO_{2}}$(110) surfaces
in terms of structure optimizations, electronic structure analyses, 
{\em ab initio} thermodynamics calculations of surface phase diagrams, 
and {\em ab initio} molecular dynamics simulations. 
All computations rely on density functional theory in the
generalized gradient approximation (PBE) and account for on--site Coulomb
interactions via inclusion of a Hubbard correction, PBE+U,
where U is computed from linear response theory.
This approach is validated by investigating the interaction
between TiO$_{2}$(110) surfaces and typical probe species
($\mathrm{H}$, $\mathrm{H_{2}O}$, $\mathrm{CO}$). 
Relaxed structures and binding energies are compared to both
data from the literature and plain PBE results, 
thus allowing 
 the performance of the PBE+U approach
for the specific purpose
to be verified. 
The main focus of the study is on the properties of gold-promoted titania
surfaces and their interactions with $\mathrm{CO}$. 
Both PBE+U and PBE optimized structures of
$\mathrm{Au}$ adatoms adsorbed on stoichiometric and reduced
$\mathrm{TiO_{2}}$ surfaces 
are  
computed, along with
their electronic structure.
The
charge rearrangement induced by the adsorbates 
at the metal/oxide contact are 
also
analyzed in detail and discussed.
By performing PBE+U {\it ab initio} molecular dynamics
simulations, it is demonstrated that 
the diffusion of $\mathrm{Au}$ adatoms
on the stoichiometric surface
is highly anisotropic.
The metal atoms migrate either along the top of the bridging oxygen rows, or
around the area between
these rows, from one bridging
position to the next along the [001] direction.
No translational motion perpendicular
to this direction is observed.
Approximate {\it ab initio} thermodynamics predicts that under $\mathrm{O}$--rich conditions,
structures obtained by substituting a $\mathrm{Ti_{5c}}$ atom
with an $\mathrm{Au}$ atom are thermodynamically stable 
over
 a wide range of
temperatures and pressures that
are relevant to applications in the realm of catalysis.
Finally, it is shown that $\mathrm{TiO_{2}}(110)$ surfaces containing positively
charged Au~ions activate molecular $\mathrm{CO}$, whereas a single negatively
charged $\mathrm{Au^{-\delta}}$ species bound to an $\mathrm{O}$ vacancy only 
weakly interacts with $\mathrm{CO}$. 
%
Despite this,
 the calculations predict that the reactivity of gold nanoparticles 
nucleated at $\mathrm{O}$ vacancies can
be 
recovered
for cluster sizes as small as Au$_{2}$.

\end{abstract}


\maketitle

\section{Introduction}
Titania, $\mathrm{TiO_{2}}$, is a metal oxide of both fundamental interest
and technological importance~\cite{rev5,diebold,sauer,rev3,rev4}.
%
It is used in
several key technologies
including 
pigments, coatings, electronic devices, implants, 
 gas sensors,
  photochemical reactions, and
catalysis 
~\cite{regan,matthey,imagawa}. 
%
One of the most important properties of titania is
that it can be easily 
reduced (see e.g. Ref.~\onlinecite{kowalski} for a concise presentation),
strongly affecting
its chemical properties in general and its reactivity
in particular~\cite{roger-piss}.
One way to reduce the $\mathrm{TiO_{2}}$ surface is to remove surface 
oxygen 
atoms,
thereby creating 
$\mathrm{O}$ vacancies.
%
The removal of an
$\mathrm{O}$ atom gives rise to two excess electrons and the appearance of new
electronic states within the band gap at about 
0.7 to 0.9~eV 
below the conduction band 
edge, thus
creating an $F$--center~\cite{henrich,diebold,Campbell,sauer}.
By this process, 
two substrate $\mathrm{Ti^{4+}}$ ions change formally 
to a $\mathrm{Ti^{3+}}$ oxidation state;
see Ref.~\onlinecite{PMD} for recent literature and 
a detailed picture of
the (de--)localization dynamics of the excess electrons.
Alternatively, the $\mathrm{TiO_{2}}$ surface can be reduced by hydroxylation
of surface $\mathrm{O}$ atoms via adsorption of hydrogen~\cite{henrich1,kunat,suzuki,fujino,YC08}. 
The interaction of
$\mathrm{TiO_{2}}$ with water is an important process which has to be taken into 
account, since it occurs easily,
even in 
well--controlled
UHV~experiments. 
The adsorption of water on $\mathrm{TiO_{2}}$ has been investigated 
extensively, 
both
experimentally and
theoretically\cite{roger-piss,selloni-review-2010,kowalski,hugenschmidt,brinkley,desegovia,goniakowski,gillan,bates,gillan1,truong,langel,zhang1,zhang2,zhang3,perron,lui2010,ExW7,ExW2,ExW3}
; 
in particular see 
Refs.~\onlinecite{roger-piss,selloni-review-2010}
for the most recent reviews of this literature. 
%

Most relevant to catalysis is the interaction of $\mathrm{Au}$ and $\mathrm{CO}$
with stoichiometric or reduced $\mathrm{TiO_{2}}$ surfaces and, in particular, the interaction
of titania--supported gold particles with CO molecules.
A detailed 
understanding  
of the process of $\mathrm{CO}$ adsorption
is required 
to best comprehend its
wide variety of
applications~\cite{sauer,diebold}, such as $\mathrm{CO}$ oxidation at low
temperature~\cite{Campbell}, the water
gas shift reaction, and $\mathrm{CO}$ 
hydrogenation~\cite{fu,sato}.
%
In a recent paper~\cite{JCP}, we investigated the interaction of
$\mathrm{CO}$ with the stoichiometric $\mathrm{TiO_{2}(110)}$ surface using a
combination of density functional theory (DFT) and post Hartree-Fock methods. 
For a single CO molecule in
the (4 $\times$ 2) surface unit cell of our slab, we 
found
that the upright position above the fivefold
coordinated Ti sites, $\mathrm{Ti_{5c}}$, remains the preferential adsorption
geometry,
even without enforcing symmetry.
On the reduced titania surface, results from 
temperature--programmed
desorption (TPD) experiments~\cite{linsebigler,kim} 
suggested 
that, at low coverages, $\mathrm{CO}$ adsorption occurs at non-adjacent
$\mathrm{Ti_{5c}}$ sites. 
These findings 
were 
supported by various calculations~\cite{ferrari,vittadini,pillay1}.
However, earlier studies 
implicated 
bridge--bonded
oxygen vacancies 
as 
adsorption sites for $\mathrm{CO}$~\cite{rocker,rocker1}, 
a conclusion 
corroborated by some theoretical investigations~\cite{mene,koba} as well. 
%

The seminal
 work of Haruta and coworkers~\cite{haruta} has shown that the 
low--temperature
oxidation of molecular $\mathrm{CO}$ can be efficiently
catalyzed by highly dispersed $\mathrm{Au}$ nanoparticles supported on
$\mathrm{TiO_{2}}$ surfaces~\cite{parker,grunwaldt,liu,mills}. 
It is now recognized that gold nanoclusters,
prepared in different ways and supported on various metal oxides, are able to
catalyze a number of reactions~\cite{haruta2,bond}, and that the size of the gold
particles substantially affects the catalytic activity. 
The gold clusters should be smaller than about 5~nm for high catalytic activity to occur,
suggesting the key importance of metal/support interfacial
interactions on a nanometer scale. 
Extensive studies of the
$\mathrm{Au/TiO_{2}}$ system link the peculiar catalytic activity of gold
nanoparticles on titania to several factors: 
high concentration of 
low--coordination
sites~\cite{lopez,lopez1}, 
quantum size effects of 
two--layer
$\mathrm{Au}$ islands~\cite{valden}, 
active perimeter sites of the nanoparticles~\cite{haruta1}, 
and
charge transfer between the gold particles and the supporting oxide~\cite{liu,molina}.
%
%
%
%
%
%
%

Over the past decade, DFT--based
calculations have
been extensively employed to study the interaction between
gold and the $\mathrm{TiO_{2}(110)}$
surface~\cite{matthey,YWG,giordanoetal,lopn,vijay,shi1,pillay,iddir,anke,graciani,shi,cretien}.
Most of the existing theoretical studies provide information on stable
adsorption sites of $\mathrm{Au}$ on the stoichiometric and reduced titania
studies, while less effort has been devoted to the study of the
$\mathrm{O}$--rich 
$\mathrm{Au}$/$\mathrm{TiO_{2}(110)}$
system~\cite{shi,cretien} and the diffusion of $\mathrm{Au}$ adatoms on the
stoichiometric and reduced $\mathrm{TiO_{2}(110)}$ surface~\cite{matthey,iddir,anke}. A
wide variation in the 
lowest--energy
positions of $\mathrm{Au}$ on titania are
reported in the literature~\cite{lopn}, which can be 
explained in part
by considering that $\mathrm{Au}$ can diffuse rather easily 
on the stoichiometric surface~\cite{iddir}. 
The potential energy surface (PES) of a single $\mathrm{Au}$ adatom
deposited on the stoichiometric $\mathrm{TiO_{2}(110)}$ surface or adsorbed
into a surface $\mathrm{O}$ vacancy has been explored using static
calculations~\cite{iddir}. 
It has been shown that $\mathrm{Au}$ migration 
on the stoichiometric surface is 
two--dimensional, with a relatively flat profile.
%
%
In the scenario where an
$\mathrm{Au}$ atom 
is substituted
 for
a surface $\mathrm{Ti_{5c}}$ 
site, 
it has been demonstrated
that the
$\mathrm{Au}$ atom is capable of weakening bonds of 
surface oxygens with the oxide~\cite{cretien}.

Most of the density functional theory studies available in 
the
literature
dealing with defects and/or molecules adsorbed on titania substrates  
using reasonably sized supercells make use of local (LDA) or semilocal (GGA) functionals.
Despite widespread use, such functionals are known to often 
(but not 
always~\cite{kowalski})
 fail to predict 
qualitatively correct electronic structures for reduced transition metal oxides, 
due to the self--interaction error inherent in the functionals. 
To partially 
correct
 for the self--interaction error, different computational methods can be used: 
perturbative 
many--body
theories such as 
``GW''~\cite{gw,gw1,gw2}, 
LDA plus dynamical mean field theory (DMFT)~\cite{dmft}, 
pseudo  
self--interaction--correction
schemes (pSIC)~\cite{psic}, 
LDA plus~U~\cite{lda+1}, 
and other methods that rely
on hybrid functionals~\cite{becke}. 
Recently, GGA+U approaches have
been applied with promising results in studies of intrinsic
electron transport in $\mathrm{TiO_{2}}$ bulk~\cite{deskins2,dupuis} and in the
investigation of the charge (de--)localization dynamics induced by surface oxygen
vacancies on the (110) surface of TiO$_2$ in the rutile structure~\cite{PMD}.
Within the GGA+U approach, the electronic structure is partially corrected for the self--interaction
error by 
adding a Hubbard term acting on the $\mathrm{Ti}$--3{\it d} orbitals.
This approach has the advantage of not adding much computational overhead to 
a standard GGA calculation in the plane wave~/ pseudopotential 
framework,
thus enabling one to use fairly large supercells
in order to allow 
structural relaxation to occur
or to carry out {\em ab initio} dynamics. 
%

In this article, the PBE+U formalism is employed to investigate the structural and
electronic properties of gold-supported $\mathrm{TiO_{2}(110)}$ surface
catalysts and their reactivity towards $\mathrm{CO}$.
As a first step, 
the PBE+U approach is applied to study the interaction between the
stoichiometric or reduced
$\mathrm{TiO_{2}(110)}$ surface and small probe species: $\mathrm{H}$,
$\mathrm{H_{2}O}$, and $\mathrm{CO}$. 
The PBE+U structures and binding energies are compared to
both 
the 
corresponding plain PBE results and reference data in the literature 
in order to assess the
applicability of the
 PBE+U approach for the present purpose. 
When it comes to gold on titania, 
our PBE+U {\em ab initio} molecular dynamics simulations extend existing static
relaxations and nudged elastic band mappings of the PES and demonstrate that
Au~adatoms diffuse 
in a highly directional manner
on the stoichiometric surface.
The metal atoms migrate easily,
either along on top of the bridging oxygen rows or around the area between these
rows, from one bridging position to the next,
along the [001]~direction.
The relative thermodynamic stability of
different $\mathrm{TiO_{2}(110)}$ structures is furthermore studied by employing the
formalism of approximate {\it ab initio} thermodynamics.
Our calculations greatly extend existing studies and show that under $\mathrm{O}$--rich
conditions, the thermodynamically most stable structure is the defective
$\mathrm{Au@V_{Ti5c}}$ surface structure
(obtained by substituting a surface $\mathrm{Ti_{5c}}$ atom with an
$\mathrm{Au}$ adatom), while under $\mathrm{Ti}$--rich conditions,
the $\mathrm{Au}$ adatoms are preferentially adsorbed at
$\mathrm{O}$ vacancies.

The remainder of the paper is organized as follows: 
In Section~\ref{sec:methods} the model system, the methods 
and the computational details are summarized.
The PBE+U formalism is validated 
in Section~\ref{small_molecules}
by studying the interaction of the $\mathrm{TiO_{2}(110)}$ surfaces 
with a set of well-studied adsorbates such as
$\mathrm{H}$, $\mathrm{H_{2}O}$ and $\mathrm{CO}$. 
The main part of the paper is Section~\ref{gold} which presents novel insights into 
the structures, electronic properties, thermodynamics and dynamics of 
gold-promoted rutile $\mathrm{TiO_{2}(110)}$ surfaces and their interaction 
with $\mathrm{CO}$.
The concluding Section~\ref{conclusions} summarizes the main results
and puts them in a broader perspective.

\section{Methods and computational details}
\label{sec:methods} 

The $\mathrm{TiO_{2}(110)}$ surfaces 
were
modeled by
four $\mathrm{O-Ti_{2}O_{2}-O}$ trilayer (4x2)~supercell slabs 
separated by more than 10~\AA\ 
of vacuous space normal to the surface.
The bottom of the slab was
passivated with pseudohydrogen atoms of nuclear charge
$+4/3$ and $+2/3$ in order to achieve 
well--converged
results.
%
This is our so--called ``standard setup'' which 
has been previously carefully constructed~\cite{kowalski} by performing extensive tests
on the convergence of surface energies as well as hydrogen and water 
adsorption energies, with respect to both the number of relaxed outermost trilayers
and the thickness of the slab itself
(see tables and graphs in \cite{kowalski} for detailed comparisons). 
The system size employed in our calculations, corresponding to 208 atoms for the
stoichiometric slab, belongs to the largest systems used so far 
in order to model the surface, in particular when it comes to
performing {\em ab initio} molecular dynamics.
In order to further check the convergence we optimized
two five trilayer slabs (one with an empty surface $\mathrm{O}$ vacancy 
and one with an Au adatom 
at
 this vacancy) 
and confirmed that the resulting spin density and excess charge localization
is the same as reported herein for our ``standard setup''. 

The 
gradient--corrected
Perdew-Burke-Ernzerhof functional (PBE)~\cite{pbe} 
was employed to describe semilocally the 
exchange--correlation
effects. 
The 
spin--polarized
Kohn-Sham equations were solved in
the plane wave~/ pseudopotential framework using
Vanderbilt's ultrasoft pseudopotentials~\cite{vanderbilt}
with a cutoff of 25~Ry using the $\Gamma$--point.
The $\mathrm{Ti}$ pseudopotential was constructed from an ionic 
3{\it d}$^{1}$4{\it s}$^{2}$ configuration and the
3{\it s} and 3{\it p} semicore electrons were treated as full valence states. 
It is 
well established
that adding a Hubbard U~term acting on the 
$\mathrm{Ti}$--3{\it d} orbitals greatly improves the quality of LDA or GGAs in 
describing the electronic structure of both oxidized and reduced 
titania surfaces~\cite{thornton,pacchioni,morgan,morgan1,%
filippone,CHS08,deskins2,dupuis,PMD}. 
Following our previous 
work~\cite{PMD},
we 
used
a 
self--consistent
linear response formalism~\cite{CG05,KC06} 
to compute the Hubbard 
term, 
which turns out to be
U~$= 4.2$~eV for this particular setup;
the occupations of the {\it d}~orbitals were calculated using atomic--like 
wave function projectors. 
%
%
It will not have escaped attention
that our value of the U~parameter is larger than
that
recently derived 
by Mattioli et al.~\cite{mattioli} (i.e. U~$=3.25$~eV). 
This 
can be attributed
 to the different {\it d}-orbitals used as projectors for 
the integration of the {\it d}-orbital occupation numbers.
%
The U~value of $3.25$~eV obtained in~\cite{mattioli} was derived using the
{\it d}-orbital of the neutral Ti~atom as a projector, whereas our value, 
U~$=4.20$~eV, is computed by using Ti$^{+1}$ as 
a
reference,
which in our opinion more closely resembles the charge state of Ti in the TiO$_{2}$(110) surface. 
%
At this point it should be noted
that we
also
 obtained U~$=3.20$~eV when using 
the {\it d}-orbital of the neutral Ti atom as the projector instead,
which is in agreement with the U~value reported in~\cite{mattioli}.
Similarly, performing the calculations of FeO Pickett et al.~\cite{pickett} 
showed that the calculated Hubbard energy strongly depends on the choice of {\it d}-orbital.
Using the {\it d}-orbital of the neutral Fe atom they obtained U~$=4.6$~eV 
whereas using the Fe$^{+2}$ dication yields a substantially larger value of U~$=7.8$~eV.
%
This is a well-known and still poorly understood shortcoming
of the U~parameter derivation procedures that use atomic-like {\it d}-orbitals
as projectors for the integration of the occupations of the {\it d}-orbitals in
solids,  
which, in turn,
is an input for the computation of the Hubbard correction to LDA/GGA
density functionals~\cite{CG05}.
%
The static optimizations for the different $\mathrm{TiO_{2}(110)}$ surface structures
were
carried out using the {\tt Quantum} {\tt Espresso}~\cite{QE} code.
All structures were relaxed by minimizing the atomic forces, where convergence
was assumed 
to have been achieved
when the maximum component of the residual forces on the ions was
less than 0.02~eV/\AA . 
%
Here, only the lowest tri–layer atoms
were constrained to their equilibrium positions while all
other atoms 
were
 free to move during optimization. 
%
All $\it{ab}$ $\it{initio}$ molecular dynamics (AIMD) simulations~\cite{dominik} 
were carried out using the same 
spin--polarized
PBE+U approach, together
with the Car--Parrinello propagation scheme,~\cite{carparrinello}
using a fictitious electron mass of 700~a.u. and a time step of 0.145~fs. 
%
Our in-house modified version of the {\tt CPMD}~\cite{cpmd} code was used for this purpose.

The adsorption energy $E_{\rm ads}^{\rm H}$ per H~atom on the stoichiometric
$\mathrm{TiO_{2}(110)}$ surface is computed from
%
\begin{equation}
E_{\rm ads}^{\rm H} = \frac{1}{N_{\rm H}} \Big[ E_{\rm tot}^{\rm
  H-ads}(N_{\rm H})- \Big(
E_{{\rm tot}}^{\rm slab-\rm{TiO_{2}}}+\frac{N_{\rm H}}{2}
E_{\rm tot}^{{\rm H}_2} \Big) \Big]
\enspace , 
\end{equation}
where $E_{\rm tot}^{\rm H-ads}(N_{\rm H})$ is the total energy of the
slab saturated with $N_{{\rm H}}$ H~adatoms; here $E_{{\rm tot}}^{\rm slab-\rm{TiO_{2}}}$ 
is the energy of the stoichiometric slab, which we take as a reference, and
$E_{\rm tot}^{\rm H_{2}}$ is the energy of a H$_2$ molecule.
When dealing with $\mathrm{CO}$, $\mathrm{H_{2}O}$, and 
Au,
the adsorption 
energies on stoichiometric and reduced
$\mathrm{TiO_{2}(110)}$ surfaces 
were
calculated according to
\begin{eqnarray}
E_{{\rm ads}}= E_{{\rm tot}}^{\rm sub + \rm X }-\left( E_{{\rm tot}}^{\rm sub} + 
E_{{\rm tot}}^{\rm X} \right) 
\enspace , 
\end{eqnarray}
%
where $E_{{\rm tot}}^{\rm sub + \rm X }$, $E_{{\rm tot}}^{\rm sub}$, and
$E_{{\rm tot}}^{\rm X}$ are the total energies of the combined system, the
($\mathrm{Au}$/)$\mathrm{TiO_{2}(110)}$ surface in
a certain oxidation state, and the isolated $\rm{X}$ adsorbate,
respectively.
The adsorption energies 
were
calculated with and without
inclusion of the Hubbard U~term
correction
 to the standard density functional,
i.e. using the plain PBE and the PBE+U approaches.
%
The $\mathrm{O}$-vacancy formation energy 
was
calculated using 
%
%
%
%
%
\begin{eqnarray}
E_{\rm{V}}^{\rm{O}}=
E_{\rm{tot}}^{\rm{O-V}} - \left(E_{\rm tot}^{\rm slab -\rm{TiO_{2}}}
 - \frac{1}{2}E_{\rm tot}^{\rm{O_{2}}} \right)
\enspace ,
\end{eqnarray}
%
%
where $E_{\rm{slab}}^{\rm{O-V}}$ and
$\frac{1}{2}E_{\rm tot}^{\rm{O_{2}}}$ represent
the total energy of the defective system and of
the $\mathrm{O}$ atom, respectively. 
%
%
%
%
%
Because (semi)local functionals are known to overbind molecular $\mathrm{O_{2}}$, 
the total energy of the $\mathrm{O}$ atom was adjusted in the manner of our previous work~\cite{kowalski}.

%
In order to analyze the thermodynamic stability of our different structures in
the
presence of $\mathrm{H}$ 
adatoms,
we employ the formalism of approximate
$\it{ab}$ $\it{initio}$
thermodynamics~\cite{thermo,thermo1,reuter,thermo2,thermo4}
by assuming that the surfaces can exchange
$\mathrm{H}$ atoms with a surrounding gas phase. Assuming thermodynamic
equilibrium, the most stable surface composition at a given temperature $T$ and
pressure $p$ is given by the minimum of the surface Gibbs free energy.
Since we are only interested in the relative stabilities of
surface structures, we 
directly compute
 the differences in the surface Gibbs
free energies $\Delta G_{\rm ads} (T,p)$ between the defective and the ideal surface
according to
\begin{eqnarray} 
\label{eq3.9}  
\begin{split}
\Delta G_{\rm ads} (T,p)= & 
\frac{1}{A} \Bigg[ E_{\rm tot}^{\rm H-ads}(N_{\rm H})
- \\ &  \Big( E_{{\rm tot}}^{\rm slab-\rm{TiO_{2}}}+ \Delta N_{\rm H}\mu_{{\rm{H}}}(T,p) 
 \Big) \Bigg]
\enspace , 
\end{split}
\end{eqnarray}
where $A$ is the surface area, $\Delta N_{\rm H}$ is the difference in the
number of $\mathrm{H}$ atoms between the two surfaces, and
$\mu_{{\rm{H}}}(T,p)$ is the chemical potential representing the Gibbs free
energy of the gas phase with which the $\mathrm{H}$ atoms are
exchanged. 
Assuming that all differences in entropy and volume contributions
in $\Delta G_{\rm ads} (T,p)$ are negligible, the Gibbs free energies 
are
approximated
by their respective total energies of our DFT slab calculations as
usual~\cite{reuter,thermo2,thermo4}. 
The upper bound for the chemical potential $\mu_{{\rm{H}}}(T,p)$ is
given by the total energy of its most stable elemental phase~\cite{thermo1}, that is,
molecular hydrogen ($\frac{1}{2}E_{\rm tot}^{\rm H_{2}}$). This upper bound is
taken as the zero of our energy scale by using $\Delta
\mu_{{\rm{H}}}=\mu_{{\rm{H}}}(T,p)-\frac{1}{2}E_{\rm tot}^{\rm H_{2}}$. 
%
%
%
%

In a similar way, the effect of temperature and pressure on the relative stability of the
$\mathrm{Au}$/$\mathrm{TiO_{2}(110)}$ surface structures is studied by
employing the formalism of approximate $\it{ab}$ $\it{initio}$
thermodynamics~\cite{thermo,thermo1,reuter,thermo2,thermo4}. 
%
%
The free energy of formation of the $\mathrm{Au}$/$\mathrm{TiO_{2}(110)}$ surface structures
$\Delta_{\rm ads} G(T,p)$ is assumed to depend on the temperature and pressure
only via the oxygen chemical potential $\mu_{{\rm O}}(T,p)$ given by
\begin{eqnarray}
\label{eq4}
\mu_{\rm{O}}(T,p)=\mu_{\rm{O}}(T,p^{0})+\frac{1}{2}kT\ln\bigg(\frac{p}{p^{0}}\bigg)
\enspace . 
\end{eqnarray}
Equation~\eqref{eq4} represents the thermodynamics reservoir of the
$\mathrm{O_{2}}$ environment that is in contact with 
the
surface under
consideration. 
The free energy differences will be calculated as a function of
$\Delta \mu_{\rm{O}}(T,P)=\mu_{\rm{O}}(T,p)-\mu_{\rm{O}}(T=0\mathrm{~K},p^{0})$, 
corresponding to changes of the oxygen chemical potential with respect 
to a zero reference state. 
%
The latter is set to the total energy of the $\mathrm{O}$ atom at
$T=0$~K, $\mu_{\rm{O}}(T=0\mathrm{~K},p^{0})=1/2E_{\rm tot}^{\rm{O_{2}}}=0$. 
%
%
%
%
Assuming thermodynamic equilibrium of the surfaces with an
$\mathrm{O_{2}}$ gas phase, the chemical potential can be converted into a
pressure scale for different temperatures by using experimental thermochemical
reference data or by applying the ideal gas
equation~\cite{reuter,thermo2,thermo4}.
%
%
%
%
Vibrational and rotational entropic contributions to $\mu_{\rm{O}}(T,p)$
are included by means of thermodynamic tables as described in Ref.~\onlinecite{reuter}. 
Under these assumptions and neglecting entropic contributions of the solids
involved, the free energy of formation as a function of pressure and
temperature assumes the 
expression
%
%
%
%
%
%
%
\begin{eqnarray}
\label{eq5}
\begin{split}
\Delta G_{\rm ads} (T,p) & = \frac{1}{A}[E_{{\rm tot}}^{\rm sub + \rm X}-E_{{\rm
    tot}}^{\rm slab-\rm{TiO_{2}}}+ N_{{\rm{O}}}^{{\rm{v}}}\mu_{{\rm{O}}}(T,p) +\\
 & \quad N_{{\rm{Ti}}}^{{\rm{v}}}[E_{{\rm{TiO_{2}}}}^{{\rm{bulk}}}-2\mu_{{\rm{O}}}(T,p)]-\mu_{{\rm{Au}}}]
\enspace ,
\end{split}
\end{eqnarray}
%
%
%
where $E_{{\rm{TiO_{2}}}}^{{\rm{bulk}}}$ is the energy of a formula unit 
of the $\mathrm{TiO_{2}}$ bulk phase. 
The quantities $N_{\rm{O}}^{\rm{v}}$ and $N_{\rm{Ti}}^{\rm{v}}$ 
represent
the number of $\mathrm{O}$ or $\mathrm{Ti}$ vacancies that are present in
the structure under consideration. 
%
Therefore,
the energy cost for the formation of surface defects
is taken into account in Eq.~\eqref{eq5} via the chemical potential of
$\mathrm{O}$ atoms and of bulk $\mathrm{TiO_{2}}$. 
Finally, the chemical potential of $\mathrm{Au}$, $\mu_{{\rm{Au}}}$, is set 
to be the total energy per atom of the bulk $\mathrm{Au}$ crystal.
The upper bound for the chemical potential $\mu_{\rm{O}}$ is given by the total energy
of its most stable elemental phase, that is, 
molecular oxygen $(\frac{1}{2}E_{\rm tot}^{\rm{O_{2}}})$. 
This upper bound is 
used to define 
$\Delta \mu_{\rm{O}}(T,P)=\mu_{\rm{O}}-\frac{1}{2}E_{\rm tot}^{\rm{O_{2}}}$. 
A lower bound for $\Delta \mu_{\rm O}$ is given by minus half 
of the formation energy of bulk $\mathrm{TiO_{2}}$,
i.e. $E_{\rm{f}}^{\rm{TiO_{2}}}=E_{\rm{bulk}}^{\rm{Ti}}+E_{\rm tot}^{\rm{O_{2}}}-E_{\rm{bulk}}^{\rm{TiO_{2}}}$,  
for which we have taken
the theoretical value of 4.8~eV from our PBE calculations;
here $E_{\rm{bulk}}^{\rm{TiO_{2}}}$ and $E_{\rm{bulk}}^{\rm{Ti}}$ are the
energies of one bulk unit cell of $\mathrm{TiO_{2}}$  and metallic $\mathrm{Ti}$,
respectively~\cite{reuter,thermo2}.


%

%
Finally, the bonding charge density has been evaluated using the 
expression
%
\begin{eqnarray}
\Delta \rho (\vec{r}) =
\rho_{\rm sub + \rm{X}} - ( \rho_{\rm{sub}} + \rho_{\rm{X}})
\end{eqnarray}
where
the $\rho$'s are the respective valence electronic charge densities
at position $\vec{r}$ in space.
%

\section{%
Small molecules on titania: Reference calculations using PBE+U}
\label{small_molecules}
%

Computationally efficient implementations of DFT based on local/semilocal LDA/GGA
density functionals predict rather delocalized defect levels for excess electrons 
in the case of reduced transition metal oxides in general, and for titania in particular.
Thus, more sophisticated techniques such as hybrid
functionals or GGA+U approaches are necessary 
to properly account for the strong correlation effects of these 
{\it d}-electrons,
resulting in localization of the excess charge on
 3{\it d}-orbitals
 of reduced Ti~atoms.
%
However, before using PBE+U to investigate the properties of gold on titania, 
it is 
necessary
 to confirm that this approach
does not 
destroy
 the
  agreement 
between previously reported plain PBE results
(mainly optimized structures and relative energies) 
and experimental observations.
%
The purpose of this section is therefore twofold: 
first,
 to validate the PBE+U approach 
 using
  a test set that probes the physics and chemistry
of titania surfaces interacting with adspecies relevant to heterogeneous catalysis;
and, second, 
 to check how PBE+U performs compared to plain PBE investigations
for the very same systems. 
Thus, the interaction of $\mathrm{H}$, $\mathrm{H_{2}O}$, and $\mathrm{CO}$ with
the $\mathrm{TiO_{2}}$(110) substrate is investigated
in the following 
section
using the PBE+U approach where a large set of 
both experimental and previous theoretical data are available. 
%

%
Many theoretical and experimental studies have been devoted to understanding
the interaction between hydrogen
atoms~\cite{kowalski,kunat,suzuki,fujino,YC08,divalentin} or
water~\cite{roger-piss,selloni-review-2010,kowalski,brinkley,desegovia,%
goniakowski,gillan,bates,gillan1,truong,langel,zhang1,zhang2,zhang3,%
perron,lui2010,hugenschmidt,ExW7,ExW2,ExW3} 
and the TiO$_2$(110) surface.
In experiments, hydrogen atoms, when adsorbed on TiO$_2$(110), stick to the bridging oxygens
and a maximum surface saturation limit of $\sim 0.7$~ML is observed~\cite{YC08}. 
These findings have been confirmed on purely theoretical grounds within the framework 
of standard GGA calculations using the PBE functional~\cite{kowalski}.  
In--depth
{\it ab initio} thermodynamics considerations reveal a maximum saturation level
of hydrogen on this oxide surface   
of about 60--70\%, 
in excellent agreement with the above--mentioned experimental observations. 
Adsorption of hydrogen on the stoichiometric surface results in its reduction
by introducing one electron  
per adsorbed H~atom into the substrate.
We first consider the interaction of $\mathrm{H}$ with the stoichiometric
$\mathrm{TiO_{2}(110)}$ surface.
The hydroxylated surface has been investigated for a wide range of H~coverages
using both PBE+U and PBE.  
In agreement with both plain PBE calculations and experimental  results,
we find that $\mathrm{H}$ atoms preferentially adsorb on top of surface 
$\mathrm{O_b}$ 
atoms of the stoichiometric titania surface, leading to the
formation of $\mathrm{OH}$ groups with $\mathrm{O}$--$\mathrm{H}$ bond
lengths of $\sim 1$~\AA . 
%
In addition, our PBE+U calculations predict that one
electron per $\mathrm{H}$ atom adsorbed on the surface is transferred to the
substrate, leading to the formation of $\mathrm{Ti^{3+}}$ ions (see Table~\ref{tab2}).
%
The newly-formed $\mathrm{OH}$ groups are 
found to be 
tilted
by about 20--50$^\circ$ in opposite [1$\bar{1}$0] 
directions,
as previously reported. 
The PBE and PBE+U values of the adsorption energies
$E_{\rm{ads}}^{\rm{H}}$ per $\mathrm{H}$ atom at different coverages are
reported in Table~\ref{tab2}.
The ground state configuration of the fully hydroxylated titania surface  
yields 
(2$\times$1) symmetry, but at room temperature 
$\mathrm{OH}$ groups will be fully 
disordered 
with respect to their axes 
because of the tiny barrier that 
must
be overcome in order to flip their orientation.
%
%
%

\begin{figure}[h]
\begin{center}
\begin{tabular}{cc}
\subfigure{\includegraphics[angle=0,clip=true,scale=0.35]{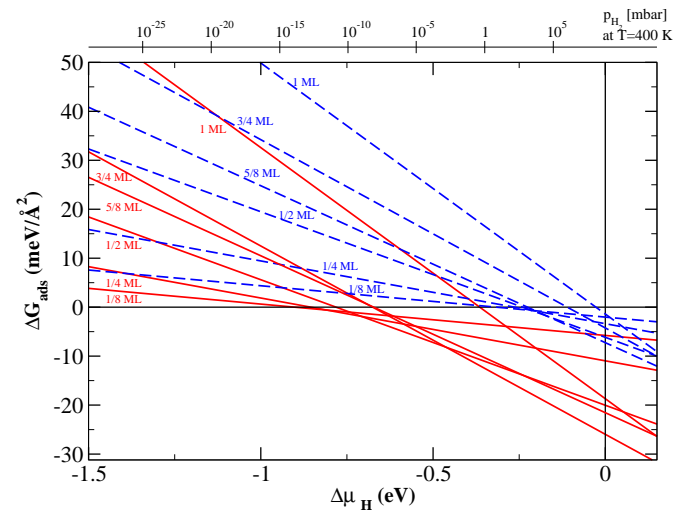}}
\end{tabular}
\caption{%
%
%
%
%
  Free energy $\Delta G_{\rm ads}(T,p)$ for $\mathrm{H}$ adsorption on $\mathrm{TiO_{2}}$(110) 
  stoichiometric surfaces with different hydrogen coverages as a function of the hydrogen
  chemical potential $\Delta \mu_{\rm{H}}$. 
  Conversion to hydrogen partial pressures $p_{\rm{H_{2}}}$ (upper axis) has been
  carried out at $T=400$~K (see text). 
  The red and blue lines represent the PBE+U and PBE results, respectively. \label{fig3}}
\end{center}
\end{figure}

At all  
coverages,
the 
PBE+U adsorption energies are found to be significantly lower,
by about $-0.3$ to $-0.6$~eV, compared to the PBE
data  
in
Table~\ref{tab2}.
However, the same stability trend 
obtained by employing the PBE+U functional 
is obtained using the standard PBE functional,
i.e. a decrease 
of the adsorption energy with increasing coverage.
Nevertheless, even at full monolayer coverage, the adsorption energy per atom 
is still significant and negative when using PBE+U, 
whereas it is close to zero according to PBE. 
Clearly, such total energy considerations 
need to be supplemented 
with 
{\em ab initio} thermodynamics in order to check the thermodynamic
stability of the surface at different saturation levels. 
The results are summarized in terms of the surface free energy diagram 
in Fig.~\ref{fig3}.                                  
Both PBE+U and PBE predict that 
the fully hydrogenated surface will
be thermodynamically unstable at all accessible hydrogen partial pressures. 
PBE+U and PBE also agree 
in that saturation of this surface by hydrogen 
is reached at a coverage on the order of 
70\%. 
Thus, the previous PBE predictions~\cite{kowalski} and 
agreement
with experimental observations~\cite{YC08} upon hydroxylating TiO$_2$(110)
are qualitatively confirmed.
(for reasoning, see the detailed conceptual discussion in Ref.~\onlinecite{kowalski}).
However,
significant strengthening of the adsorption 
and shifts in the surface phase diagram are observed when using 
PBE+U, 
thus localizing 
the excess charges upon introducing the Hubbard correction.

\begin{table}
\caption{\label{tab2}
Adsorption energies ${E_{\rm ads}}$ (in eV) per $\mathrm{H}$ atom for adsorption of hydrogen at different
coverages as indicated. 
In the second column the number of reduced $\mathrm{Ti^{3+}}$ 
ions which are present in the substrate is reported.}
\begin{ruledtabular}                                           
\begin{tabular}{cccc}                                           
Configuration  & ${E_{\rm ads}}$(PBE+U)   &  $\mathrm{Ti^{3+}}$ & ${E_{\rm ads}}$(PBE) \\                         
\hline                                                         
1H  & $-0.91$ & 1 & $-0.56$ \\                                         
2H  & $-0.86$ & 2 & $-0.40$ \\                                         
4H  & $-0.77$ & 4 & $-0.24$ \\                                         
6H  & $-0.68$ & 6 & $-0.15$ \\                                         
8H  & $-0.37$ & 8 & $-0.04$ \\                                         
\end{tabular}                                                  
\end{ruledtabular}                                             
\end{table}

It is well-known that the investigated surface contains a significant number of
oxygen vacancies ($\sim 5$~{\%}), not just in ill-defined industrial situations,
but even under well-controlled experimental conditions.~\cite{diebold}
Therefore, the investigation of the adsorption of water molecules on the
reduced TiO$_2$(110) surface is of great importance in the frame of addressing
the fundamental issue of dissociative {\em versus} molecular adsorption
modes~\cite{kowalski,lui2010,roger-piss,selloni-review-2010}.
%
We thus consider the interaction between $\mathrm{H_{2}O}$ and the $\mathrm{TiO_{2}}$ surface. 
We have performed systematic 
PBE+U and PBE calculations of $\mathrm{H_{2}O}$ adsorbed on the reduced $\mathrm{TiO_{2}}$ 
surface,
considering a titania surface containing a single $\mathrm{V_O}$ vacancy;
this greatly extends our  
recent comprehensive
work~\cite{kowalski} 
concerning
water on the stoichiometric surface using plain PBE. 
%
%

\begin{figure}
\begin{center}
\begin{tabular}{ccc}
\subfigure[Site
labeling]{\includegraphics[width=3.0cm]{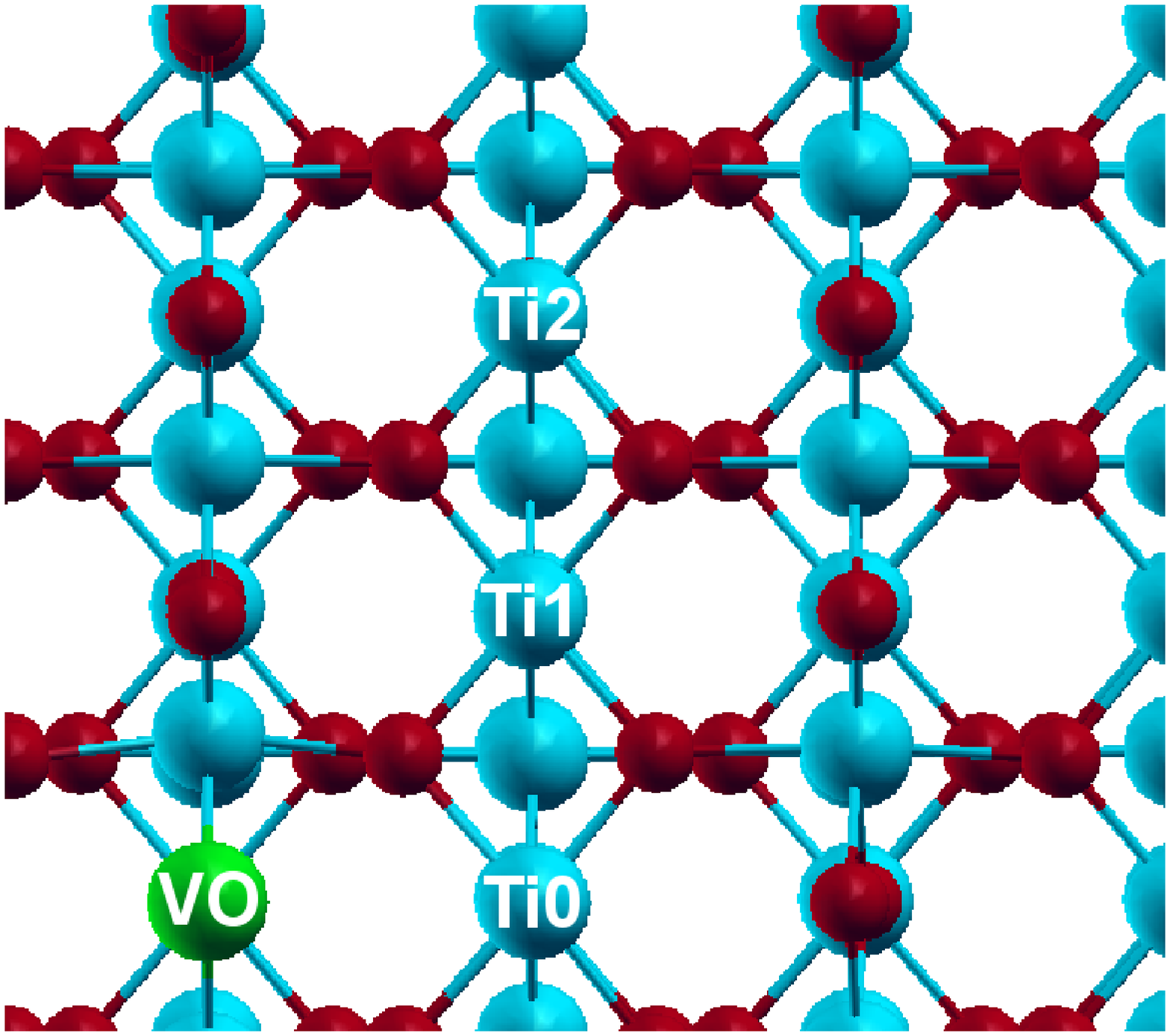}}\hspace{0.50cm}
\subfigure[$\mathrm{H_{2}O}$ at $\mathrm{O_{v}}$ site]{\includegraphics[width=3.0cm]{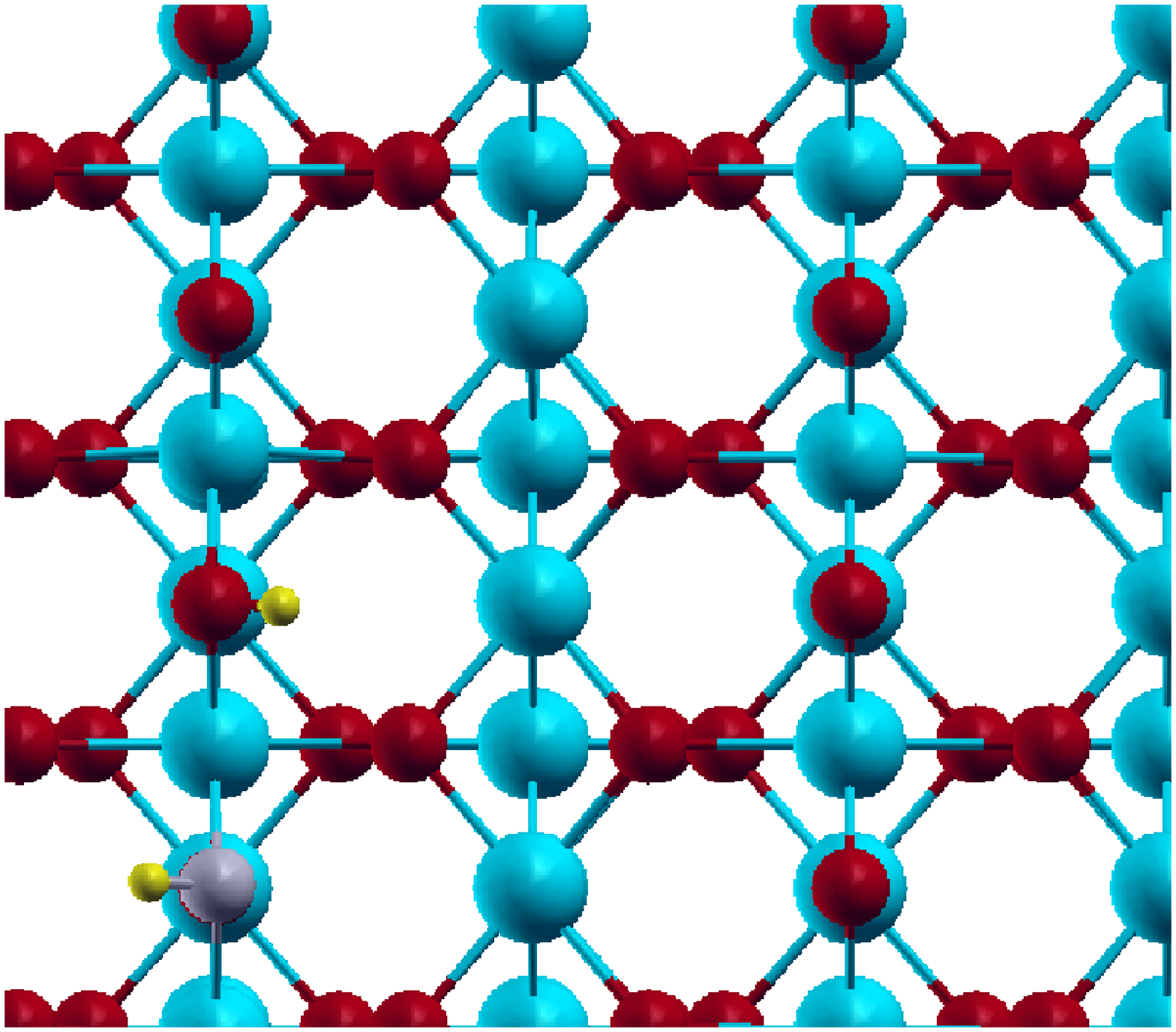}}\\
\subfigure[{\bf DB} $\mathrm{H_{2}O}$ at site {\bf 0}]{\includegraphics[width=3.0cm]{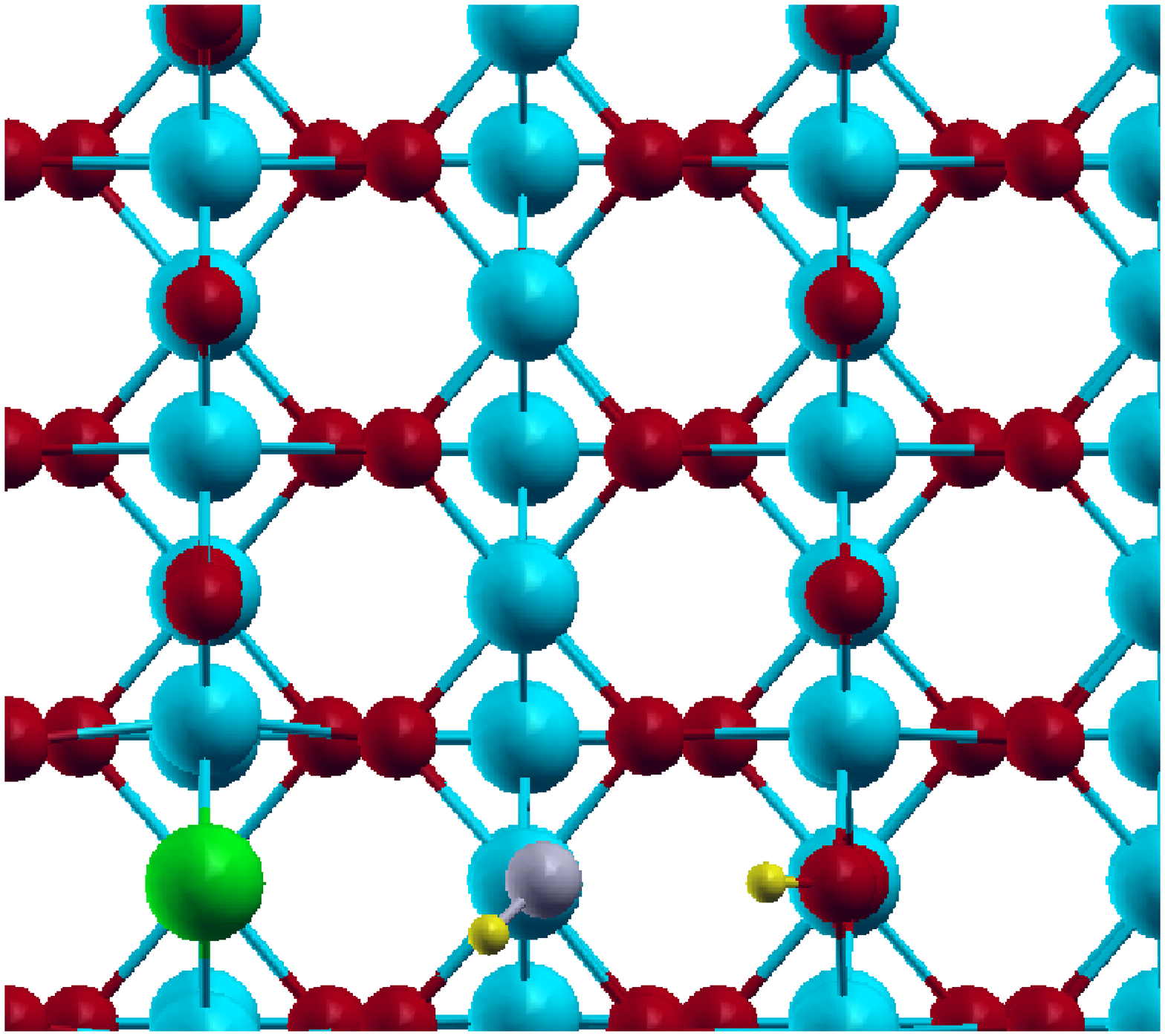}}\hspace{0.50cm}
\subfigure[{\bf DA} $\mathrm{H_{2}O}$ at site {\bf 0}]{\includegraphics[width=3.0cm]{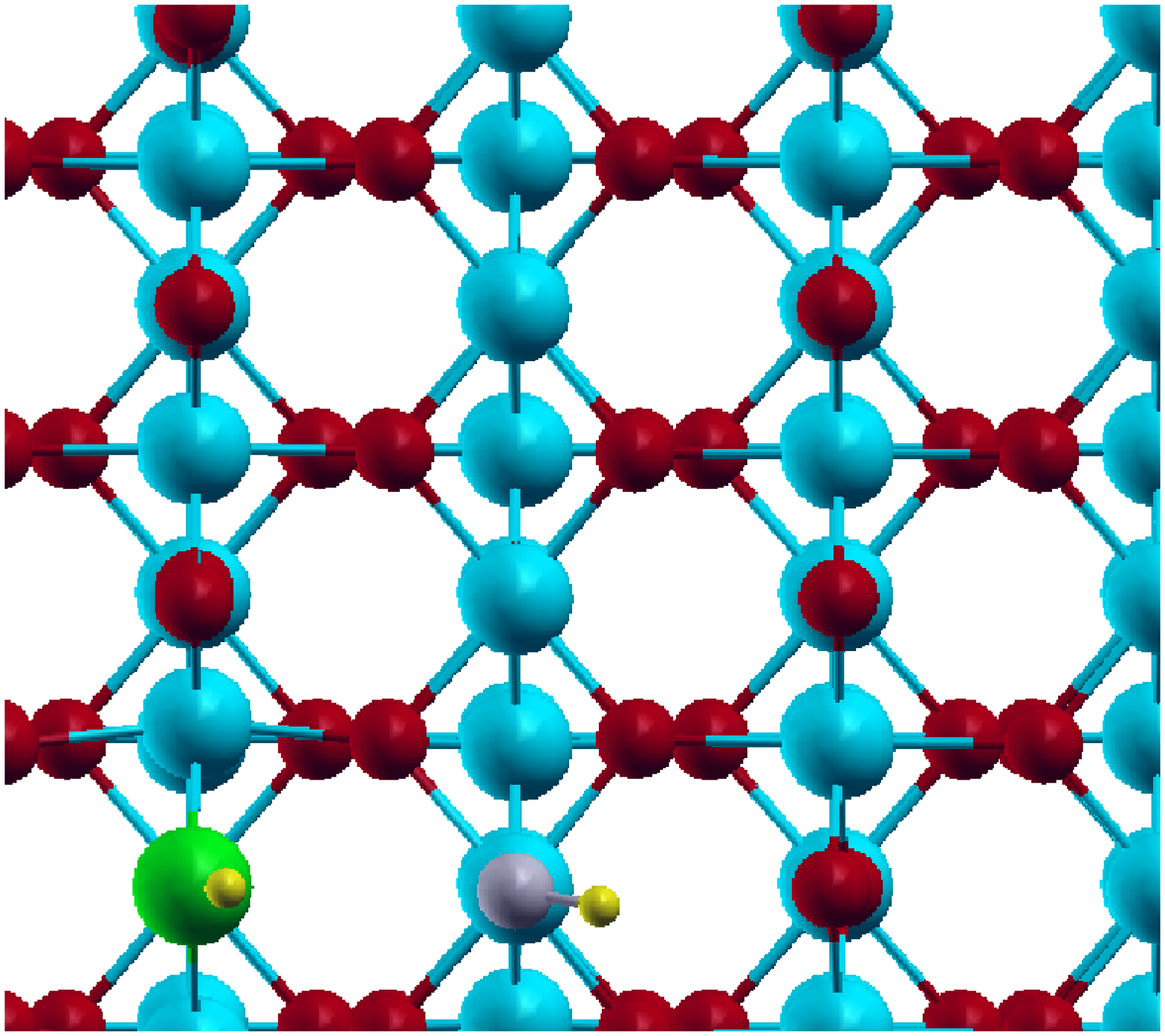}}\\
\subfigure[{\bf MA} $\mathrm{H_{2}O}$ at site {\bf 0}]{\includegraphics[width=3.0cm]{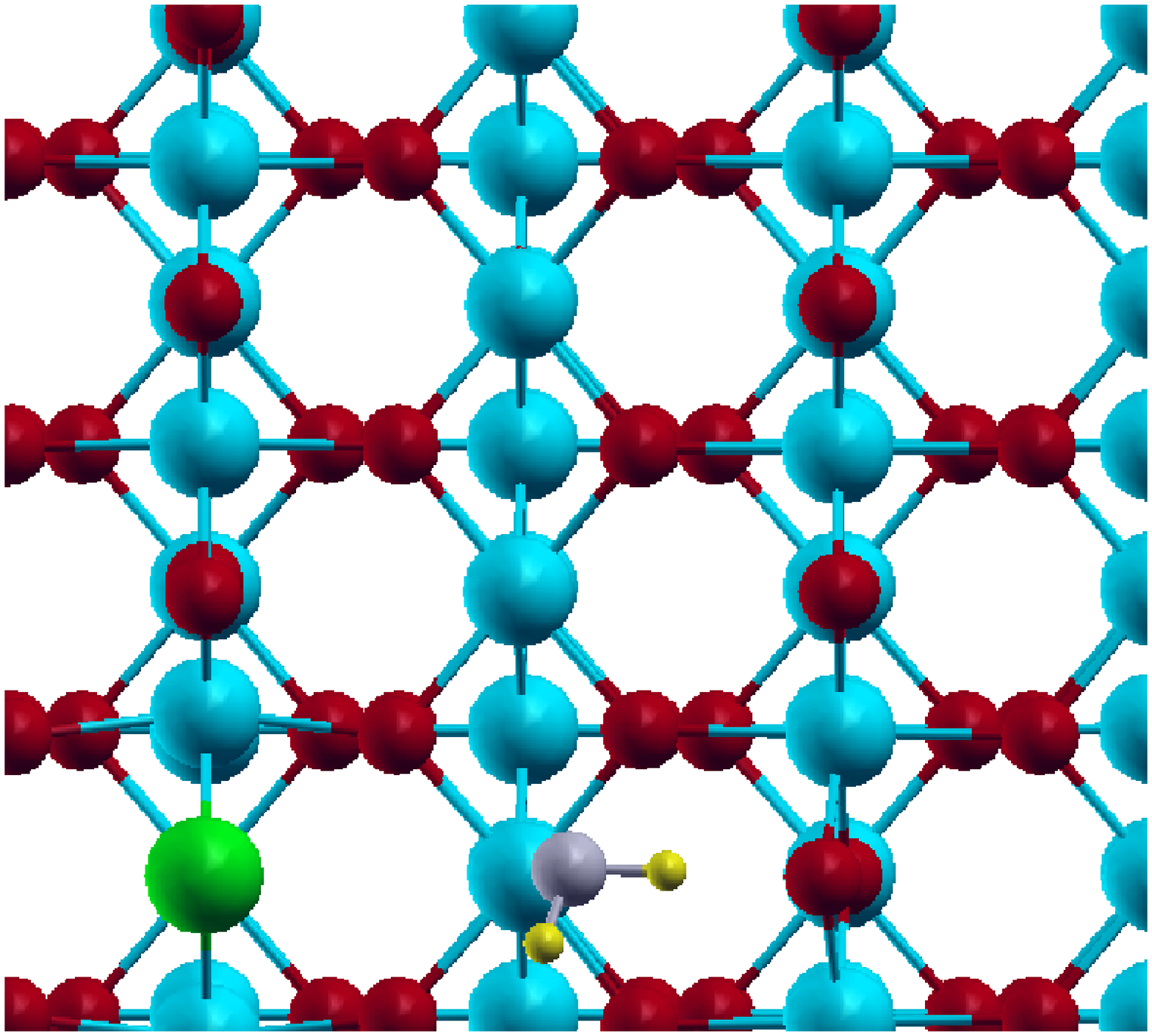}}\hspace{0.50cm}
\subfigure[{\bf MB} $\mathrm{H_{2}O}$ at site {\bf 0}]{\includegraphics[width=3.0cm]{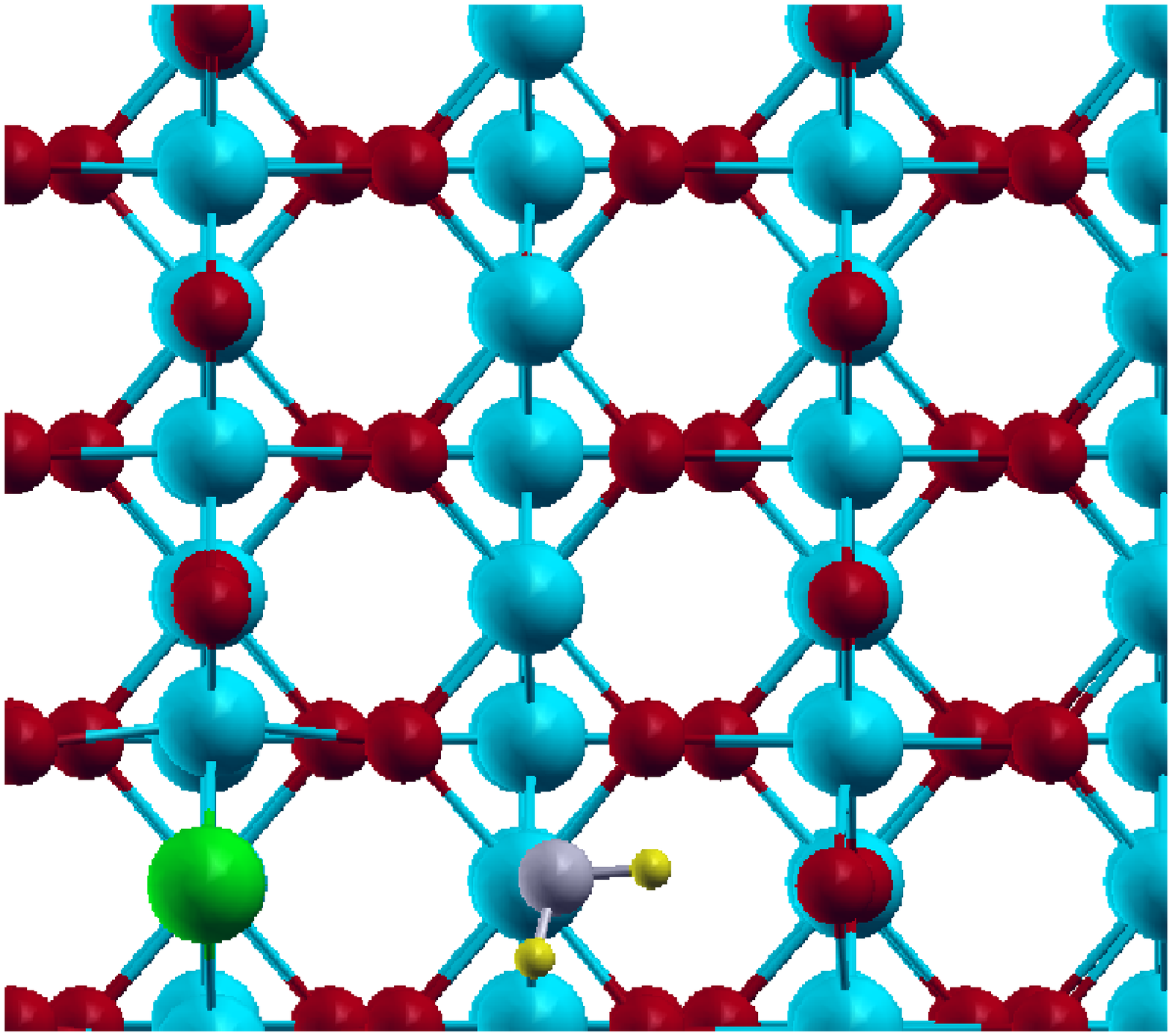}}\\
\subfigure[{\bf DB} $\mathrm{H_{2}O}$ at site {\bf 1}]{\includegraphics[width=3.0cm]{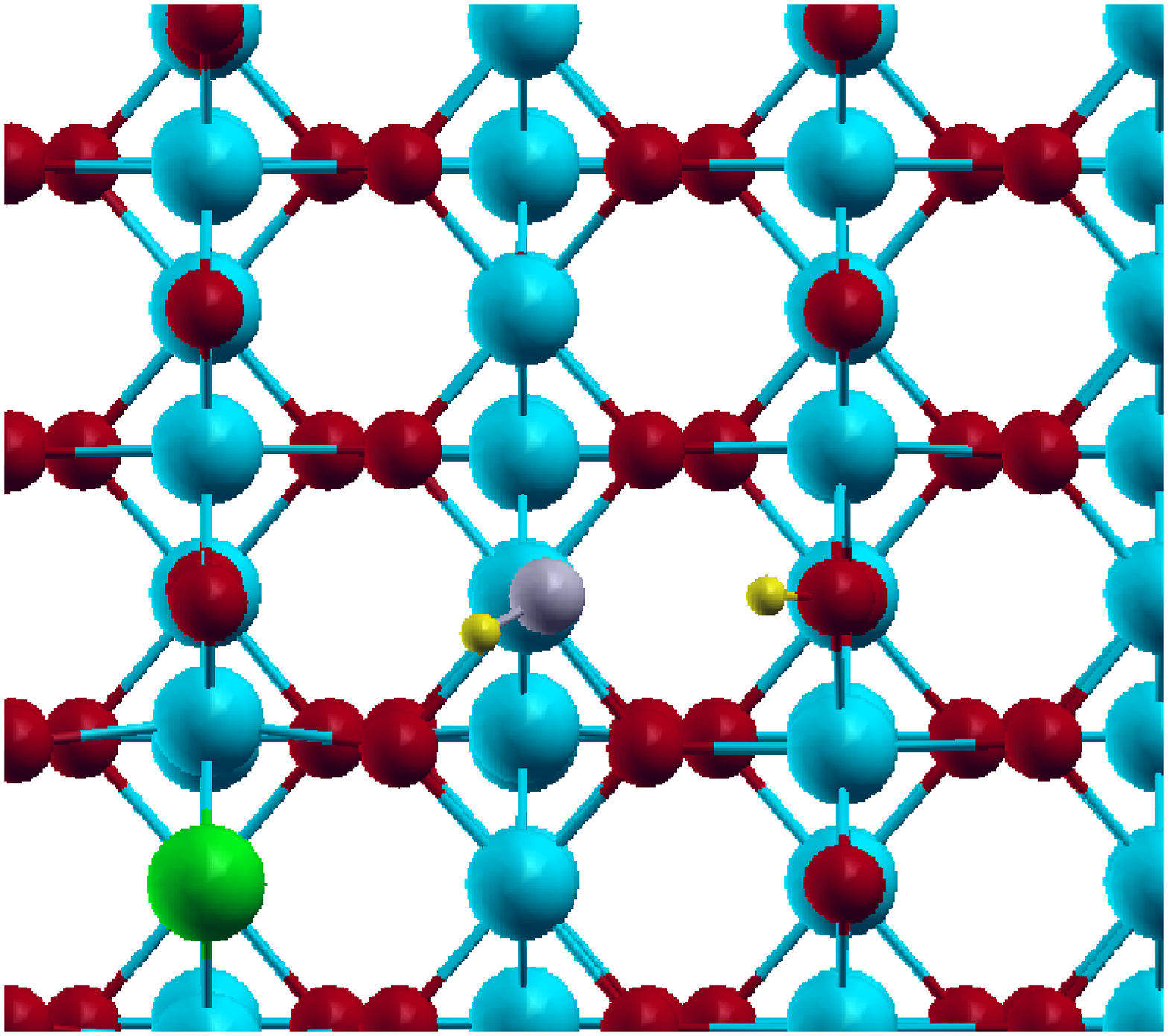}}\hspace{0.50cm}
\subfigure[{\bf DA} $\mathrm{H_{2}O}$ at site {\bf 1}]{\includegraphics[width=3.0cm]{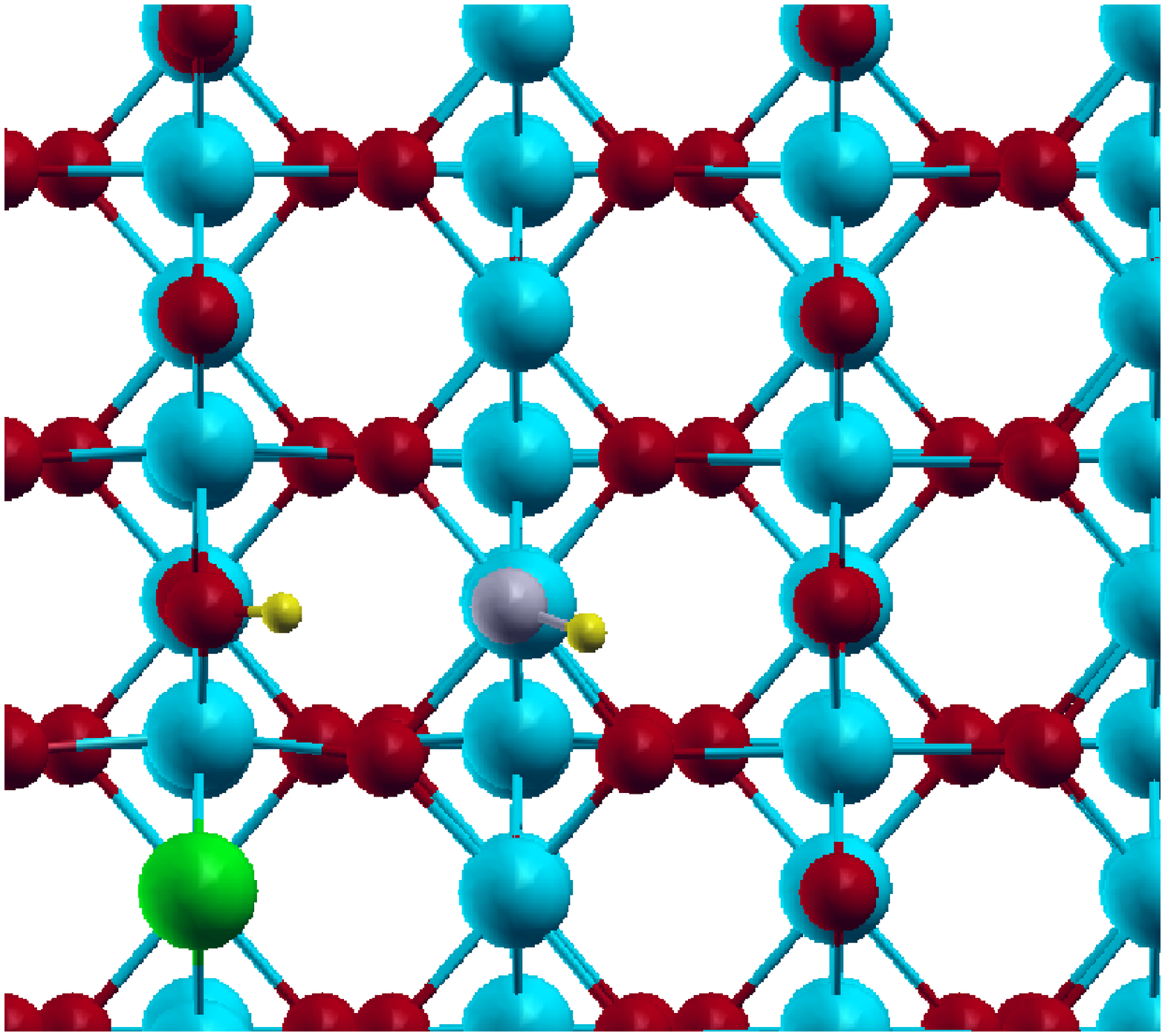}}\\
\subfigure[{\bf MA} $\mathrm{H_{2}O}$ at site {\bf 1}]{\includegraphics[width=3.0cm]{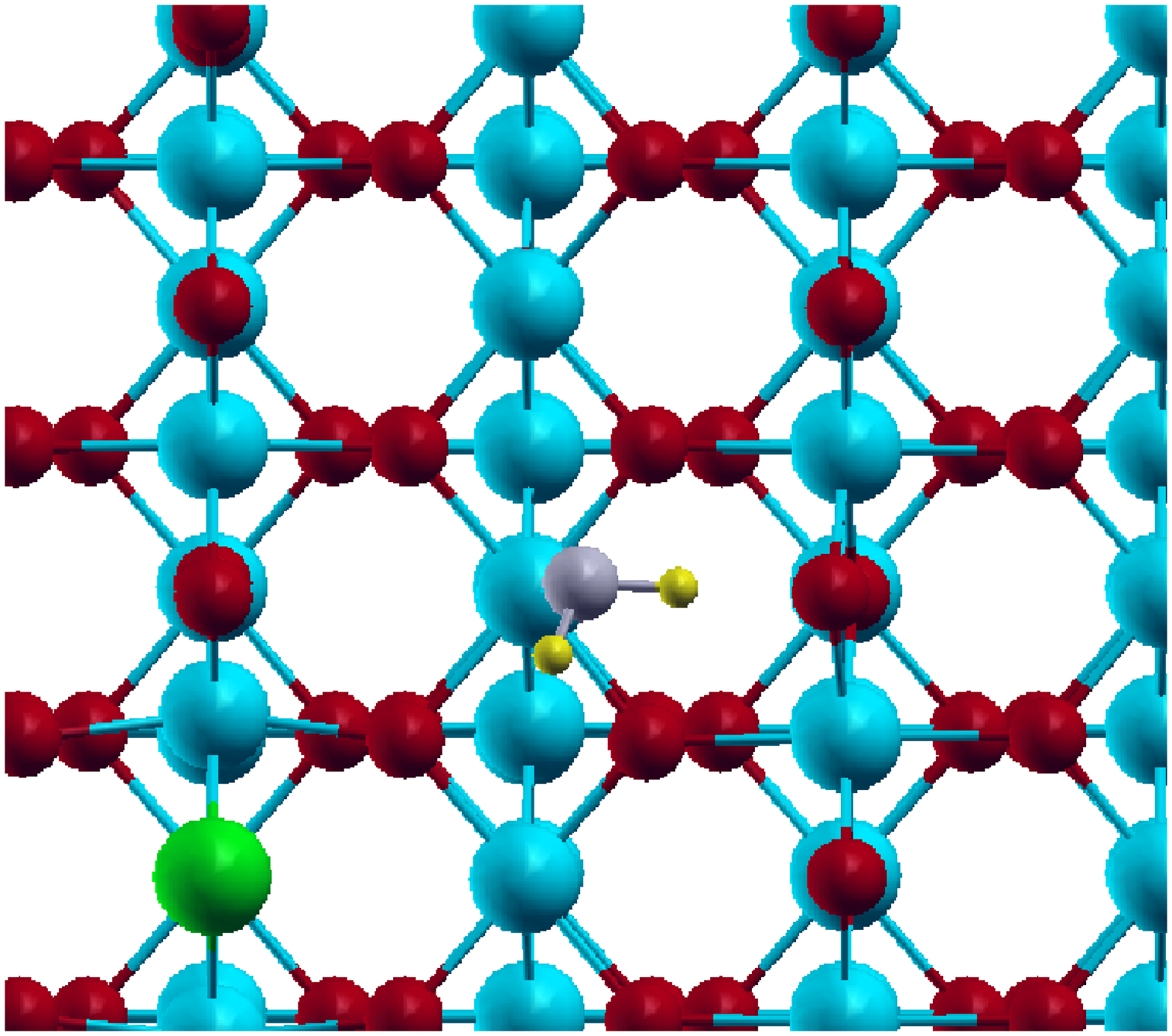}}\hspace{0.50cm}
\subfigure[{\bf MB} $\mathrm{H_{2}O}$ at site {\bf 1}]{\includegraphics[width=3.0cm]{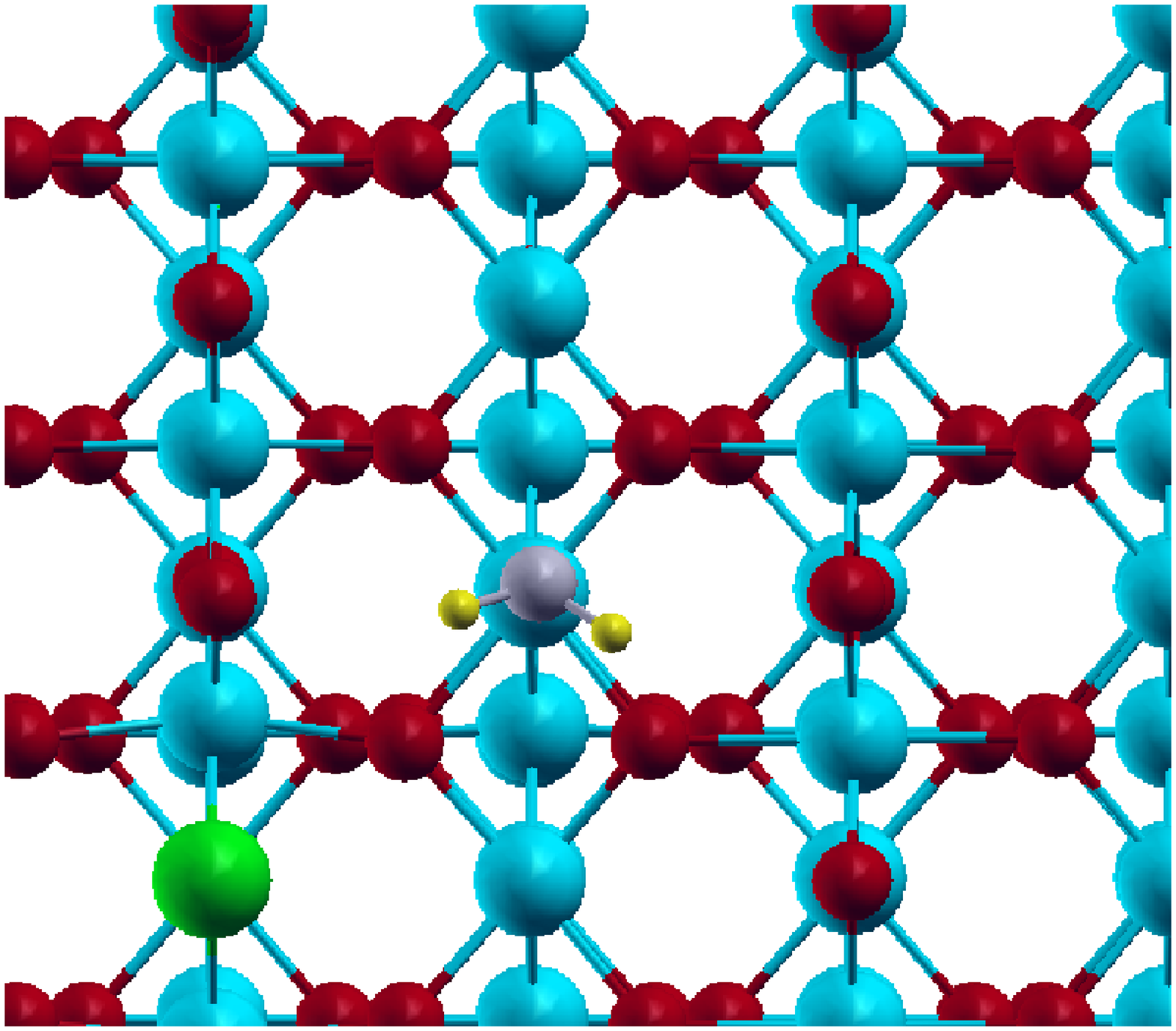}}\\
\subfigure[{\bf DB} $\mathrm{H_{2}O}$ at site {\bf 2}]{\includegraphics[width=3.0cm]{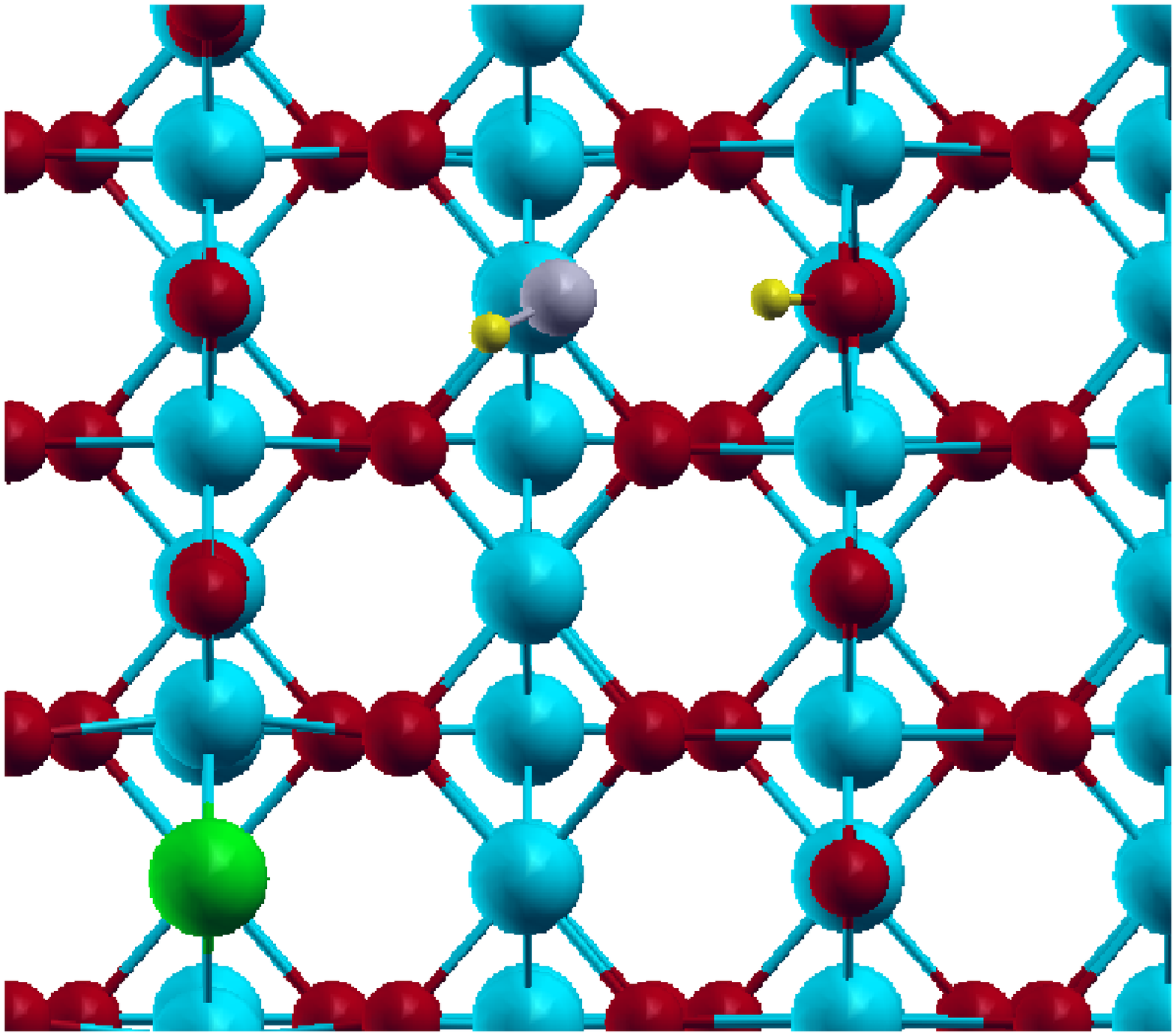}}\hspace{0.50cm}
\subfigure[{\bf DA} $\mathrm{H_{2}O}$ at site {\bf 2}]{\includegraphics[width=3.0cm]{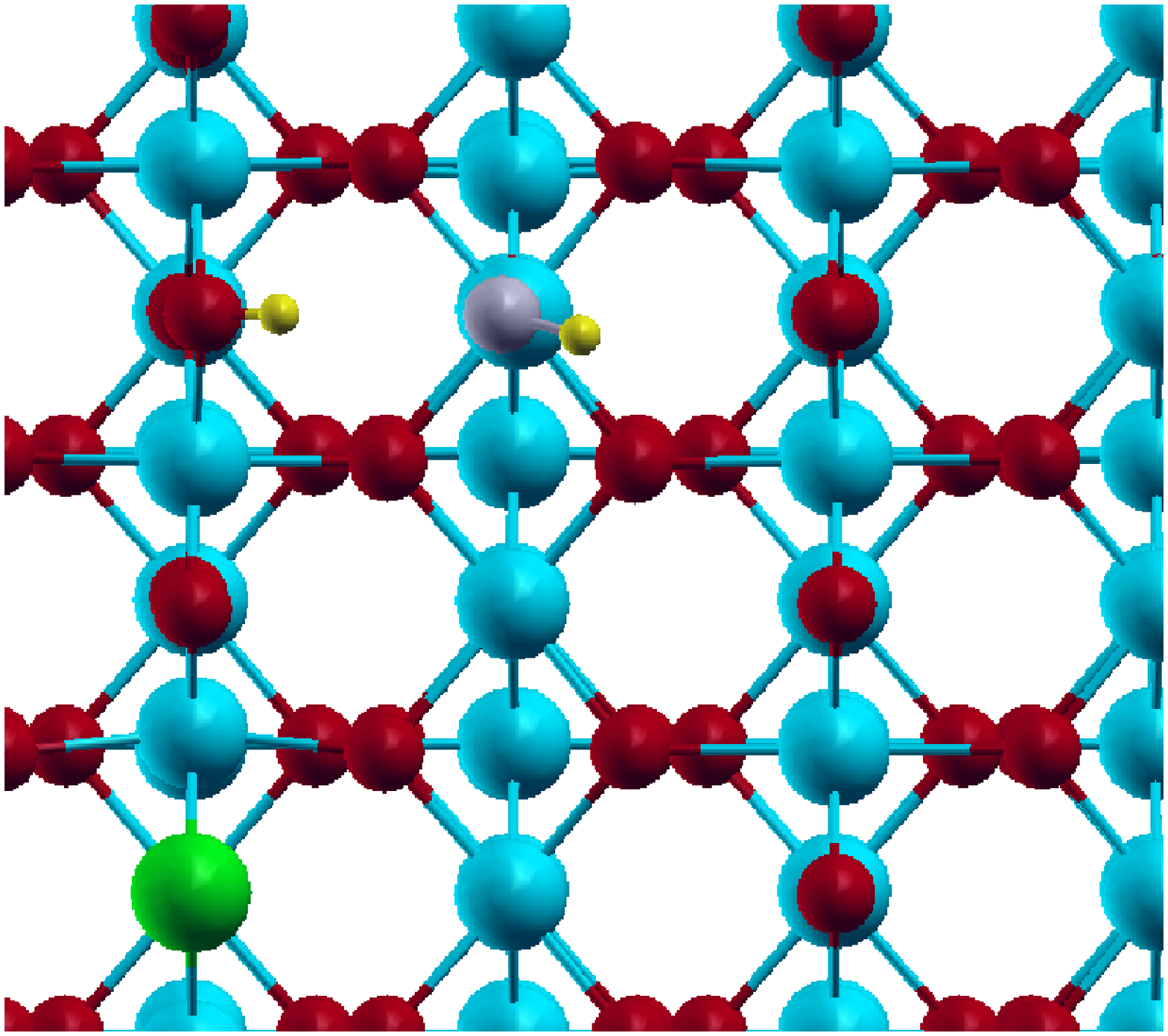}}\\
\subfigure[{\bf MA} $\mathrm{H_{2}O}$ at site {\bf 2}]{\includegraphics[width=3.0cm]{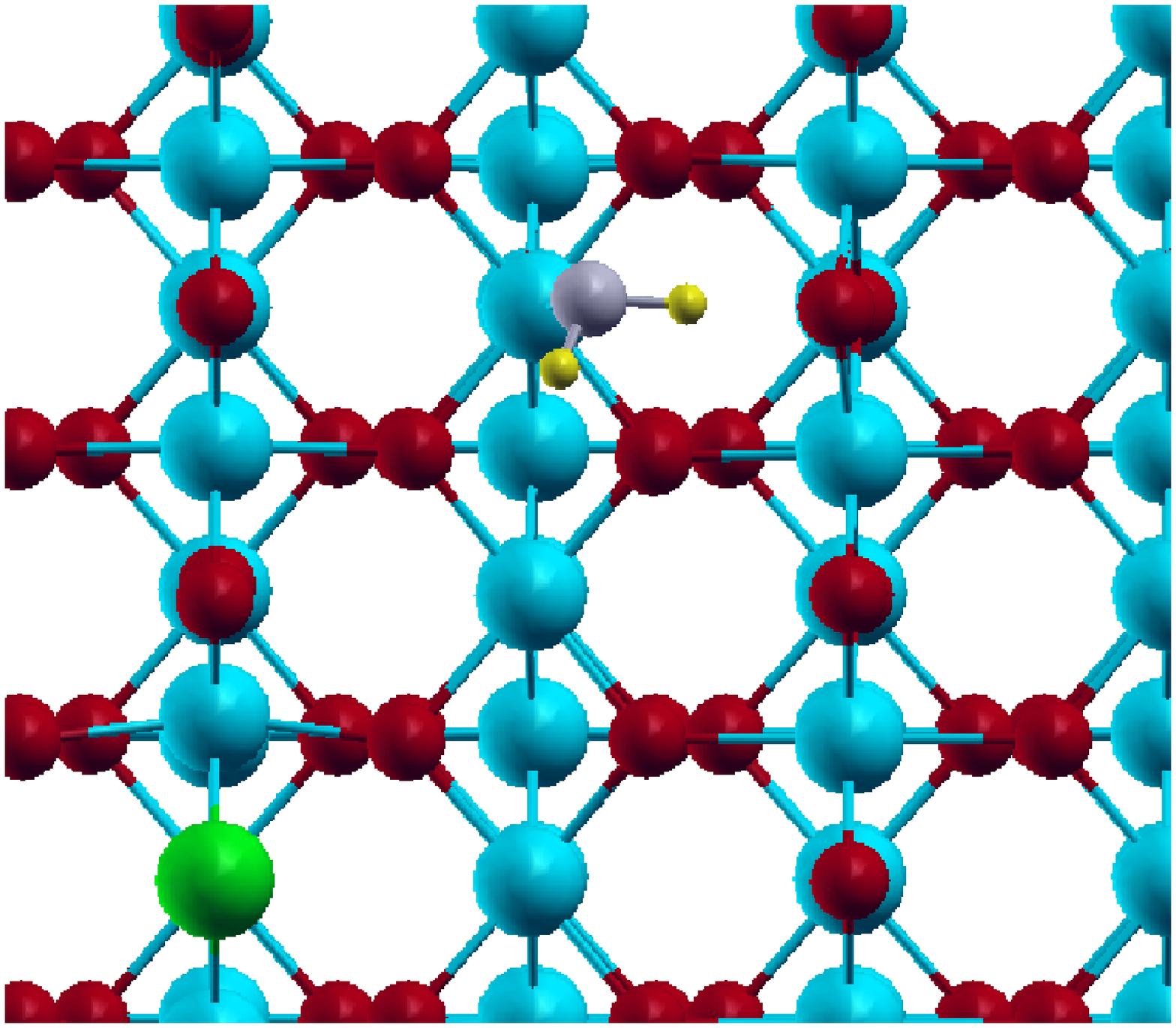}}\hspace{0.50cm}
\subfigure[{\bf MB} $\mathrm{H_{2}O}$ at site {\bf 2}]{\includegraphics[width=3.0cm]{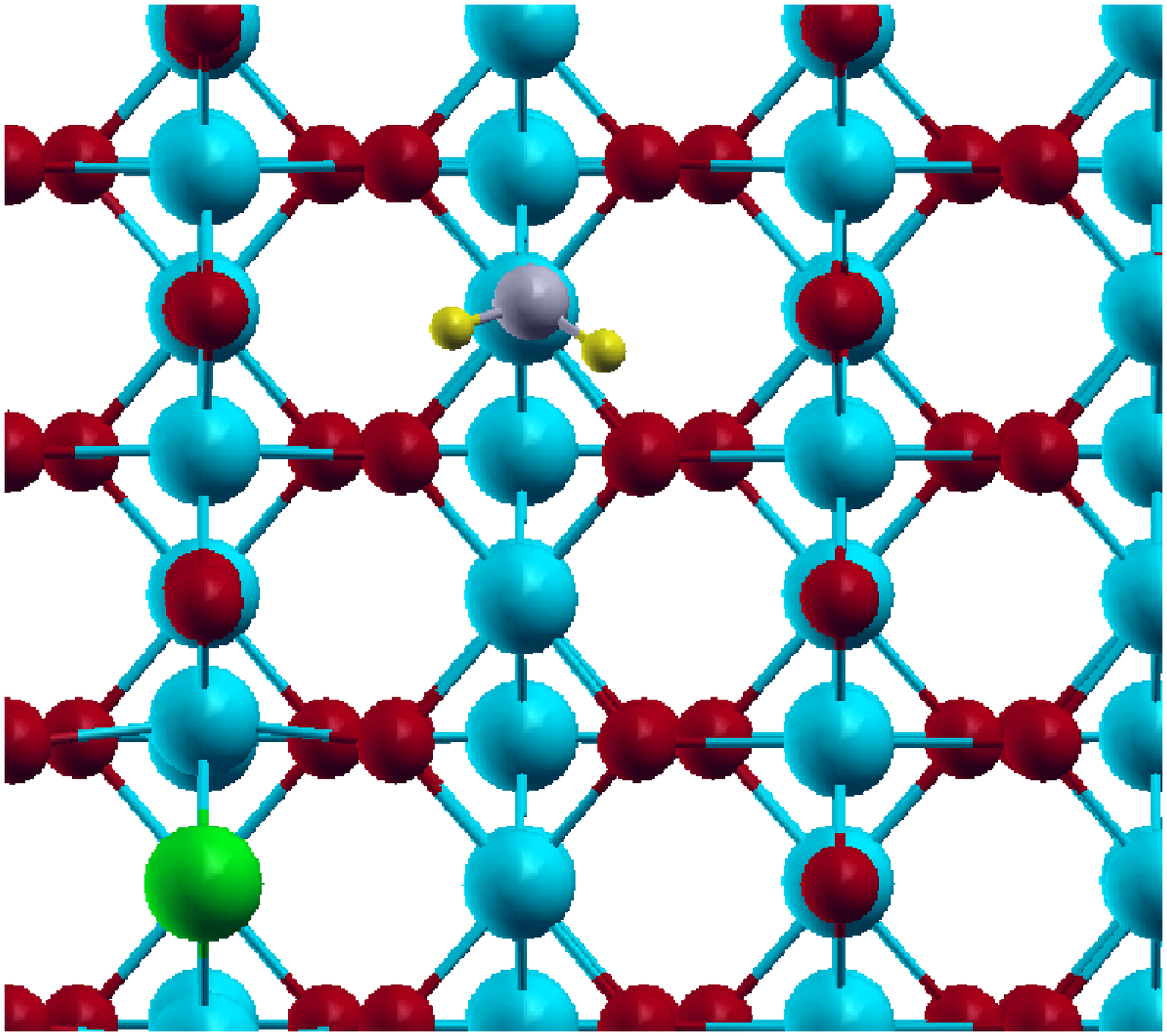}}
\end{tabular}
\caption{Ball and stick models of 
  relevant 
  configurations (see text)
  for 
  an
  $\mathrm{H_{2}O}$ molecule adsorbed either molecularly (``M'') or dissociatively (``D'') 
  in two configurations (``A'' and ``B'')
  on $\mathrm{TiO_{2}}$(110) surfaces (top view)
  at $\rm V_O$, Ti0, Ti1, and Ti2 sites 
  (see panel~(a) for site labeling) 
  obtained using the PBE+U approach. 
%
  Red, blue, violet, and yellow spheres are
  substrate $\mathrm{O}$, $\mathrm{Ti}$, water $\mathrm{O}$, and $\mathrm{H}$
  atoms, respectively, and the oxygen vacancy site $\mathrm{V_O}$ is
  highlighted using a green 
  sphere.
  See
  Table~\ref{tab3} for corresponding adsorption energies. 
%
%
%
%
%
%
%
%
%
%
%
%
%
%
\label{fig4}}
\end{center}
\end{figure}

The specific adsorption configurations for a water molecule on the reduced surface, which
we 
consider
 in this investigation, are compiled in Fig.~\ref{fig4},
and the corresponding adsorption energies are collected in Table~\ref{tab3}.
The $\mathrm{Ti_{5c}}$ surface sites are labeled in relation to the $\mathrm{V_O}$
vacancy, where site~$\mathrm{Ti0}$  denotes a nearest--neighbor $\mathrm{Ti_{5c}}$~atom  
and
sites~$\mathrm{Ti1}$ and~$\mathrm{Ti2}$ are the 
second and third nearest-neighbour
$\mathrm{Ti_{5c}}$ sites parallel to the $\mathrm{O_b}$ row containing 
the $\mathrm{V_O}$ 
site;
see panel~(a) of Fig.~\ref{fig4}. 
In agreement with experimental  
data,
we find that on the reduced 
surface,
$\mathrm{H_{2}O}$ molecules prefer to
dissociatively adsorb at $\mathrm{V_O}$ vacancy  
sites,
leading to a
configuration with two surface $\mathrm{OH}$ groups
as shown in panel~(b) of Fig.~\ref{fig4}. 
%
We therefore end up with a stoichiometric titania surface with
two $\mathrm{H}$ atoms adsorbed on two surface $\mathrm{O_b}$ atoms.
%
Once the water molecule dissociates at the 
oxygen defect through proton
transfer to an adjacent $\mathrm{O_b}$ 
atom, 
the PBE+U (PBE) adsorption
energy is $-1.61$~eV ($-1.18$~eV). 
 A projected PDOS analysis  
reveals
that upon
dissociation of water at the $\mathrm{V_O}$  
site,
two second--layer
$\mathrm{Ti^{3+}}$ ions are present in the substrate.
This value of ${E_{ads}}$ is $\sim 0.4$~eV lower than the corresponding
plain PBE value and previously reported values,~\cite{ExW2,TV05,ExW3}
which are between $-0.94$ and $-1.1$~eV.
The 
energy
value deduced from a water desorption
peak at 520~K in TDS experiments~\cite{HE03} using the simple 
Redhead formula~\cite{RH62}  
is
about $-1.4$~eV, which is  
between the PBE+U and PBE values.
However, the estimation of desorption energies using the Redhead formula 
can be biased by as much as 
$25\%$,
which implies
that both values must be considered to be consistent with experiment.
%
We now turn our attention to
the  
adsorption and dissociation of 
water at $\mathrm{Ti_{5c}}$ sites next to $\mathrm{V_O}$. 
We anticipate that, as observed in the case of $\mathrm{H_{2}O}$ dissociatively adsorbed at the
$\mathrm{V_O}$ site, the interaction between water and 
the
$\mathrm{Ti_{5c}}$
sites does not 
further reduce
the metal oxide support. 
All PBE+U calculations predict the presence of two reduced $\mathrm{Ti^{3+}}$ ions
before and after the adsorption of $\mathrm{H_{2}O}$ at $\mathrm{Ti_{5c}}$ sites.
Water molecules can be
adsorbed either dissociatively (labeled as ``D'') or molecularly (``M'') at the various
$\mathrm{Ti_{5c}}$ sites (i.e. Ti0, Ti1, and Ti2) next to  
$\mathrm{V_O}$ (see Fig.~\ref{fig4}).
When $\mathrm{H_{2}O}$ dissociates at a $\mathrm{Ti_{5c}}$ 
site,
the resulting configuration contains 
an
$\mathrm{OH}$ group bonded to a $\mathrm{Ti_{5c}}$
atom and  
an
$\mathrm{H}$ atom bonded to a 
nearest--neighbor $\mathrm{O}$ atom
of the $\mathrm{Ti_{5c}}$ in the [1$\bar{1}$0] direction. 
We have considered two different 
topologies: first,
where the 
$\mathrm{H}$~atom coming from the dissociated water 
molecule binds to an $\mathrm{O}$ atom belonging to the $\mathrm{O_b}$ row in
which the $\mathrm{V_O}$ vacancy site is present (labeled 
configuration~``A''); 
and 
second, 
with the $\mathrm{H}$~atom bonded to $\mathrm{O}$~atoms 
belonging to an adjacent
$\mathrm{O_b}$ row parallel to the $\mathrm{O_b}$ row 
that
hosts $\rm V_O$ 
(configuration~``B''). See Table~\ref{tab3} for the corresponding adsorption energies.

%
%

Our PBE+U and PBE calculations suggest that, 
in the presence of an
oxygen vacancy,  
water molecules adsorb dissociatively or molecularly at $\mathrm{Ti_{5c}}$ sites with
${E_{\rm ads}}$ in the range 
of
$-0.45$ to $-0.92$~eV. 
Once the water molecule dissociates at a $\mathrm{Ti_{5c}}$ site, it forms a pair of
terminal hydroxyls. 
Therefore, in the dissociative case, we always end up with an $\mathrm{OH}$ group bonded to a
$\mathrm{Ti_{5c}}$ site and a protonic $\mathrm{H}$ atom transferred to an
adjacent $\mathrm{O_b}$ atom, the $\mathrm{O_b}$ atom belonging either to
the $\mathrm{O_b}$ row 
that
contains $\mathrm{V_O}$, or to the adjacent 
one 
that contains no
oxygen vacancy.
%
As shown in Table~\ref{tab3} the most stable dissociative configurations 
are 
those
with the protonic $\mathrm{H}$ atom transferred to 
an
$\mathrm{O_b}$ atom of 
an
$\mathrm{O_b}$ row parallel to
the 
row
featuring the $\mathrm{V_O}$ site (configurations denoted 
``DB'' in Table~\ref{tab3} and Fig.~\ref{fig4}). 
The PBE+U/PBE calculations show that protons prefer to
bind to $\mathrm{O_b}$ atoms 
belonging to an 
$\mathrm{O_b}$ row in absence
of $\mathrm{V_O}$ sites, with ${E_{\rm ads}}$ ranging from $-0.87$ to $-0.95$~eV.         
Otherwise, if we consider the case  
where,
after water 
dissociation at a $\mathrm{Ti_{5c}}$ site, a proton
transfers to 
an
$\mathrm{O_b}$ atom belonging to 
an
$\mathrm{O_b}$ row
that includes a $\mathrm{V_O}$ site (configurations denoted 
``DA'' in
Table~\ref{tab3} and Fig. \ref{fig4}), our calculations provide adsorption
energies  
in the range of
$-0.45$ to $-0.71$~eV; the same trend is observed with 
or
without the inclusion of  
a Hubbard U~term 
in the calculations.
We note that the structure with an $\mathrm{OH}$ group adsorbed at a $\mathrm{Ti_{5c}}$
site and 
an
$\mathrm{H}$ placed right at the $\mathrm{V_O}$ vacancy 
is unstable; 
the adsorption energy of this configuration ${E_{\rm ads}}$ is positive by $2.27$~eV. 
%
%
%

Our results show that 
on the reduced $\mathrm{TiO_{2}(110)}$ 
surface,
water prefers to adsorb dissociatively 
onto
$\mathrm{Ti_{5c}}$ sites. 
Once water
molecules are molecularly adsorbed on the reduced
surface, the binding energy of the surface is about $0.1$~eV higher than it is when water is
dissociated on the same substrate. 
These findings are in agreement with previous studies of $\mathrm{H_{2}O}$
interaction with the stoichiometric $\mathrm{TiO_{2}}(110)$ surface. 
%
However, 
as shown based on carefully converged 
calculations~\cite{kowalski},
molecular and dissociated configurations
become essentially energetically degenerate
at very low coverages, 
which explains why some studies favor molecular adsorption 
whereas others yield the dissociated state as the lowest 
energy configuration in this regime~\cite{roger-piss}.
Most 
experimental works, except a recent one~\cite{WB09} claiming 
mixed adsorption, indicate molecular adsorption only. 
This gives rise to 
the well--known
discrepancy between 
theoretical 
predictions and experimental results in 
providing a consistent and comprehensive picture of water adsorption 
on titanium dioxide surfaces, in particular at low coverages. 

\begin{table}
\caption{\label{tab3}
Adsorption energies ${E_{\rm ads}}$ (in~eV) in the case of adsorption of water on the reduced
surface for PBE+U and plain PBE reported in parentheses;
see Fig.~\ref{fig4} for labeling. 
}
\begin{ruledtabular}
\begin{tabular}{cccc}
configuration  & site~Ti0   &  site~Ti1 & site~Ti2  \\
\hline
DA  & $~~2.27$ ({ $~~1.58$}) &  $-0.71$ ({ $-0.81$})  &  $-0.45$  ({ $-0.48$}) \\
MA  &  $-0.77$  ({ $-0.79$}) &  $-0.73$ ({ $-0.76$})  &  $-0.72$  ({ $-0.72$}) \\
DB  &  $-0.87$  ({ $-0.89$}) &  $-0.92$ ({ $-0.95$})  &  $-0.89$  ({ $-0.90$}) \\
MB  &  $-0.77$  ({ $-0.80$}) &  $-0.83$ ({ $-0.81$})  &  $-0.84$  ({ $-0.82$}) \\
$\mathrm{H_{2}O}@\mathrm{V_O}$ & $-1.61$ ({ $-1.18$}) & & \\

\end{tabular}
\end{ruledtabular}
\end{table}

Last but not least, we focus on the interaction between $\mathrm{CO}$ and the
$\mathrm{TiO_{2}}(110)$ surface.
It is well known that, on the stoichiometric TiO$_2$(110) surface,
the $\mathrm{CO}$ molecule adsorbs onto
$\mathrm{Ti_{5c}}$ sites,
thus forming $\mathrm{Ti}$-$\mathrm{C}$ bonds. 
Because van der Waals dispersion interactions and  
non--local electron correlations  
significantly influence this type of 
bonding, we have previously investigated 
the interaction of CO with the stoichiometric TiO$_2$(110) surface using
a combination of DFT and post Hartree-Fock (``SCS-MP2'') methods~\cite{JCP}. 
The $\mathrm{CO}$ binding energy has been found to vary significantly 
with coverage and increases upon reaching the saturation limit. The SCS-MP2 $E_{ads}$
computed values 
are  $-0.20$~eV for the
full saturated surface and $-0.36$~eV for a
single CO molecule adsorbed on the surface. 
The PBE adsorption energy of a single $\mathrm{CO}$ molecule, $-0.32$~eV, is close to 
the SCS-MP2 value of  
$-0.36$~eV,
and both energies are in accord with the 
experimental value obtained by means of thermal desorption spectroscopy. 
This demonstrates that PBE as such is able to describe the
interaction of $\mathrm{CO}$ with the ideal (110) rutile 
surface in the limit of low coverages, which remains unaltered when using
PBE+U, which gives $-0.31$~eV for 
the
adsorption energy. 

\begin{figure}[h]
\begin{center}
\begin{tabular}{cc}
\subfigure{\includegraphics[angle=0,scale=0.30]{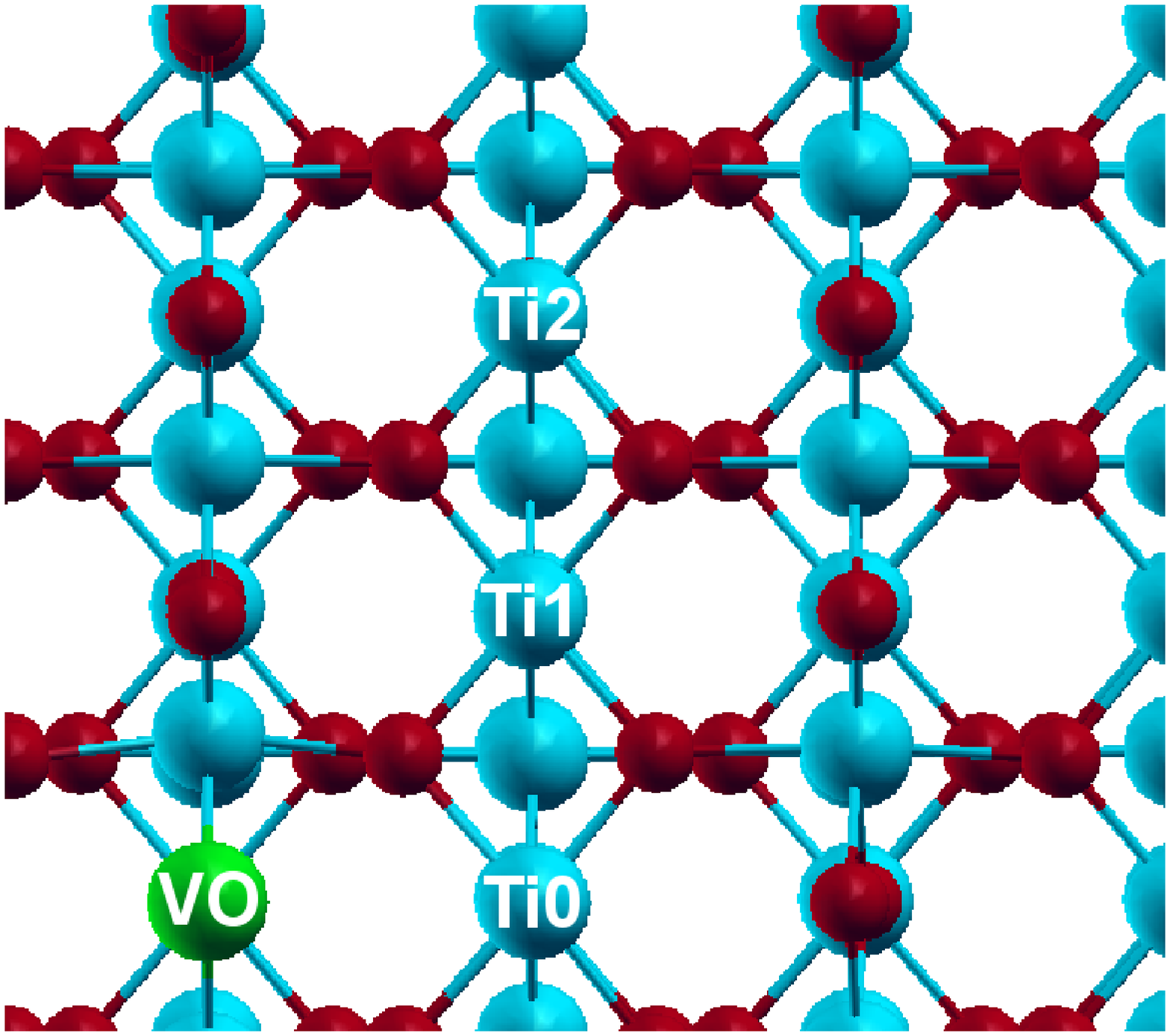}}
\end{tabular}
\caption{%
Ball and stick model of the $\mathrm{TiO_{2}}$(110) surface 
(top view),
with site labeling. 
Red and blue spheres are $\mathrm{O}$ and $\mathrm{Ti}$ atoms, respectively,
and the oxygen vacancy site, $\rm V_O$, 
within the bridging oxygen row, $\rm O_b$, 
is highlighted using a green sphere.
%
%
%
%
\label{fig1}}
\end{center}
\end{figure}

\begin{table}
\caption{
Adsorption energies ${E_{\rm ads}}$ (in eV) of a $\mathrm{CO}$ molecule
  adsorbed on the reduced rutile $\mathrm{TiO_{2}}$(110) surface
  at different 
  sites,
  labeled according to Fig.~\ref{fig1}. 
  In the
  second
  column, 
  the number of reduced $\mathrm{Ti^{3+}}$ ions 
  present in the substrate is reported. \label{tab1} }
\begin{ruledtabular}
\begin{tabular}{llccccc}
 & Configuration  & ${E_{\rm ads}}$(PBE+U) & $\mathrm{Ti^{3+}}$  & ${E_{\rm ads}}$(PBE)  \\
\hline
 & $\mathrm{CO @ V_O}$   & $-0.32$  & 2  & $-0.29$  \\
 & $\mathrm{CO @ Ti0}$   & $-0.22$  & 2  & $-0.28$   \\
 & $\mathrm{CO @ Ti1}$   & $-0.29$  & 2  & $-0.30$   \\
 & $\mathrm{CO @ Ti2}$   & $-0.33$  & 2  & $-0.29$   \\

\end{tabular}
\end{ruledtabular}
\end{table}

\begin{figure*}
\begin{center}
\begin{tabular}{cc}
\subfigure{\includegraphics[angle=0,scale=0.5]{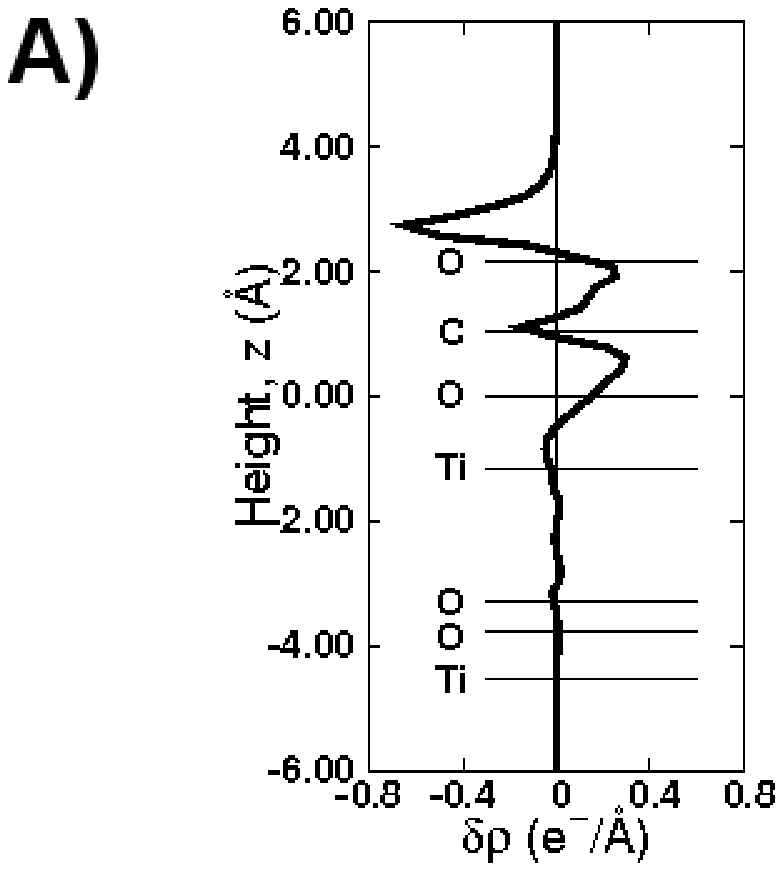}}\hspace{0.50cm}
\subfigure{\includegraphics[angle=0,scale=0.20]{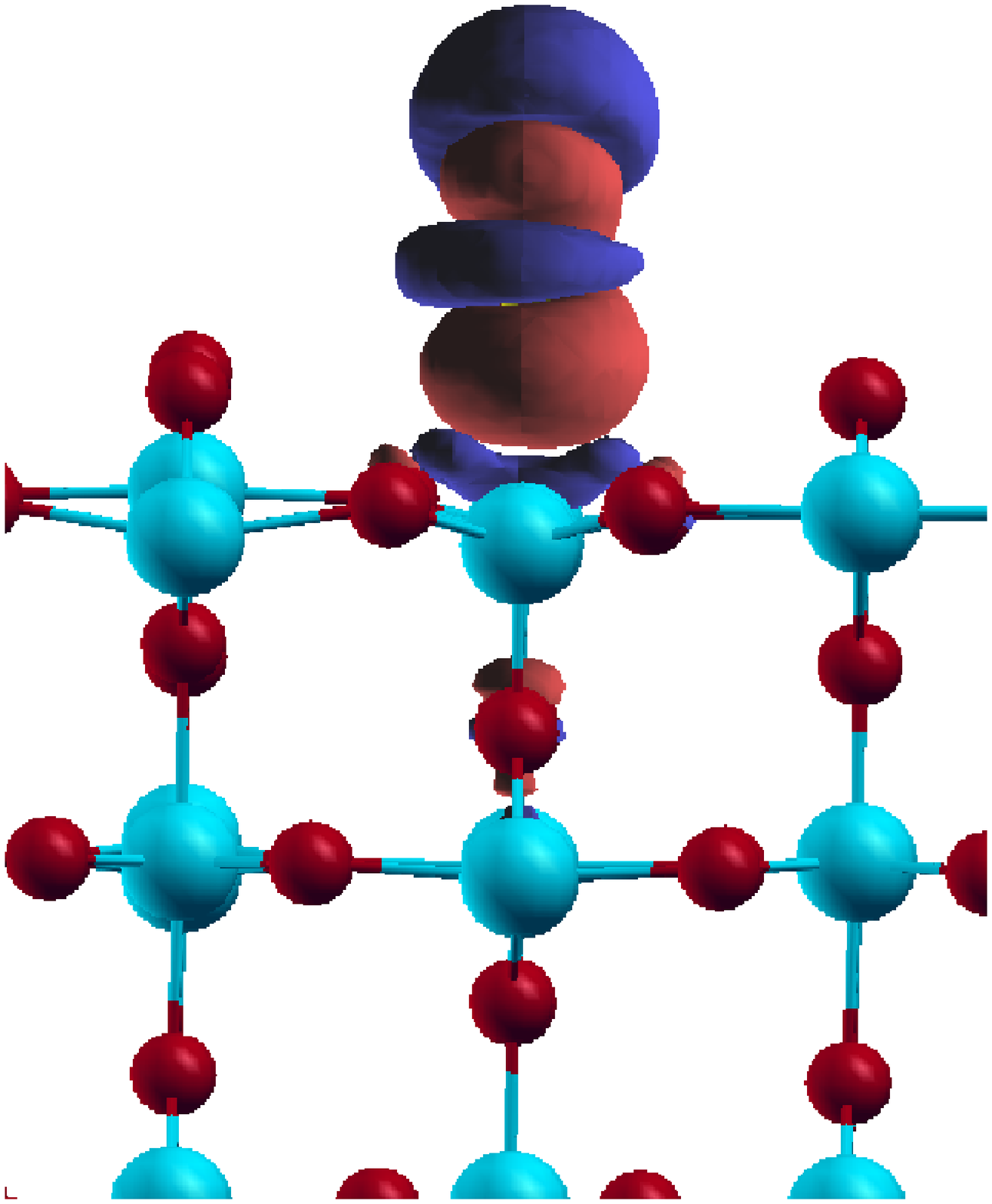}}\hspace{1.00cm}
\subfigure{\includegraphics[angle=0,scale=0.35]{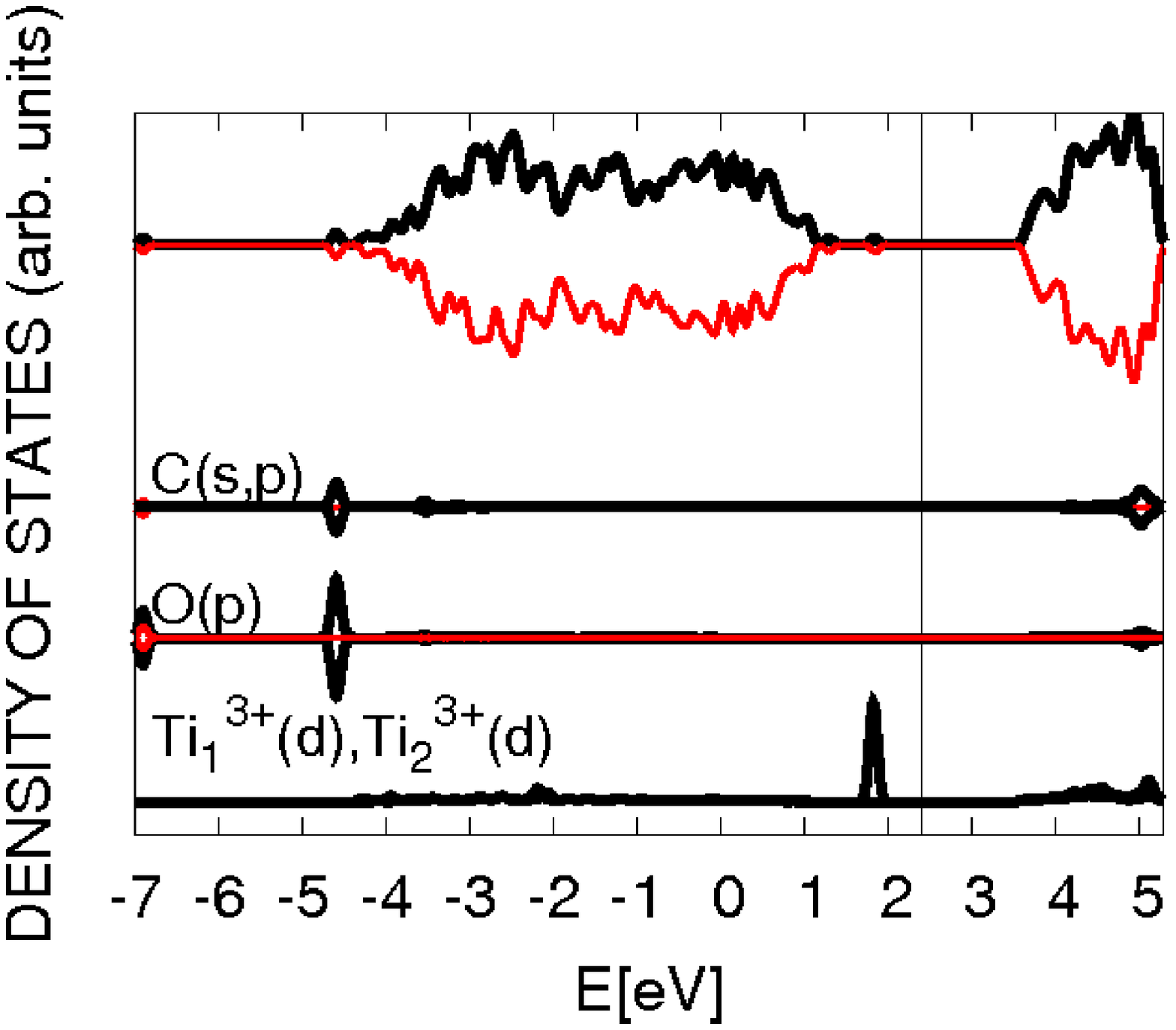}}
\end{tabular}
\caption{%
Electronic structure analyses (based on the PBE+U approach) 
of molecular $\mathrm{CO}$ adsorbed
on the reduced $\mathrm{TiO_{2}}$(110) surface at the Ti2~site 
with respect to the oxygen vacancy, $\rm V_O$, located in 
the row of bridging oxygens, $\rm{O}_b$ 
(see Fig. \ref{fig1} and Table \ref{tab1}). 
The left panel represents the bonding charge $\delta\rho (z)$  integrated in planes
perpendicular to the surface and plotted as a function of the height from
the surface. 
The central panel 
displays the bonding charge $\Delta\rho (\vec{r})$ at an isovalue of
$\pm 0.06$~$|e|$/\AA$^{3}$
where electron accumulation and depletion are
represented by red and blue areas, respectively. 
The right panel shows the total DOS and 
atom--resolved
projected DOS (PDOS) 
as indicated;
here, energy values are with respect to the Fermi level, which is marked by a solid vertical line.
\label{fig2}
}
\end{center}
\end{figure*}

We now turn our attention to $\mathrm{CO}$ adsorbed onto the $\mathrm{O}$ vacancy site, 
$\rm V_O$,  as well as on the 
fivefold coordinated
surface sites, $\mathrm{Ti_{5c}}$, 
at various distances from the vacancy. 
Note that the $\mathrm{Ti_{5c}}$ site belongs to the Ti~row next to the bridging $\rm{O}_{b}$ row containing $\rm V_O$ (see Fig.~\ref{fig1}).
In each case, the $\mathrm{CO}$ molecule is placed  
perpendicular to the substrate, with the carbon atom pointing toward the surface. 
All
 the structures are fully relaxed according to our aforementioned
convergence criterion.
The computed adsorption energies (${E_{ads}}$)
are compiled in Table~\ref{tab1}. 
Interestingly, we do not see a significant variation in the adsorption energies 
computed for different structures. 
The PBE+U (PBE) adsorption energy for the $\mathrm{CO}$ molecule adsorbed at the surface
$\mathrm{O_b}$ vacancy site is $-0.32$~eV ($-0.29$~eV). 
$\mathrm{CO}$ adsorption at  
sites~$\mathrm{Ti1}$
and $\mathrm{Ti2}$ results in adsorption energies 
 of $-0.29$~eV ($-0.30$~eV) and $-0.33$~eV ($-0.29$~eV), respectively  
(see Fig.~\ref{fig1} for 
site labeling scheme). 
The computed values of ${E_{ads}}$ are in qualitative agreement with 
previous studies~\cite{selloni,zhao}. 
We note that these values are close to 
those
 obtained when $\mathrm{CO}$ is adsorbed at the $\rm V_O$
vacancy site ($-0.32$~eV) and on the stoichiometric surface~\cite{JCP},   
i.e. $-0.31$~eV ($-0.32$~eV) for PBE+U (PBE).
When adsorbed at site~$\mathrm{Ti0}$, the adsorption energy of the
$\mathrm{CO}$ molecule, computed with PBE+U, results in a distinctly higher value 
for
${E_{ads}}$ of $-0.22$~eV in agreement with Ref.~\onlinecite{zhao}, while
the corresponding PBE value is $-0.28$~eV, comparable to the adsorption at
sites~$\mathrm{Ti1}$ and $\mathrm{Ti2}$. 
Our PBE+U results therefore confirm previous findings that $\mathrm{CO}$ molecules
weakly interact with the reduced $\mathrm{TiO_{2}}$(110) oxide
surface, ${E_{ads}}$ being of 
the order of about $-0.3$~eV 
(see Refs.~\onlinecite{JCP,zhao,selloni}).
The calculations indicate that CO adsorbs at 
both $\rm V_O$ vacancies and $\mathrm{Ti_{5c}}$ sites, but 
while PBE calculations give similar energy values for $\mathrm{CO}$ adsorption at
$\rm V_O$ vacancies and at $\mathrm{Ti0}$, $\mathrm{Ti1}$, and $\mathrm{Ti2}$ sites, the
inclusion of a Hubbard U~term suggests that the adsorption of $\mathrm{CO}$
at $\mathrm{Ti0}$ sites, the 
sites
facing the $\rm V_O$ vacancy, is discouraged. 
In this case, 
the adsorption energy ${E_{ads}}$ 
is found
to be $\sim 0.1$~eV
higher 
when
compared to the adsorption energy values at sites $\mathrm{Ti1}$ and $\mathrm{Ti2}$.
Upon $\mathrm{CO}$ adsorption on the reduced  
surface, the charge redistribution that results from attaching the molecule does not further reduce
the oxide support: 
the PBE+U calculations yield two $\mathrm{Ti^{3+}}$ ions before and after adsorption,
which results in an insignificant change in the adsorption energies when 
switching from PBE to PBE+U calculations. 
This behavior is confirmed by the computed electronic density of states (DOS).
In Fig.~\ref{fig2},  
we depict the electronic DOS and the bonding
charge density $\mathrm{\Delta \rho}$ of the structure where a $\mathrm{CO}$ molecule
is attached at the $\mathrm{Ti_{5c}}$ site 
(labeled site~$\mathrm{Ti2}$ in Fig.~\ref{fig1}).
%
The electronic DOS features a peak in the band gap below the Fermi level. 
The projected DOS (PDOS) analysis reveals that this filled gap state is related 
to the charge localized on two $\mathrm{Ti}$--3{\it d}  orbitals  
that
yield
the two reduced $\mathrm{Ti^{3+}}$ sites. 
%
%
We  
conclude therefore 
that the binding of a $\mathrm{CO}$ molecule does not induce
a significant charge rearrangement at the $\mathrm{CO}$/oxide 
contact site.
However, as we will 
 show, 
the catalytic activity of the $\mathrm{TiO_{2}}$(110) 
substrate 
for efficient
$\mathrm{CO}$ oxidation is 
improved by supported
and dispersed $\mathrm{Au}$ adatoms on 
this
substrate.
%

%
In conclusion, this detailed assessment convincingly demonstrates that although similar
trends are observed in the adsorption energies with or without the inclusion
of a Hubbard U~correction, the PBE+U method is seen to significantly 
{\em improve the description of the electronic structure whenever reduction occurs.} 
%
In particular, the localization of excess charge on the titania substrate induced
by $\mathrm{O}$ vacancies ($F$-centers) or upon $\mathrm{H}$ atom adsorption on bridging
$\mathrm{O}$ atoms (hydroxylation), is correctly predicted by the PBE+U approach. 
Clearly, 
an adequate description of the electronic structure of such $\mathrm{TiO_{2}}$(110) surfaces
is crucial when dealing with metal-promoted oxide surfaces in the realm of catalysis. 

\section{Gold--promoted titania: Electronic structure, dynamics, and thermodynamics from PBE+U} 
\label{gold}

\subsection{Au adatom adsorption on the stoichiometric $\mathbf{TiO_{2}(110)}$ surface}

\begin{figure*}
\begin{center}
\begin{tabular}{ccc}
\subfigure{\includegraphics[angle=0,scale=0.5]{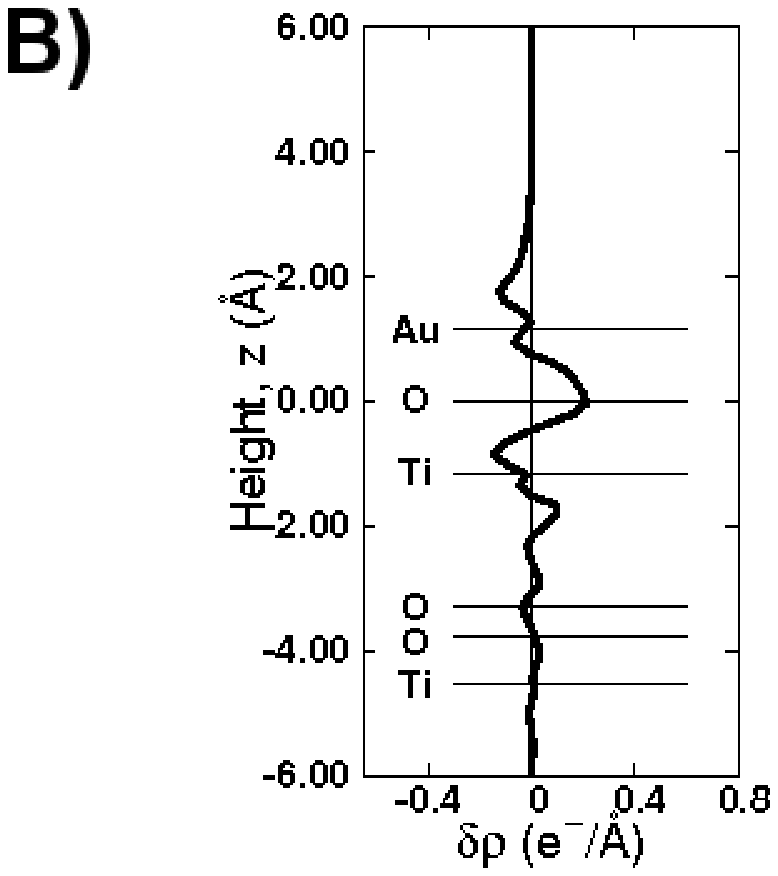}}\hspace{0.50cm}
\subfigure{\includegraphics[angle=0,scale=0.2]{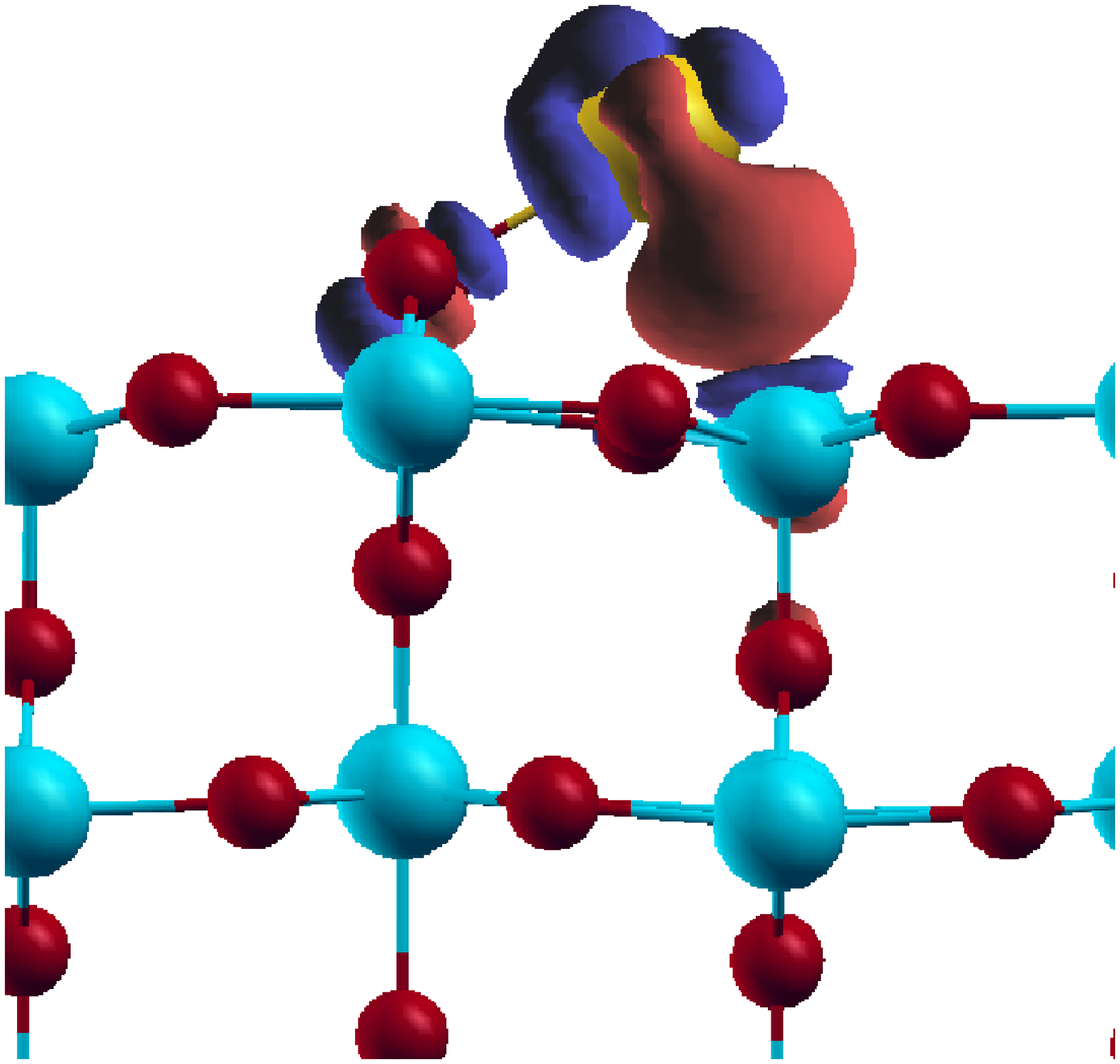}}\hspace{1.00cm}
\subfigure{\includegraphics[angle=0,scale=0.35]{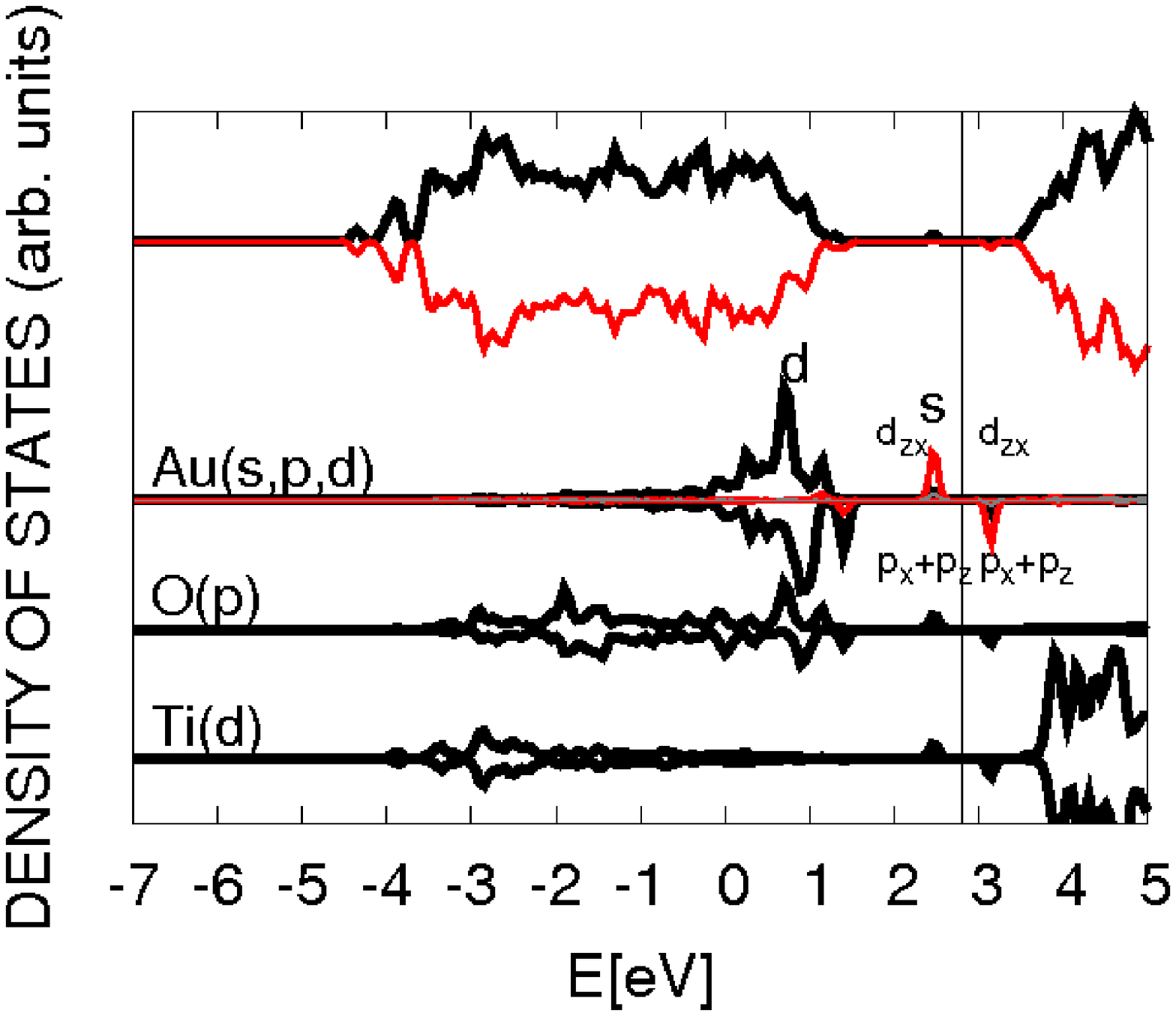}}\\
\subfigure{\includegraphics[angle=0,scale=0.5]{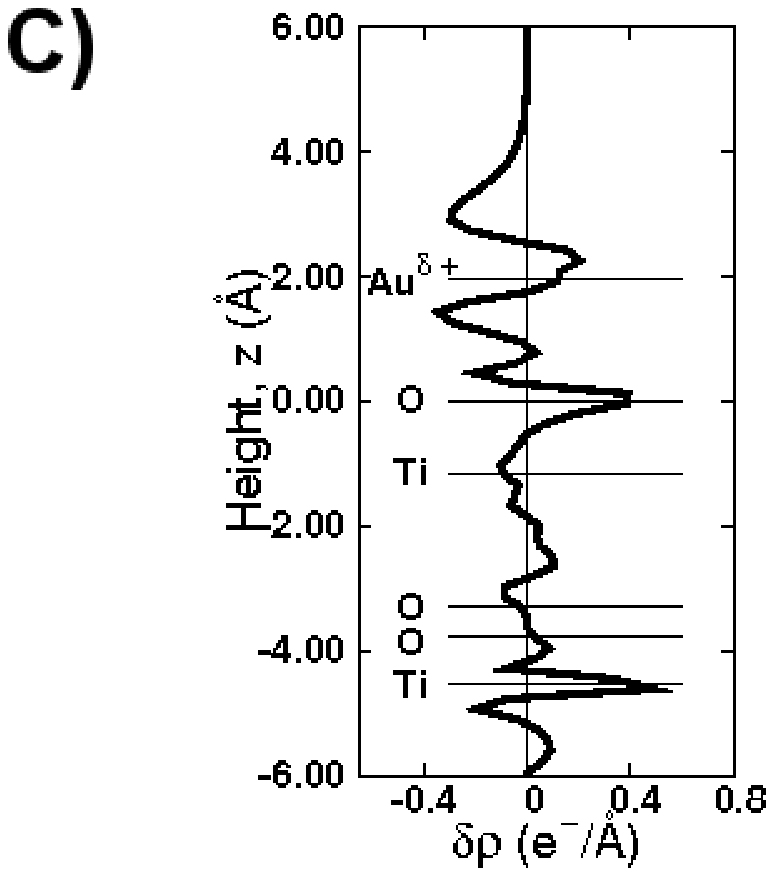}}\hspace{0.50cm}
\subfigure{\includegraphics[angle=0,scale=0.20]{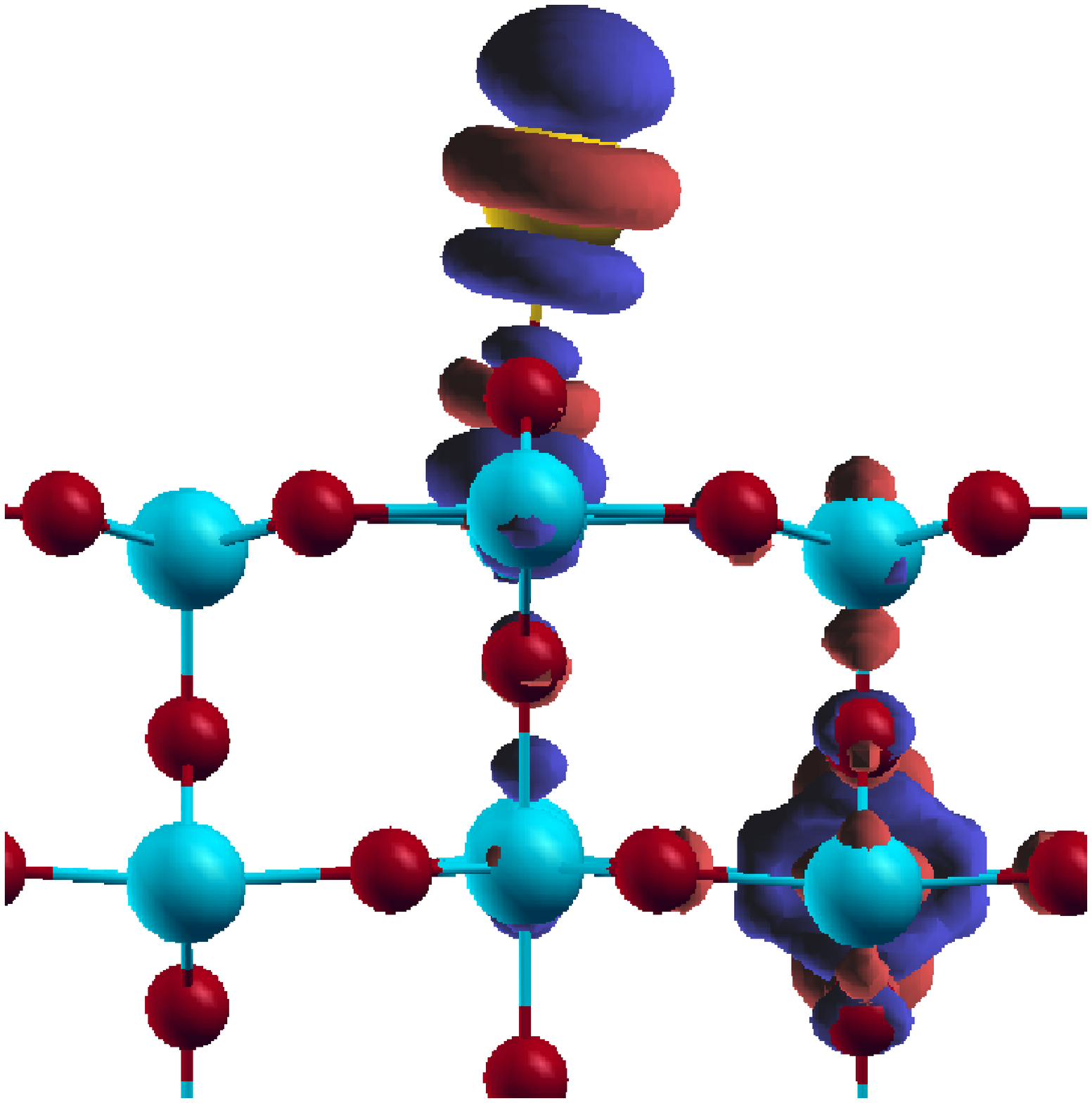}}\hspace{1.00cm}
\subfigure{\includegraphics[angle=0,scale=0.35]{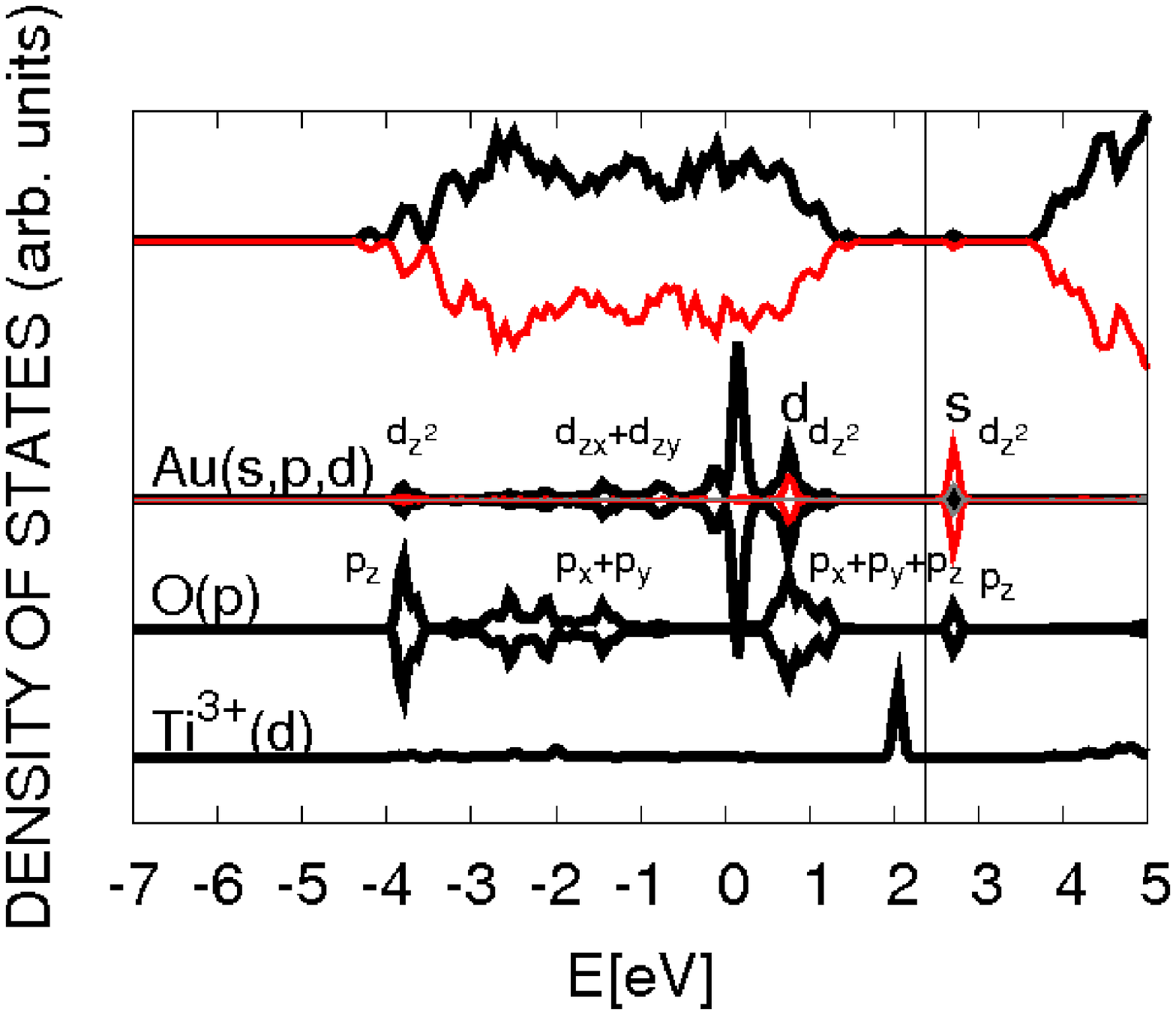}}\\
\subfigure{\includegraphics[angle=0,scale=0.5]{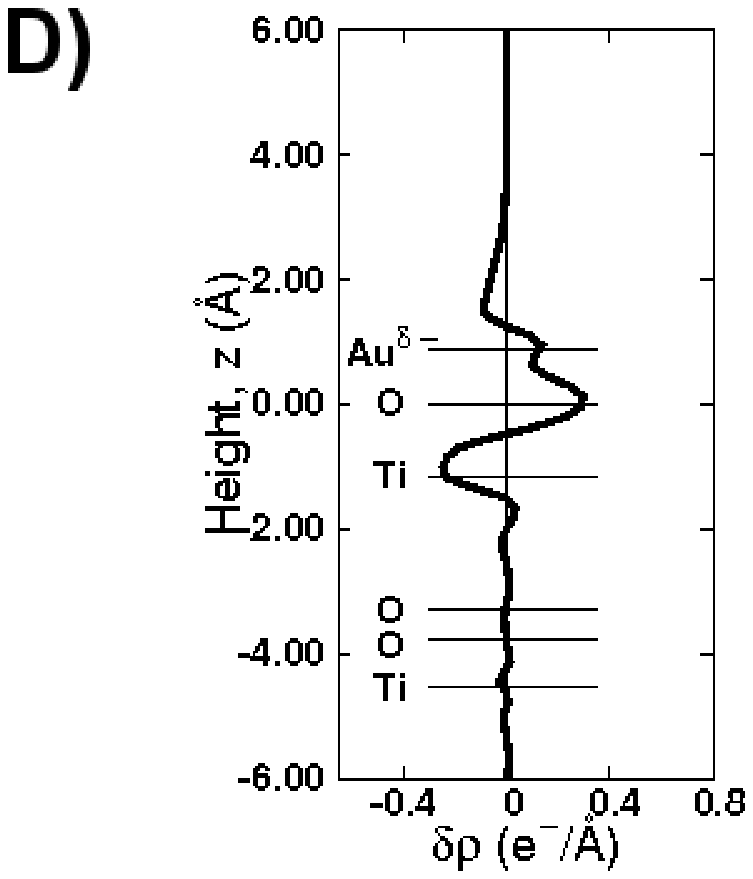}}\hspace{0.0cm}
\subfigure{\includegraphics[angle=0,scale=0.20]{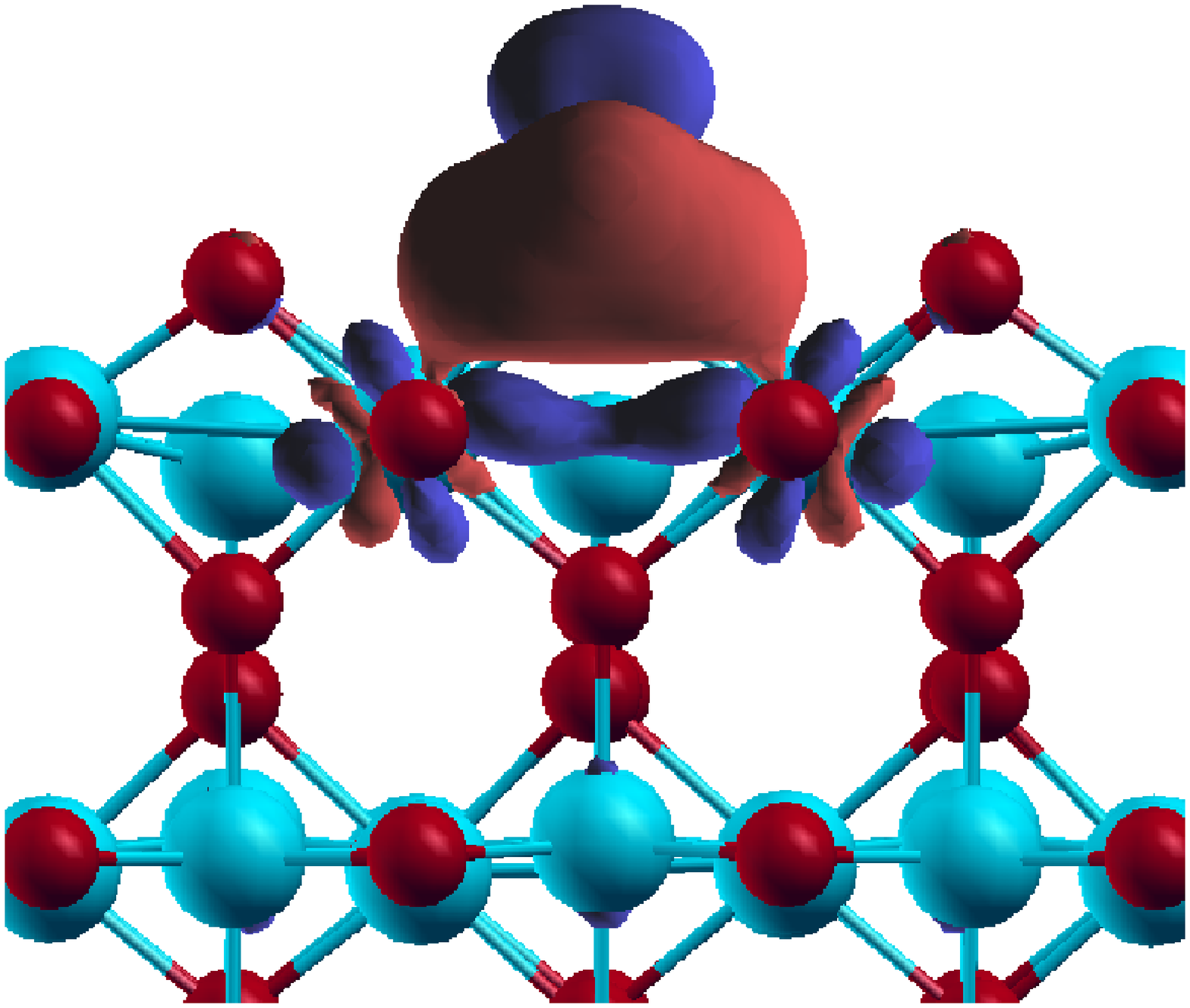}}\hspace{1.00cm}
\subfigure{\includegraphics[angle=0,scale=0.35]{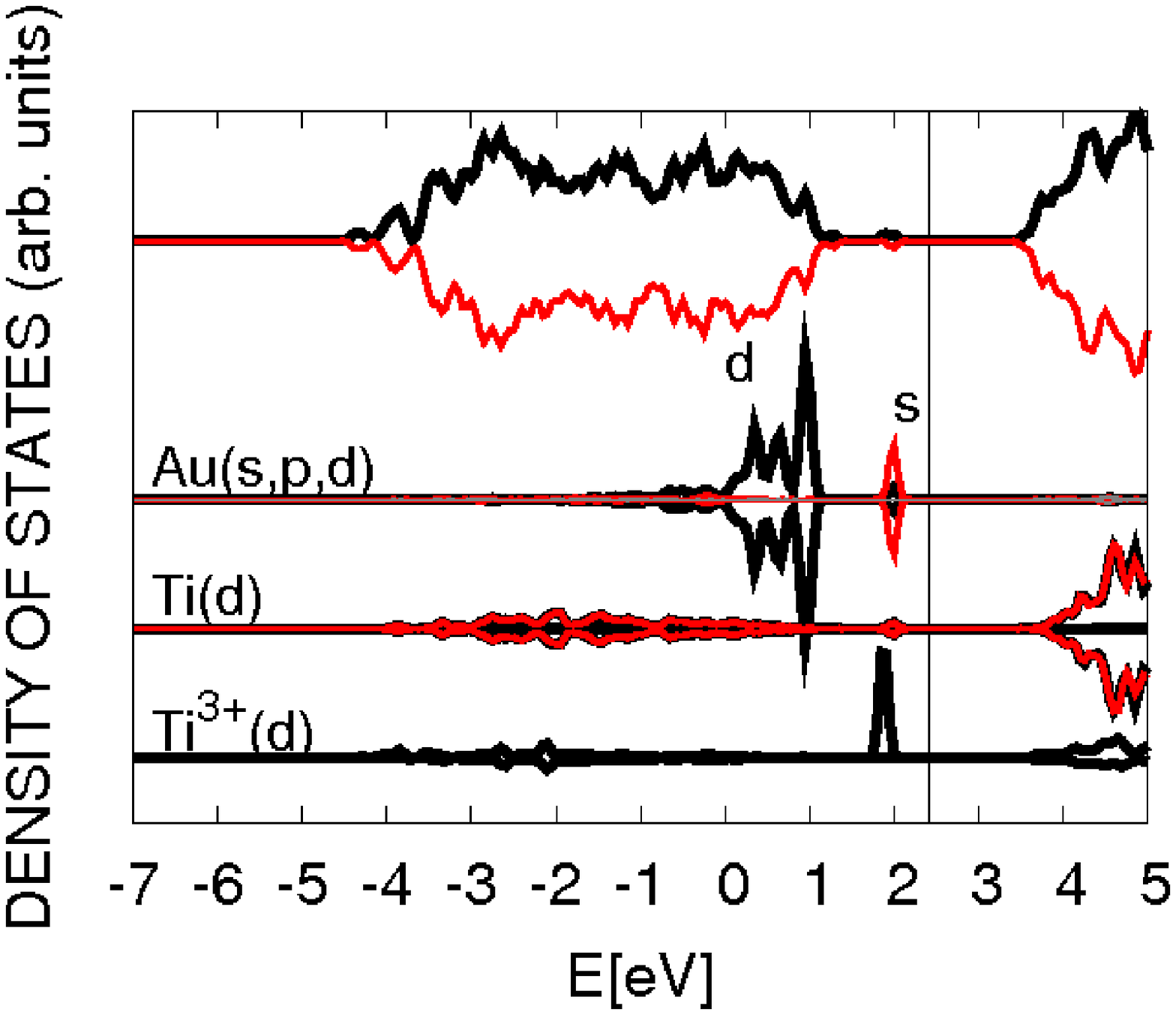}}\\
\subfigure{\includegraphics[angle=0,scale=0.5]{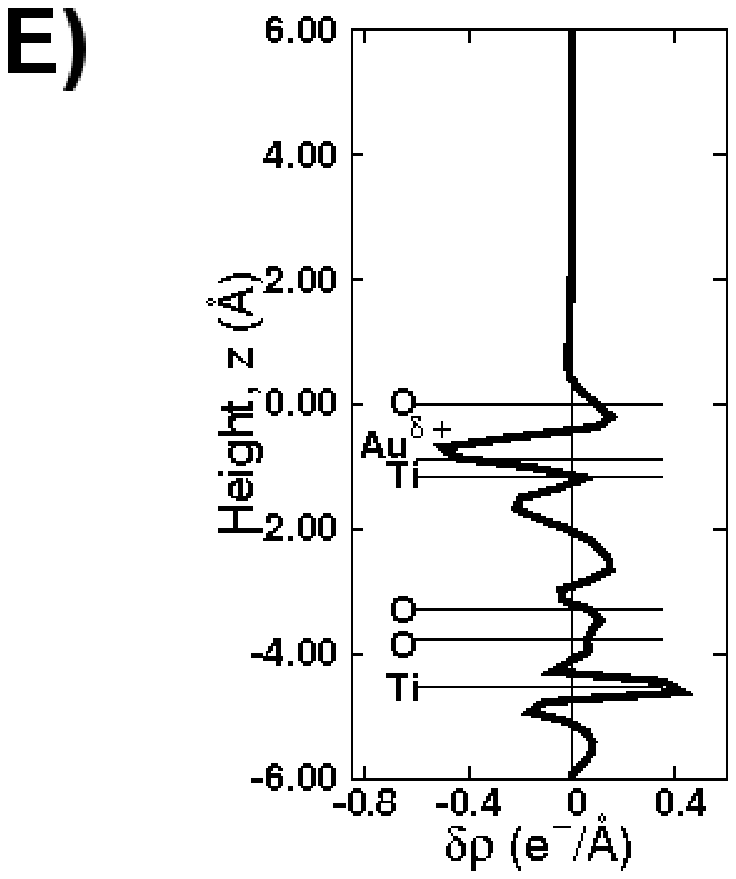}}\hspace{0.50cm}
\subfigure{\includegraphics[angle=0,scale=0.20]{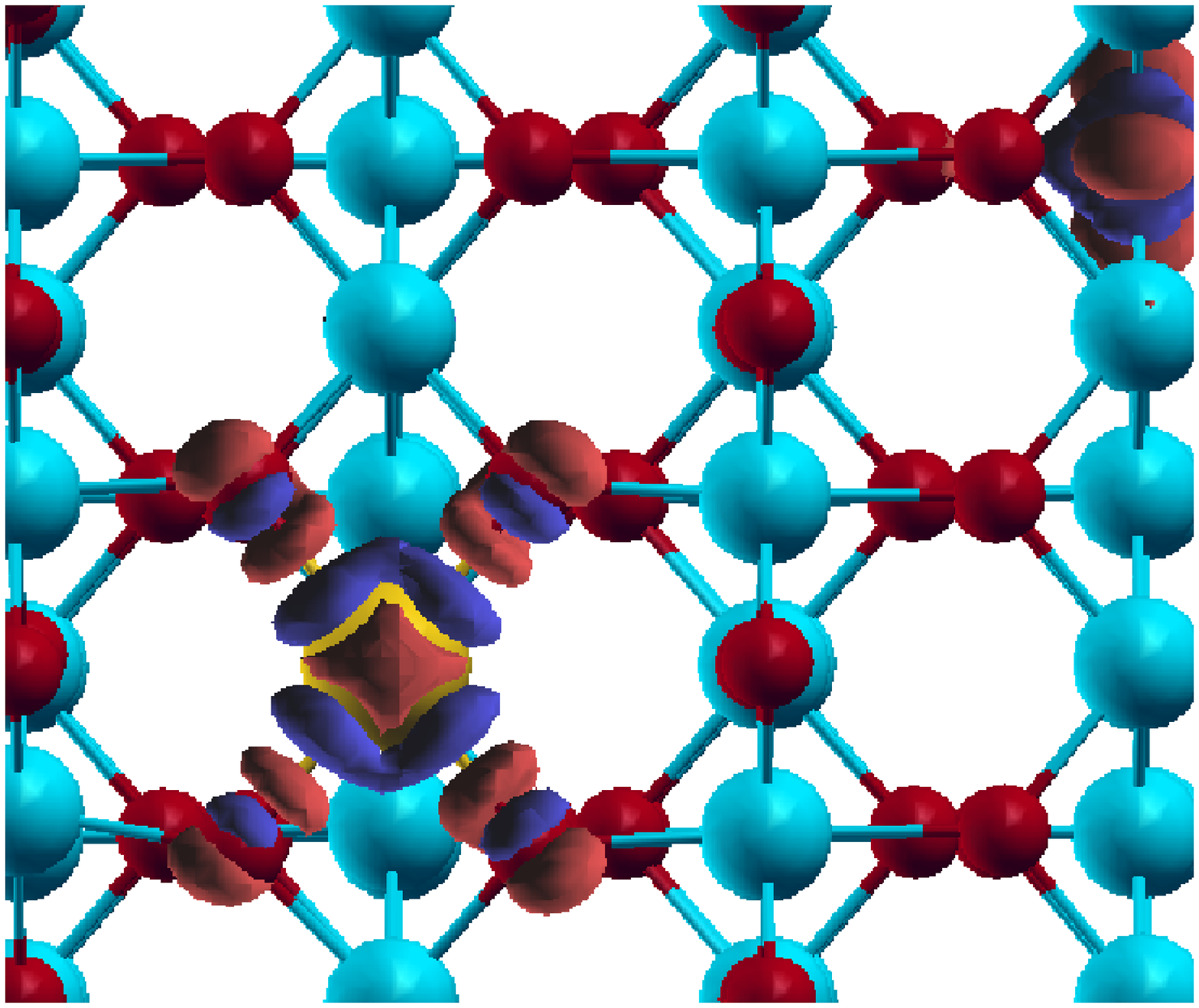}}\hspace{1.00cm}
\subfigure{\includegraphics[angle=0,scale=0.35]{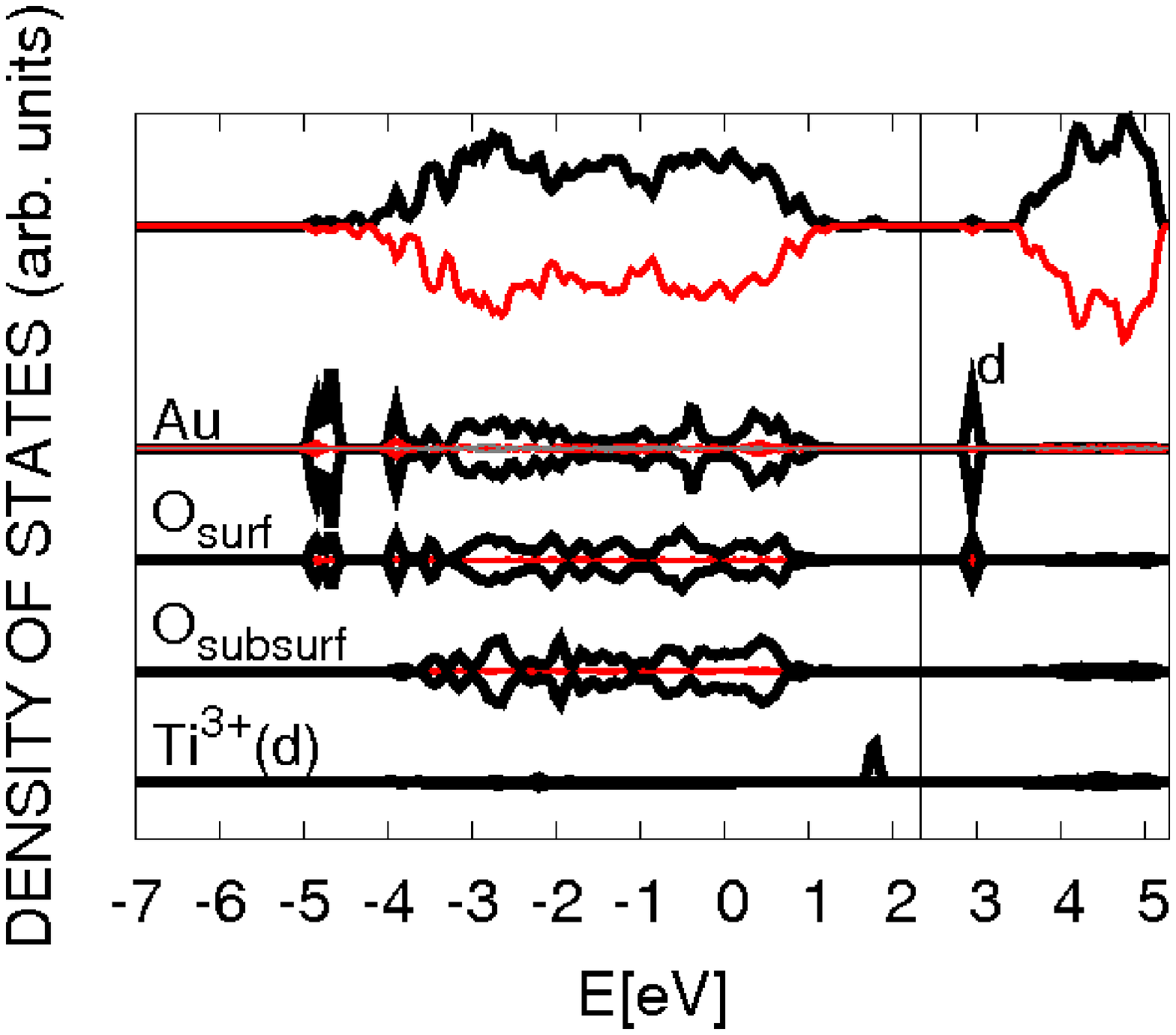}}
\end{tabular}
\caption{%
Electronic structure analyses (based on the PBE+U approach)
of 
an
$\mathrm{Au}$ adatom
(B)
  supported by the stoichiometric $\mathrm{TiO_{2}}$(110) surface in the
  bridge position, 
(C)
  supported by the stoichiometric $\mathrm{TiO_{2}}$(110) surface in the top
  position, 
(D)
  adsorbed on a surface $\mathrm{V_O}$ vacancy on the $\mathrm{TiO_{2}}$(110)
  surface, and 
(E)
  substituting a surface $\mathrm{Ti_{5c}}$ atom in the presence of a surface
  $\mathrm{V_O}$ vacancy 
  in a bridging position
  on the $\mathrm{TiO_{2}}$(110)
%
%
%
 surface (see Fig.~\ref{fig9}). 
Left panels represent the bonding charge $\delta\rho (z)$  integrated 
over
planes
perpendicular to the surface and plotted as a function of the height from
the surface.
Central panels display the bonding charge $\Delta\rho (\vec{r})$ at an isovalue of
$\pm 0.06$~$|e|$/\AA$^{3}$
where electron accumulation and depletion are
represented by red and blue areas, respectively.
Right panels 
show the total DOS and atom resolved projected DOS (PDOS) 
as indicated,
where 
energy values are with respect to the Fermi level, which is marked by a solid vertical line.
%
\label{fig5}}
\end{center}
\end{figure*}

Having shown that the PBE+U formalism performs well for
a set of reference calculations on the adsorption of $\mathrm{H}$,
$\mathrm{H_{2}O}$, and $\mathrm{CO}$ on stoichiometric and 
reduced $\mathrm{TiO_{2}}(110)$ surfaces, and that it significantly 
improves the description of reduced titania surfaces, 
we now 
progress
 to investigating the interactions between gold and 
$\mathrm{TiO_{2}}$(110) surfaces. 
%
The adsorption or substitution of gold 
induces strong charge rearrangements at the $\mathrm{Au}$/oxide 
contact, which affects
the electronic structure. 
%
Of particular interest in the realms of metal/support interactions and
heterogeneous catalysis is the 
oxidation state of $\mathrm{Au}$ adatoms, which is determined
by the site 
where
the metal atom is adsorbed, as well as by the stoichiometry of
the supporting oxide.
%
As in previous GGA studies~\cite{vijay,pillay,matthey,graciani}, 
two stable adsorption sites of a single $\mathrm{Au}$ adatom on the
stoichiometric $\mathrm{TiO_{2}}$(110) surface have been identified,
the two structures differing  by only $\sim 0.1$~eV in energy. 
The most stable adsorption site for 
an
$\mathrm{Au}$ adatom deposited on this
titania surface is a
bridge site between 
an
$\mathrm{O_b}$ and a $\mathrm{Ti_{5c}}$ atom
as depicted in 
the
central panel of Fig.~\ref{fig5} {\bf (B)}.
The computed PBE+U (PBE) adsorption energy and the
$\mathrm{Au}$--$\mathrm{O}$~/ $\mathrm{Au}$--$\mathrm{Ti}$ bond lengths are
$-0.58$~eV ($-0.41$~eV) and 2.30 (2.39)~/  2.79 \AA\ 
(2.88 \AA\ )
respectively, 
%
%
%
in agreement with previous
studies~\cite{Campbell,vijay,pillay,matthey,graciani} based on 
standard GGA calculations;
see
Table~\ref{tab4} for a summary. 
%

The bonding charge density analysis reveals that
0.11~$|e|$ are transferred from the metal atom to the oxide 
substrate,
thus 
indicating a very weak oxidation of $\mathrm{Au}$. The excess charge in the
substrate is mostly localized around the $\mathrm{O_b}$ bonded to the
$\mathrm{Au}$ adatom.
This value of the charge
transfer has been obtained by integrating the bonding charge density on planes
parallel to the surface from the center of the vacuum region to the center of
the $\mathrm{O}$--$\mathrm{Au}$--$\mathrm{Ti}$ bond (see 
left panel 
of
Fig.~\ref{fig5} {\bf (B)}). 
As demonstrated by the PDOS analysis 
shown in the right panel
of Fig.~\ref{fig5} {\bf (B)}, in this
configuration all the $\mathrm{Ti}$ ions belonging to the substrate preserve 
their formal oxidation state $\mathrm{Ti^{4+}}$.
%

The second identified stable site is a top 
site (see central panel of Fig.~\ref{fig5} {\bf (C)}),
where the $\mathrm{Au}$ adatom is adsorbed on top of 
an
$\mathrm{O_b}$ atom. 
The corresponding PBE+U (PBE)  adsorption energy and the
$\mathrm{Au}$--$\mathrm{O}$ bond length are $-0.48$~eV ($-0.29$~eV) and 2.00~\AA\
(2.16~\AA ), 
%
%
%
akin to previous GGA studies~\cite{vijay,graciani}. 
Again, a net charge transfer from metal to 
surface, leading to a positively charged $\mathrm{Au^{\delta +}}$ ion, is observed. 
In this case, however, the magnitude of the charge 
transfer, 0.35~$|e|$, is more significant, i.e. three times larger than in the 
previous case. 
The excess charge in the substrate is now mostly localized around the
surface $\mathrm{O_b}$ atom bound to the $\mathrm{Au^{\delta +}}$ 
and a 
second--layer
$\mathrm{Ti}$ ion which reduces 
$\mathrm{Ti^{4+}}$ $\rightarrow$ $\mathrm{Ti^{3+}}$ (see Fig.~\ref{fig5} {\bf (C)}). 
The reduced $\mathrm{Ti^{3+}}$ 
ion
is located 
at a site
adjacent to the
$\mathrm{Au^{\delta +}}$ adatom in the second subsurface
layer under the $\mathrm{Ti_{5c}}$. 
These findings are corroborated by the computed DOS plotted in Fig.~\ref{fig5} 
{\bf (C)},
which displays two features in the band gap. 
The projected DOS analysis reveals that the filled 
state 
below the Fermi level and closest to the valence band results from the charge
transferred from the metal to the substrate being localized on a second--layer 
reduced $\mathrm{Ti^{3+}}$ atom. 
The unoccupied level closest to the conduction band is
instead related to the $\mathrm{Au}$--$\mathrm{O}$ bonding. 
The 6{\it s} levels of $\mathrm{Au}$ are partially empty and 
are located 
above the Fermi 
level, 
leading to the $\mathrm{Au}$ oxidation.
%

In summary, the PBE+U calculations predict two lowest--energy configurations for
$\mathrm{Au}$ adsorption on the stoichiometric $\mathrm{TiO_{2}}$(110)
surface: a bridge site with the $\mathrm{Au}$ adatom adsorbed between 
$\mathrm{O_b}$ and $\mathrm{Ti_{5c}}$ atoms and a top site with the
$\mathrm{Au}$ adatom adsorbed on top of 
an
$\mathrm{O_b}$ atom. 
Once adsorbed 
at
the bridge 
site,
a very weak oxidation of the $\mathrm{Au}$ adatom is
observed. 
On the other 
hand,
the adsorption process of $\mathrm{Au}$ on top of 
an
$\mathrm{O_b}$ atom induces a net charge transfer from the adsorbate to the
substrate,
leading to the formation of a distinctly positively 
charged $\mathrm{Au^{\delta+}}$ species
where about a third of an electron is transferred from the metal atom to the
oxide substrate. 
A qualitatively similar scenario has been observed in recent studies 
of the related $\mathrm{Au/CeO_{2}}$ system~\cite{Ce,Ce1,Ce2}. 
%
However, 
unlike
titania, 
with
ceria the charge transfer involved in
the adsorption of $\mathrm{Au}$ on the stoichiometric oxide surface 
always leads
to the
reduction of a substrate $\mathrm{Ce}$ ion. 
In addition, the fact that the excess charge 
$\delta -$, stemming from Au in the present case,
localizes on a 
second--layer
$\mathrm{Ti}$ ion is in line with our recent findings
on reduced titania surfaces~\cite{PMD}.

\begin{table}
\caption{%
Adsorption energies ${E_{\rm ads}}$ (in~eV) of $\mathrm{Au}$ and $\mathrm{CO}$ species on
  stoichiometric and reduced
  rutile $\mathrm{TiO_{2}}$(110) surfaces and
  on the $\mathrm{Au}$/$\mathrm{TiO_{2}}$ metal/support system;
  see text for labeling. 
  In the second
  column,
  the number of reduced $\mathrm{Ti^{3+}}$ ions 
  present in the substrate is reported.  \label{tab4} }
\begin{ruledtabular}
\begin{tabular}{ccccccc}
   &  &${E_{\rm ads}}$(PBE+U) &$\mathrm{Ti^{3+}}$  & ${E_{\rm ads}}$(PBE)  \\
\hline
%
 & $\mathrm{Au@O(bridge)}$   & $-0.58$   &  0   & $-0.41$    \\
 & $\mathrm{Au@O(top)}$      & $-0.48$   &  1   & $-0.29$   \\
 & $\mathrm{Au@V_{O}}$ & $-1.54$   & 1 & $-1.57$   \\
 & $\mathrm{Au_{2}@V_{O}}$ &  $-1.17$   & 2 & $-1.19$   \\
 & $\mathrm{Au@V_{Ti{5c}}}$ &$-6.38$  &0 &  $-6.21$ \\
\hline
\hline
 &  $\mathrm{CO@Au@O(bridge)}$& $-2.27$ & 1 & $-2.17$  \\
 &  $\mathrm{CO@Au@O(top)}$ &$-2.66$ & 1 & $-2.33$   \\
 & $\mathrm{CO}$-$\mathrm{Au@V_{O}}$ &$-0.41$ &1 & $-0.32$  \\
& $\mathrm{CO}$-$\mathrm{Au_{2}@V_{O}}$ &$-1.22$ &2 & $-1.00$ \\
\end{tabular}
\end{ruledtabular}
\end{table}

\subsection{Au adatom diffusion on the stoichiometric $\mathrm{TiO_{2}(110)}$ surface}

Several previous theoretical studies~\cite{matthey,iddir,anke} explored 
the PES of a single $\mathrm{Au}$
adatom deposited
on the stoichiometric $\mathrm{TiO_{2}(110)}$ surface or adsorbed 
onto
a $\mathrm{V_O}$ vacancy site  
using static 
calculations, 
including 
nudged elastic band (NEB) mappings~\cite{neb}. 
%
A key finding of these investigations is 
that the PES for $\mathrm{Au}$ migration is 
quite flat, with low energy barriers.
%
This indicates that
$\mathrm{Au}$ might diffuse rather easily on the stoichiometric surface.
The estimated values of the energy barriers~\cite{matthey,iddir,anke} agree
with the experimental observation of facile $\mathrm{Au}$ diffusion 
on the oxide surface even at temperatures as low as 140~K, 
as well as
the estimates for the binding energy of 0.5~eV and small migration barriers
of 0.07~eV (see Refs.~\onlinecite{parker,Campbell}). 
%
%
%

Inspired by these 
findings,
we decided to perform explicit dynamics using unconstrained 
{\it ab initio} molecular dynamics~\cite{dominik} in order to reveal 
the mechanism of 
diffusion of an $\mathrm{Au}$ adatom on the
stoichiometric $\mathrm{TiO_{2}}$(110) surface. 
In order to probe the dynamics more 
efficiently,
the temperature of the
simulations 
was
set to $T = 900$~K using the Car-Parrinello scheme~\cite{carparrinello}
to propagate the system 
consistently using
the PBE+U functional. 
The selected temperature is far above ambient yet sufficiently low 
so as to not decompose the surface. Thus,
the phonon dynamics is accelerated and
the sampling of the PES is enhanced on the
picosecond AIMD time scale. 
%
As starting configurations for the AIMD simulations, we employed
one
%
%
structure in which the $\mathrm{Au}$ adatom is adsorbed on
top of 
an
$\mathrm{O_b}$ atom 
(see Fig.~\ref{fig5}~{\bf (C)} and Table \ref{tab4}) 
and 
a second structure
where it
is adsorbed in a bridge position between 
an
$\mathrm{O_b}$ atom
and a $\mathrm{Ti_{5c}}$ atom
(see Fig.~\ref{fig5}~{\bf (B)} and Table \ref{tab4}).
After equilibrating the structures at 300~K for several 
picoseconds,
the system
was heated 
to the target temperature of 900~K for 
the present analysis.
%

\begin{figure*}
\begin{center}
\begin{tabular}{ccc}
\subfigure{\includegraphics[width=5cm]{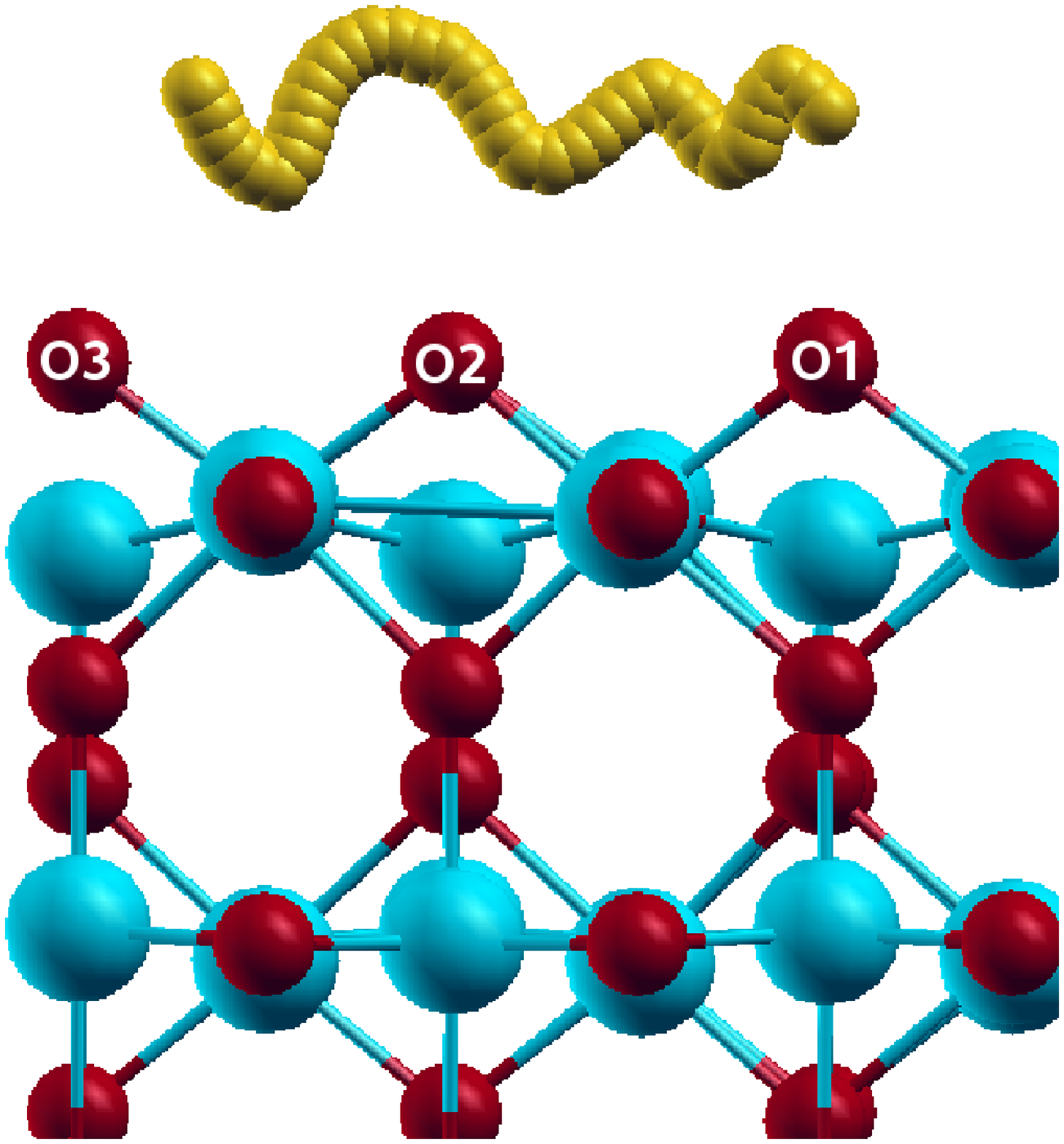}}\hspace{0.50cm}
\subfigure{\includegraphics[width=4cm]{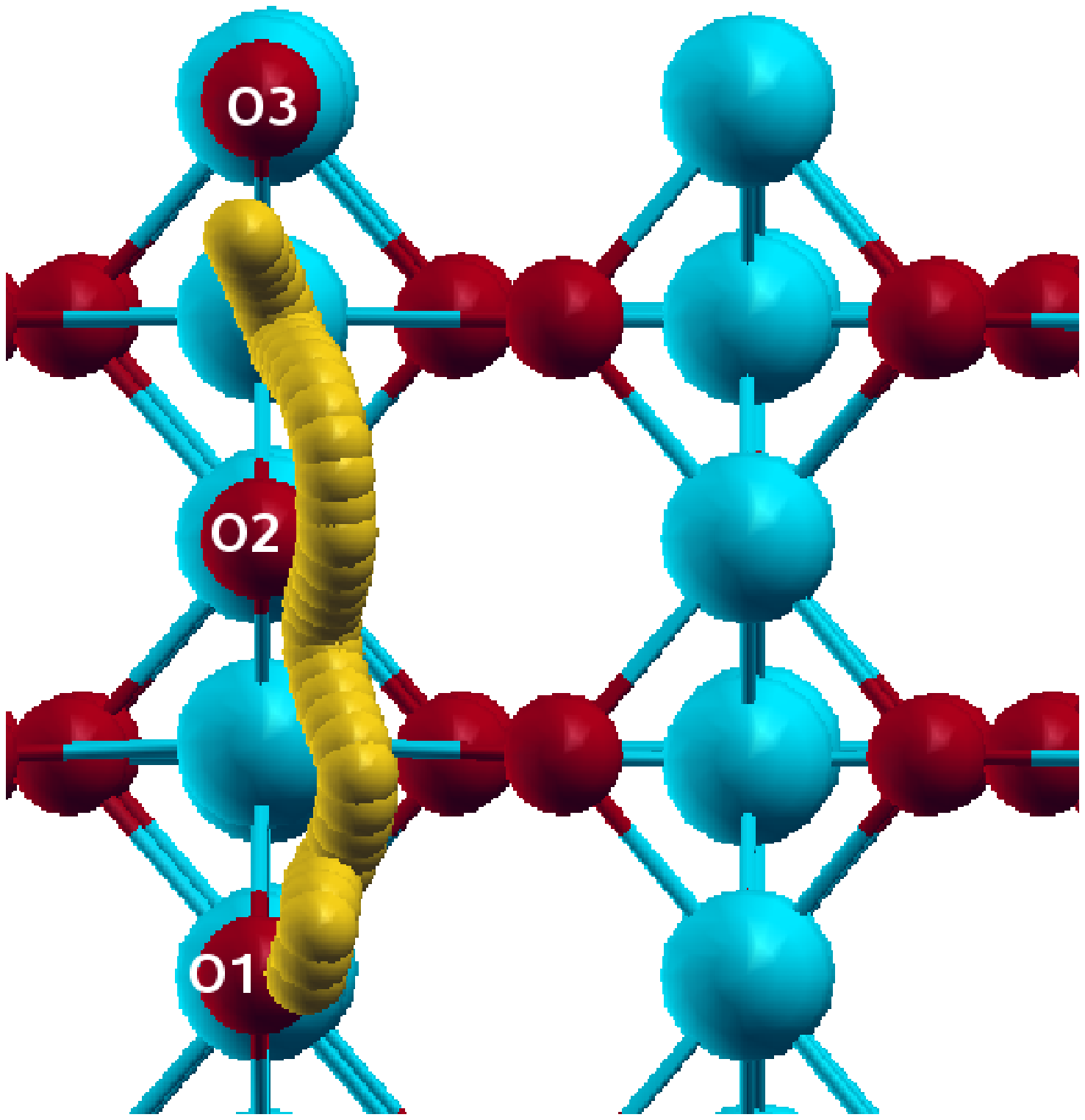}}\hspace{0.50cm}
\subfigure{\includegraphics[width=7cm]{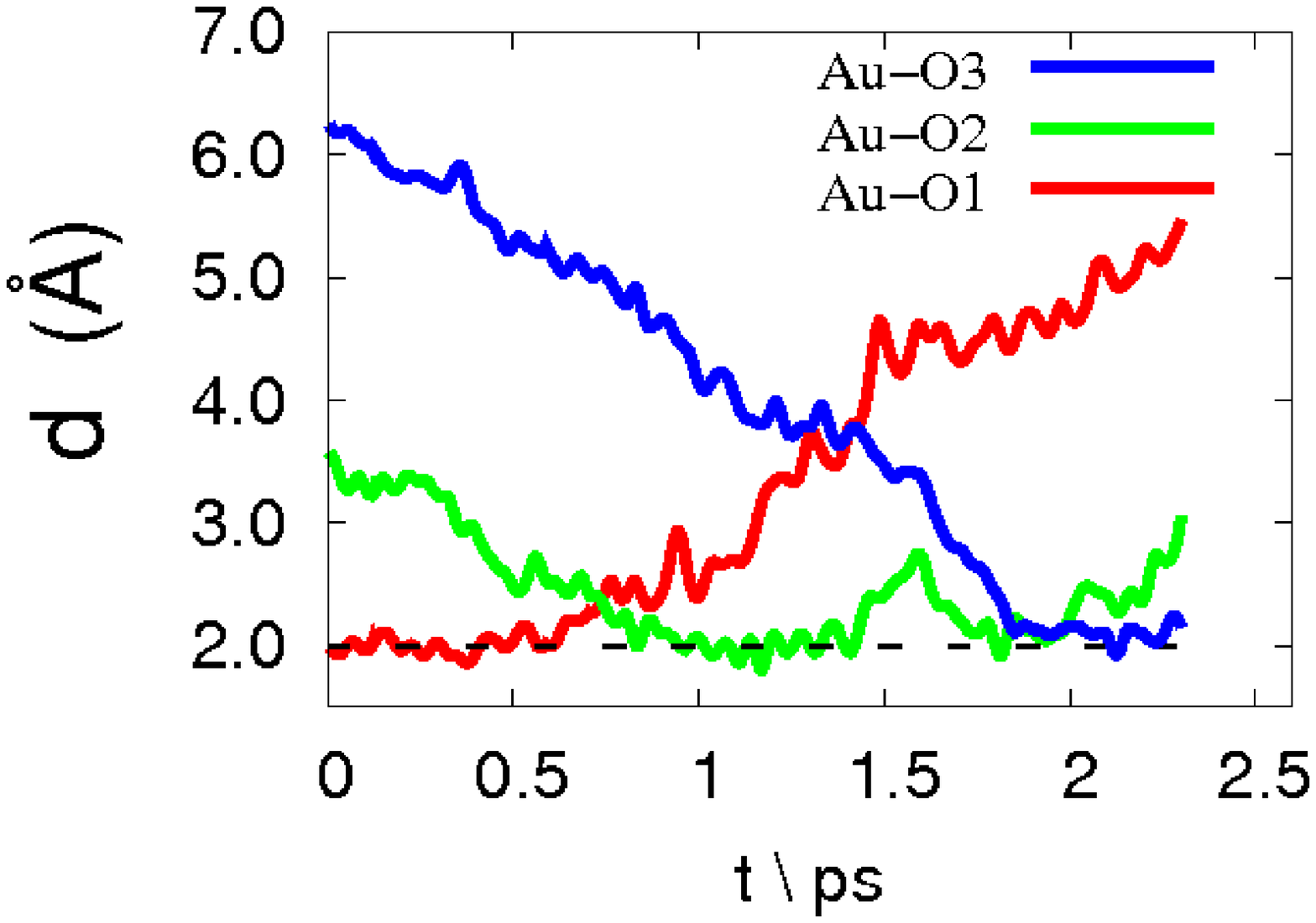}}
\end{tabular}
\caption{%
Side view (left panel) and top view (central panel) of the diffusion path 
of the $\mathrm{Au}$ adatom on the stoichiometric
$\mathrm{TiO_{2}}$(110) surface at 900~K. 
The $\mathrm{Au}$ atom, 
initially adsorbed on top of the bridging oxygen atom 
$\mathrm{O1}$,  
diffuses along the $\mathrm{O_b}$ bridging row by hopping 
from one 
oxygen to the next,
$\rm O1 \to O2 \to O3$,  
as visualized by the worm--like trajectory. 
The right panel shows the time evolution of the
$\mathrm{Au}$--$\mathrm{O{\it n}}$ ($\mathrm{\it n}$=1,2,3) bond lengths;
the dashed line at 2.00~\AA\  refers to the optimized equilibrium 
$\mathrm{Au}$--$\mathrm{O}$ bond length.
%
%
\label{fig6}}
\end{center}
\end{figure*}

Let us first consider the scenario with the $\mathrm{Au}$ adatom adsorbed 
on top of a surface $\mathrm{O_b}$ 
atom,
labeled as site 
$\mathrm{O1}$ 
in Fig.~\ref{fig6}, where 
the diffusion path of the $\mathrm{Au}$ adatom on the stoichiometric
$\mathrm{TiO_{2}}$(110) surface is visualized. 
%
During
the 
simulation,
the $\mathrm{Au}$ adatom diffuses in the [001] direction 
along the row of bridging oxygen atoms, $\mathrm{O_b}$. 
Adatom diffusion is mediated by 
the $\mathrm{Au}$ atom hopping
between nearest--neighbor oxygen atoms. 
As demonstrated 
in
Fig.~\ref{fig6}, the $\mathrm{Au}$ adatom is originally bonded to the
surface bridging $\mathrm{O1}$ atom with an $\mathrm{Au}$--$\mathrm{O1}$ bond
length of $\sim 2$~\AA\ (red line); it diffuses along the $\mathrm{O_b}$
row and after $\sim 0.75$~ps reaches a configuration 
where
it is
equidistant between the $\mathrm{O1}$ and $\mathrm{O2}$ atoms. 
%
Then it jumps on top of the row's next atom, the $\mathrm{O2}$ site, where
the $\mathrm{Au}$--$\mathrm{O2}$ bond length is
$\sim 2$~\AA\ (green line). 
The $\mathrm{Au}$ diffusion proceeds along the
$\mathrm{O_b}$ row and at about 2~ps the $\mathrm{Au}$ adatom is shared
between $\mathrm{O2}$ and the next site, $\mathrm{O3}$ (blue line),
until it jumps on top of $\mathrm{O3}$ forming a bond of $\sim 2$~\AA . 
%
%
%
%
%

The charge localization and charge hopping dynamics along the adatom migration path 
is monitored by computing, as a function of time, the occupation matrix 
of each $\mathrm{Ti}$ ${\it d}$--$\mathrm{\alpha}$ and ${\it d}$--$\mathrm{\beta}$ 
spin orbital along the trajectory (same analyis as in~\cite{PMD}). 
As shown in Fig.~\ref{fig8} the excess charge donated 
by the $\mathrm{Au}$ atom to the substrate is initially localized on 
the second--layer $\mathrm{Ti2}$ site 
but transfers from there to site $\mathrm{Ti3}$ 
on the sub--picosecond time scale (at $t\sim 0.75$~ps). 
%
As seen by comparing Fig.~\ref{fig8} to Fig.~\ref{fig6},
one observes 
that $t\sim 0.75$~ps corresponds exactly to the jump of 
$\mathrm{Au}$ from the $\mathrm{O1}$ to the $\mathrm{O2}$ site.
%
%
A qualitatively similar scenario happens at about 2~ps which corresponds 
to the next hopping event of the $\mathrm{Au}$ atom 
from site $\mathrm{O2}$ to 
$\mathrm{O3}$.
Thus, the motion of the surface gold adatom along the row of 
bridging oxygen atoms, $\mathrm{O_b}$, appears to be fully correlated with the 
localization and hopping dynamics of the excess charge
injected into the oxide support 
in the second layer of Ti~atoms.
%
%
%

This dynamical scenario is distinctly different 
from what has been found recently for 
the excess charge induced by oxygen vacancies $\mathrm{V_O}$ in the bridging row
on the same substrate~\cite{PMD}.
In the presence of gold 
adatoms, the present simulations suggest a more localized configuration for the excess electron. 
This localized electron appears to preferentially populate sites in the vicinity of the $\mathrm{Au}$ atom
and to closely follow the motion of the oxidized adatom.
%
Another interesting phenomenon observed 
during
the simulation is the absence of
excess
charge on the substrate at about 1.5~ps.
%
%
Computing the spin density close to $t=1.5$~ps 
we observe
that the charge localized at
second--layer $\mathrm{Ti}$ sites disappears from
the substrate 
and goes to the
$\mathrm{Au}$ adatom,
where it sits for a fraction of 
a picosecond
before returning
to the substrate and
occupying the $\mathrm{Ti3}$ site. 
%
%
%
%
As observed in our previous work~\cite{PMD} the excess charge populating specific 
second--layer $\mathrm{Ti}$ sites and coming from the $\mathrm{Au}$ adatom 
adsorbed on top of $\mathrm{O_b}$ atoms
migrates easily by phonon--assisted (thermally activated) hopping to other
$\mathrm{Ti}$ sites.
%

\begin{figure*}
\begin{center}
\begin{tabular}{ccc}
\subfigure{\includegraphics[width=5cm]{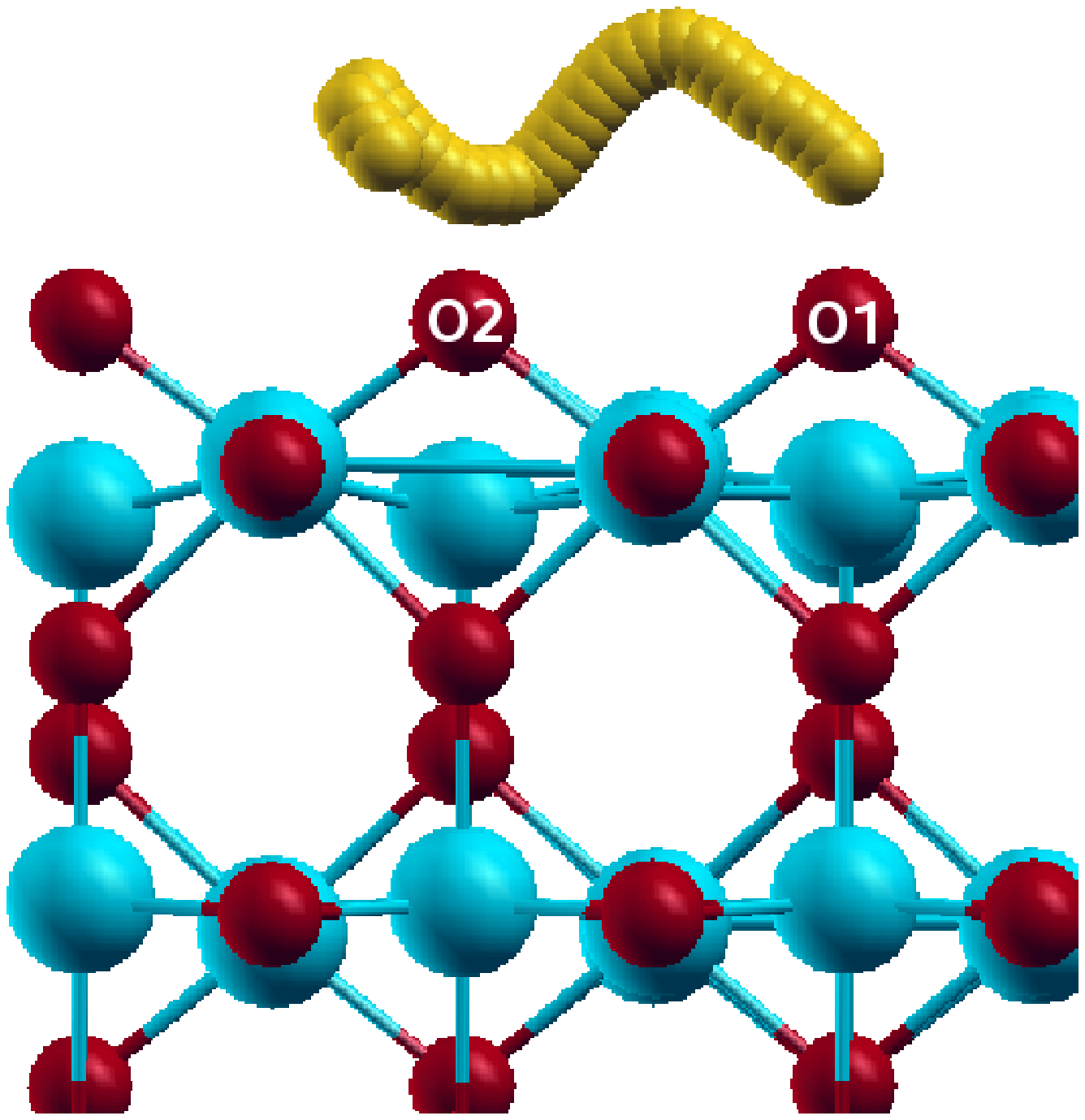}}\hspace{0.50cm}
\subfigure{\includegraphics[width=4cm]{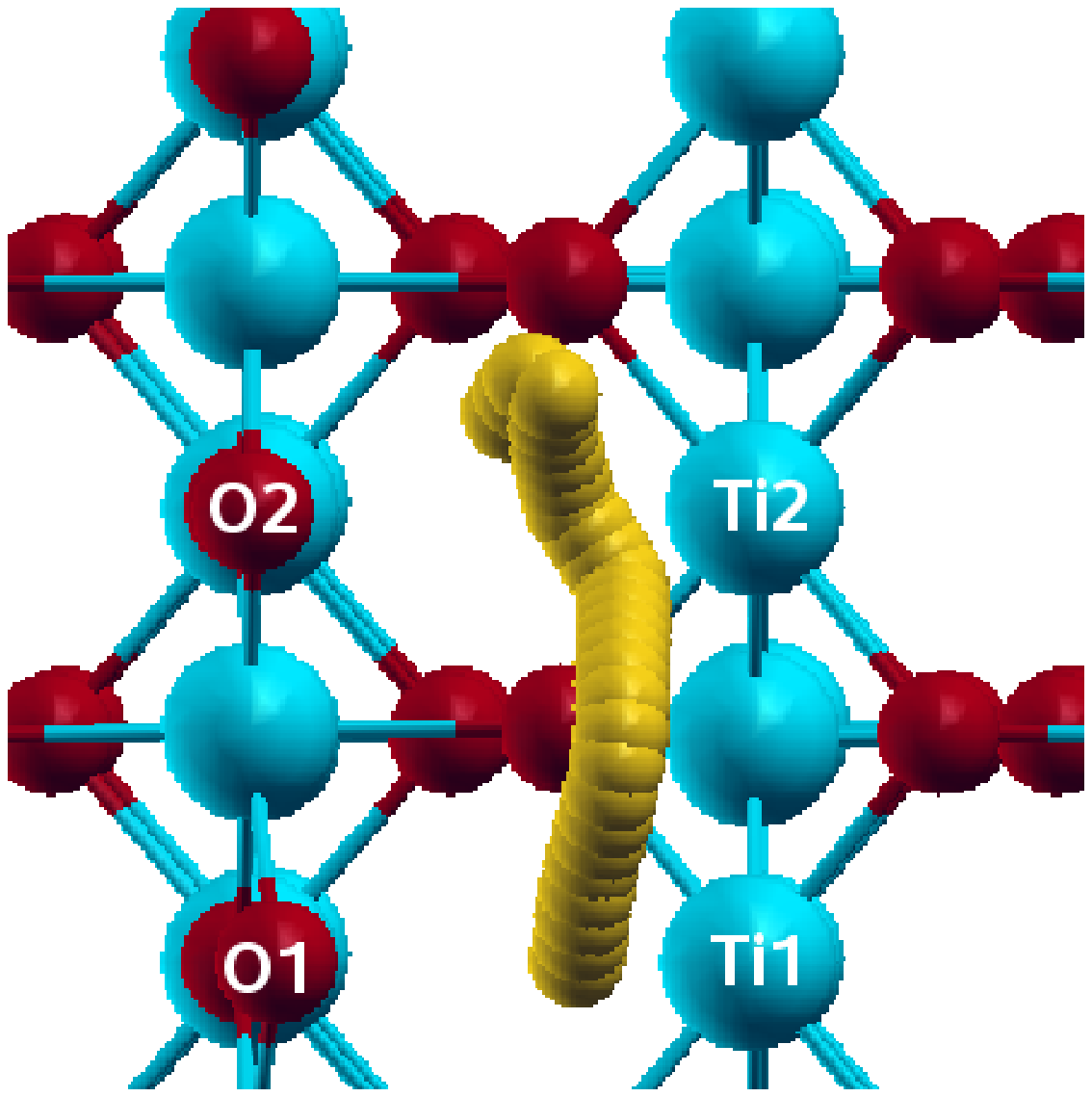}}\hspace{0.50cm}
\subfigure{\includegraphics[width=7cm]{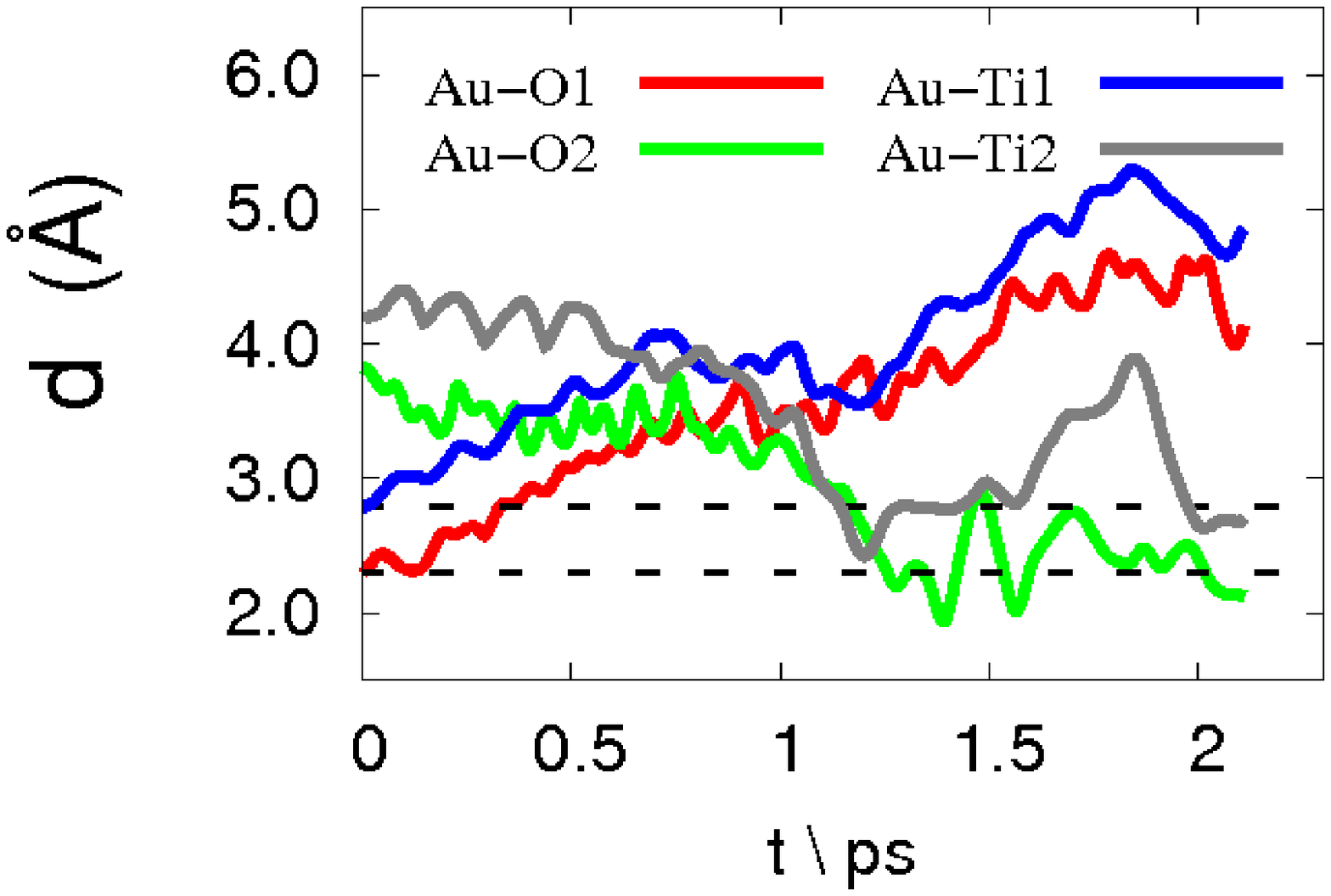}}
\end{tabular}
\caption{%
Side view (left panel) and top view (central panel) of the diffusion path 
of the $\mathrm{Au}$ adatom on the stoichiometric
$\mathrm{TiO_{2}}$(110) surface at 900~K. 
%
The $\mathrm{Au}$ atom, 
initially adsorbed on a bridge site between the 
$\mathrm{O1}$ and 
$\mathrm{Ti1}$ atoms, 
%
%
diffuses 
by hopping 
in between pairs of
nearest--neighbor $\mathrm{O_b}$ bridge and $\mathrm{Ti_{5c}}$ atoms 
as visualized by the worm--like trajectory. 
The right panel shows the time evolution of the
$\mathrm{Au}$--$\mathrm{O{\it n}}$ 
and $\mathrm{Au}$--$\mathrm{Ti{\it n}}$ 
($\mathrm{\it n}$=1,2) bond lengths;
the dashed lines at 2.30 and 2.80~\AA\  refer to the optimized equilibrium 
$\mathrm{Au}$--$\mathrm{O}$ 
and $\mathrm{Au}$--$\mathrm{Ti}$
bond lengths, respectively.
%
%
%
%
\label{fig7}}
\end{center}
\end{figure*}

\begin{figure*}
\begin{center}
\begin{tabular}{ccc}
\subfigure{\includegraphics[width=4.5cm]{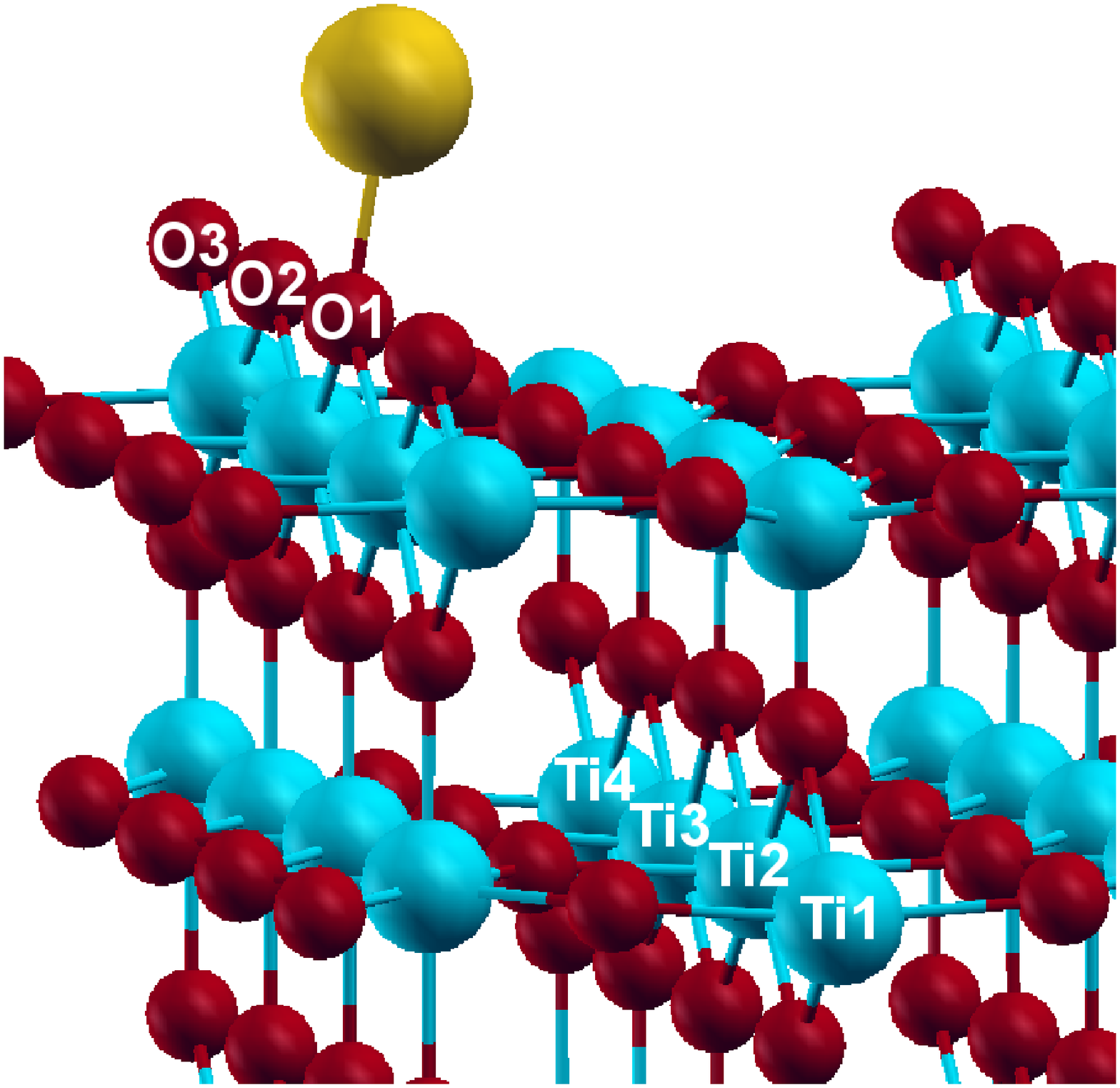}}\hspace{0.50cm}
\subfigure{\includegraphics[width=3cm]{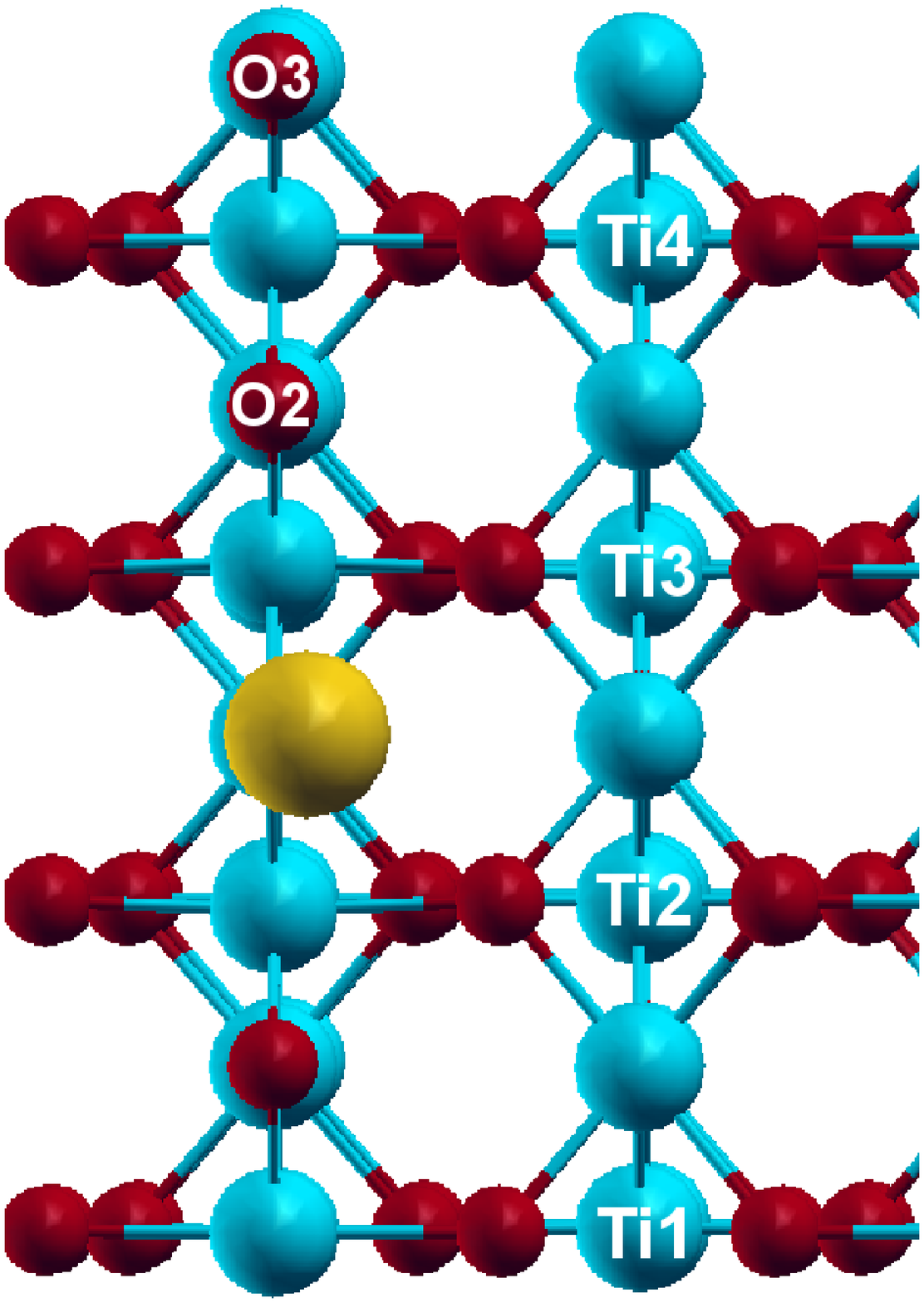}}\hspace{0.50cm}
\subfigure{\includegraphics[width=6cm]{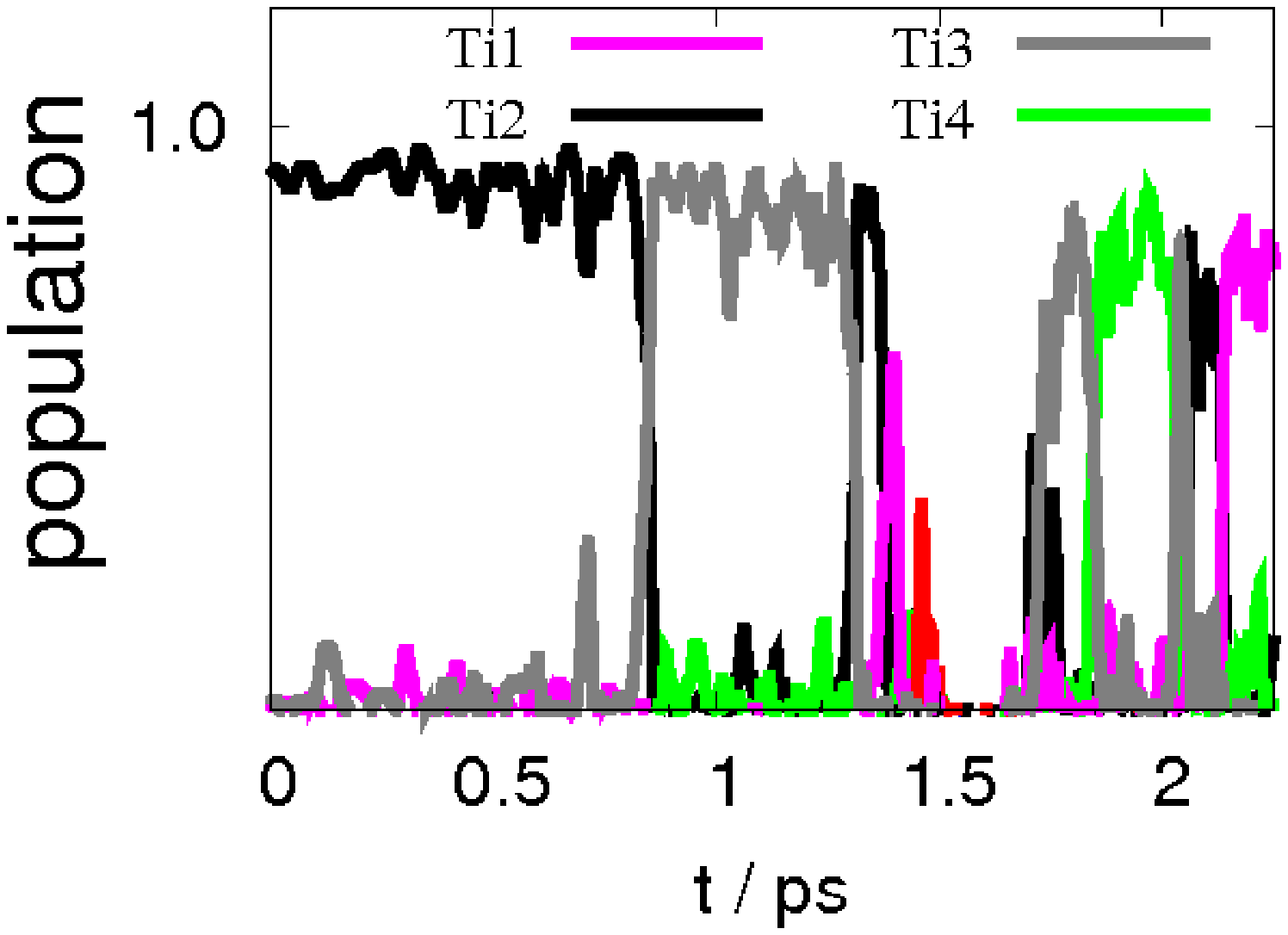}}
\end{tabular}
\caption{
%
%
%
%
Side view (left panel) and top view (central panel) of the $\mathrm{Au}$ adatom 
adsorbed on top of the surface bridging oxygen atom O1, which served
as the
initial configuration for the migration path depicted in Fig.~\ref{fig6}.
Right panel:
%
%
%
Corresponding time evolution of the fractional occupation number of particular 
$\mathrm{Ti}$--3{\it d} orbitals at specific 
sites, as indicated;
populations of about 0 and 1 correspond to
$\mathrm{Ti^{4+}}$ and $\mathrm{Ti^{3+}}$, respectively.
At 
$t=0$~ps,
the Au~adatom is bound to O1 and the excess charge is fully 
localized in the second subsurface layer at the $\mathrm{Ti2}$ site.
\label{fig8}}
\end{center}
\end{figure*}

Next we consider the situation where the $\mathrm{Au}$ adatom is initially 
adsorbed 
at
a bridge position between 
an
$\mathrm{O_b}$ atom
and a $\mathrm{Ti_{5c}}$ 
atom (see Fig.~\ref{fig7}).
There it forms two bonds with the $\mathrm{O1}$ and $\mathrm{Ti1}$ 
atoms,
with $\mathrm{Au}$--$\mathrm{O1}$ and $\mathrm{Au}$--$\mathrm{Ti1}$ bond
lengths of $\sim 2.3$~\AA\ (red line) and $\sim 2.8$~\AA , respectively.
Also in this 
case,
the gold atom originally bonded to the $\mathrm{O1}$ and 
$\mathrm{Ti1}$ atoms is found to diffuse exclusively along the [001]~direction. 
%
%
%
%
%
%
Now,
however, the gold atom hops between pairs of nearest--neighbor 
$\mathrm{O}$ and $\mathrm{Ti}$ atoms.
After 
$\sim 0.8$~ps, 
a configuration is reached in which 
the gold atom
is equidistant between
the $\mathrm{O1}$ and $\mathrm{O2}$ (and $\mathrm{Ti1}$ and $\mathrm{Ti2}$) sites
before it jumps into another bridge position between the $\mathrm{O2}$ 
(green line) and $\mathrm{Ti2}$ sites.
%
%
%
By performing static calculations,
we 
show
that $\mathrm{Au}$ adatoms 
adsorbed on bridge sites do not induce reduction of the 
substrate (see Fig. \ref{fig5}~{\bf (B)}),
thus all the $\mathrm{Ti}$ ions of the
substrate preserve their $4+$ oxidation state and a very weak oxidation 
of the $\mathrm{Au}$ adatom 
is
observed.
%
%
%
This is fully confirmed by the dynamical simulations: along the 
trajectory,
the $\mathrm{Ti}$--3{\it d} orbitals are found to be 
empty,
which implies
that no localization of charge on substrate Ti~sites is observed. 
%
%

%
%
Even at an elevated temperature of 
900~K,
we do not observe diffusion 
of the $\mathrm{Au}$ adatom in the [1$\bar{1}$0] direction, namely from 
the top of $\mathrm{O_b}$ atoms to bridge sites between $\mathrm{Ti_{5c}}$ 
and $\mathrm{O_b}$ atoms, on the timescale of picoseconds. 
This dynamics is consistent with previous findings~\cite{iddir,matthey} based 
on static or NEB~\cite{neb} calculations which predict a relatively 
high energy barrier, $\sim 0.35$~eV, for this process to happen
compared to others.            
%
%
%
In conclusion, dynamical PBE+U simulations demonstrate that Au adatoms diffuse 
highly directionally on the stoichiometric rutile (110) surface.
They can easily migrate 
either 
along the top of the bridging oxygen rows of the clean $\mathrm{TiO_{2}}$(110) surface
or around the area between 
these rows from one bridging position to the next one
along the [001]~direction.
We did not observe, on 
the
picosecond timescale, translational motion 
perpendicular to this direction, e.g. from one $\rm O_b$ row to a 
neighboring 
row
via suitable bridging positions.

%
%
%
\subsection{Au adatom adsorption on the reduced $\mathbf{TiO_{2}(110)}$ surface}

Starting with the reduced $\mathrm{TiO_{2}}$(110) surface in the presence of 
an
$\mathrm{O_b}$ surface oxygen vacancy
the $\mathrm{Au}$ metal adatom is found to adsorb 
preferentially 
at the $\mathrm{V_O}$ vacancy. 
Such surface oxygen vacancies result in stable anchoring
sites for $\mathrm{Au}$ adatoms, which bind at about 0.90~\AA\ above the
$\mathrm{O}$ vacancy site with two $\mathrm{Ti}$ nearest neighbors at
2.68~\AA\ obtained from both PBE+U and PBE.
%
The corresponding PBE+U (PBE) calculated adsorption energies
of $-1.54$~eV ($-1.57$~eV) are much larger than the binding to the
stoichiometric 
surface (see Table~\ref{tab4}),
which is consistent ($-1.6$ to $-1.8$~eV) with previous GGA calculations~\cite{anke}.
The strong adsorption of an $\mathrm{Au}$ atom at the $\mathrm{V_O}$ site
entails a strong charge rearrangement at the $\mathrm{Au}$/oxide contact. 
In 
the
presence of an isolated $\mathrm{V_O}$ 
vacancy,
the charge neutrality of the
system is maintained by the presence of two reduced $\mathrm{Ti^{3+}}$ ions. 
The bonding charge distribution (see Fig. \ref{fig5}~{\bf (D)}) 
shows that, 
upon $\mathrm{Au}$ adsorption at the $\mathrm{V_O}$ vacancy site, the 
charge transfer occurs now from the reduced substrate to the supported metal 
atom,
thus leaving a reduced surface with a single $\mathrm{Ti^{3+}}$ ion. 
This indicates that the charge transferred from the reduced substrate to the 
adsorbate comes from one of the two $\mathrm{Ti^{3+}}$ ions. 
As a result this process leads to the formation of a negatively charged
$\mathrm{Au^{\delta -}}$ adspecies. 
The analysis of the DOS (see Fig. \ref{fig5}~{\bf (D)}) 
now shows that the charge 
%
%
transferred from the substrate to the $\mathrm{Au}$ atom 
moves
toward 
the latter's
half--filled
6{\it s} 
band,
which turns out 
to be almost completely 
filled.
%
%
%
In conclusion, these calculations not only suggest a 
greatly increased
stability of $\mathrm{Au}$ adatoms 
adsorbed 
onto
$\mathrm{O}$ vacancies when compared to the stoichiometric surface
as a reference,
but also a very different chemical reactivity with respect to 
admolecules in view of their 
different charge state.
The resulting ramifications 
for
CO activation will be discussed
in Sec.~\ref{sec:co-au-tio2}. 
%

%
%
%

\subsection{Au substitutional defects on the $\mathbf{TiO_{2}(110)}$ surface: 
%
%
$\mathbf{Au_{x}Ti_{(1-x)}O_{2-\delta}}$}

\begin{figure}[h]
\begin{center}
\begin{tabular}{cc}
\subfigure{\includegraphics[angle=0,scale=0.25]{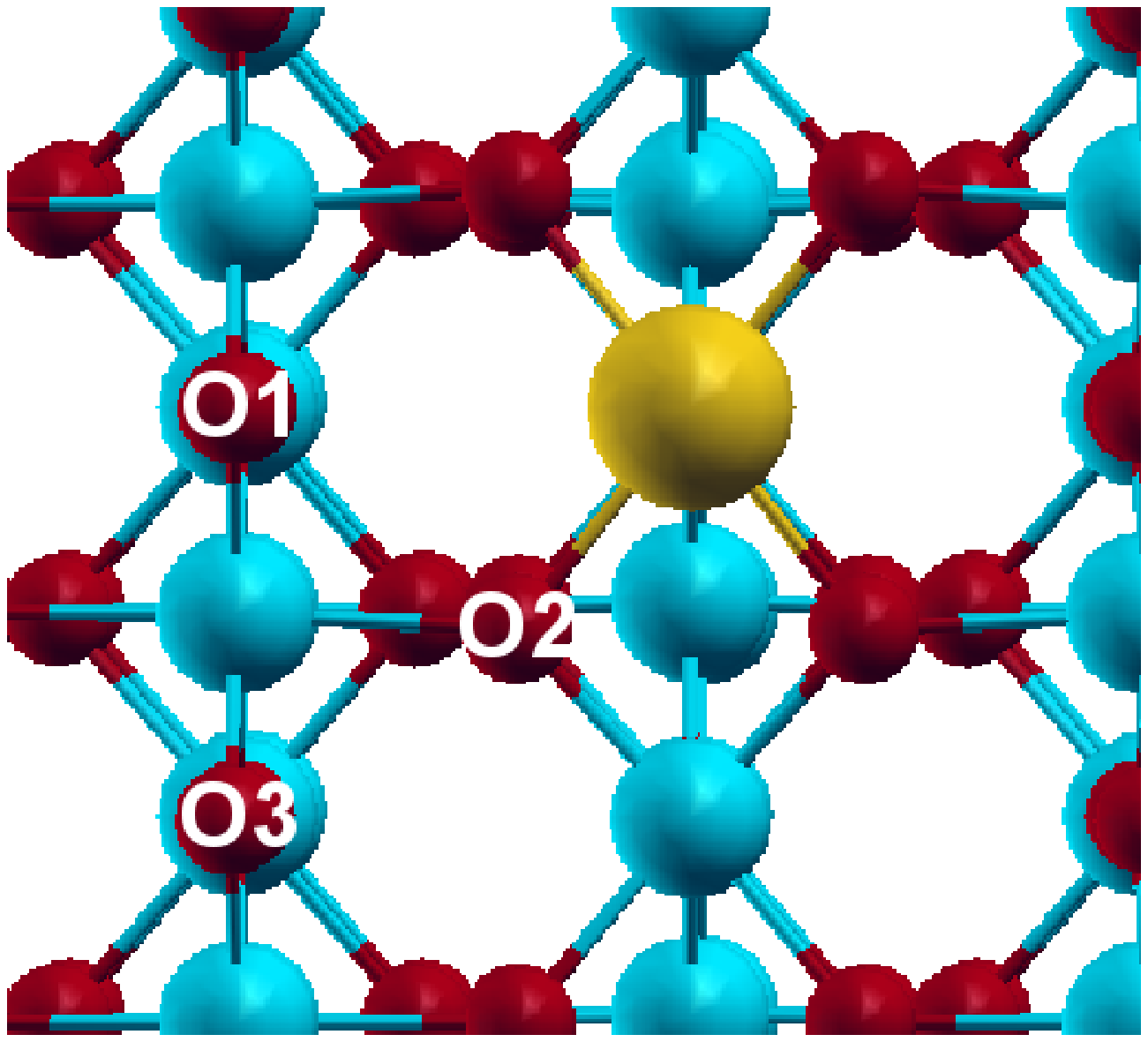}}\hspace{1.00cm}
\subfigure{\includegraphics[angle=0,scale=0.25]{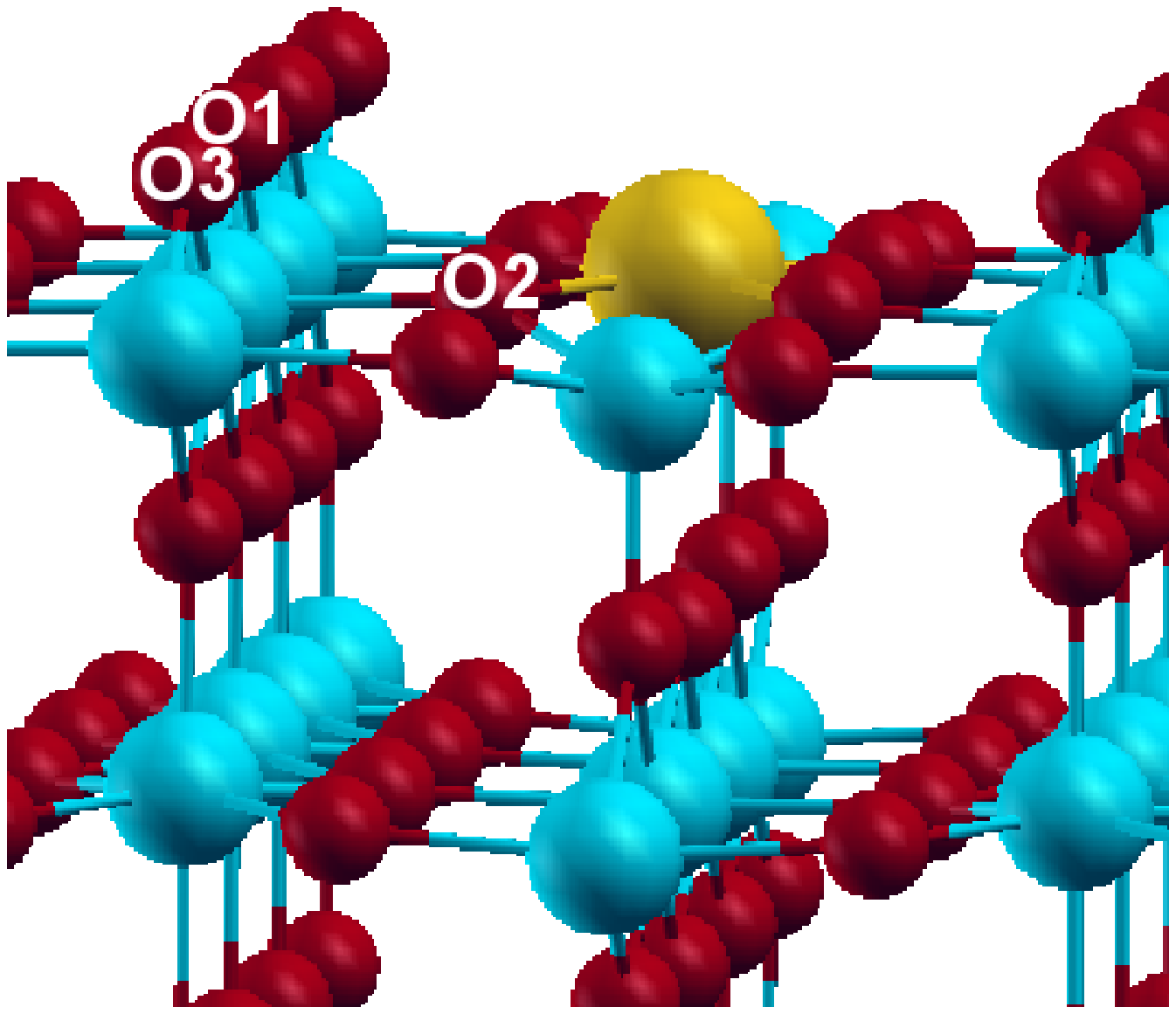}}
\end{tabular}
\caption{%
Ball and stick model of 
an
$\mathrm{Au_{x}Ti_{(1-x)}O_{2}}$ system
  obtained by substituting a surface $\mathrm{Ti_{5c}}$ atom by
  an $\mathrm{Au}$ adatom in side (left) and top (right) views.
%
%
%
\label{fig9}}
\end{center}
\end{figure}

Another reaction channel of gold interacting with titania surfaces
is 
via the chemical exchange of Ti~atoms.
%
We studied 
the scenario where an $\mathrm{Au}$ atom substitutes a surface
$\mathrm{Ti_{5c}}$ site, which we call ``$\mathrm{Au@V_{Ti5c}}$''.
%
%
%
The presence of such 
an
$\mathrm{Au}$ substitutional point defect
induces a rearrangement of the neighboring atoms, leading to the formation of a
distorted squared planar ``$\mathrm{AuO_{4}}$'' unit (see Fig.~\ref{fig9}). 
In this 
configuration,
the Au atom relaxes outward by 0.62~\AA\ and is found to be
coordinated by four surface $\mathrm{O}$ atoms 
%
%
(at about 2.0~\AA ). 
The incorporation of an $\mathrm{Au}$ atom into the titania surface does not yield a change 
in the occupation of the $\mathrm{Ti}$--3{\it d} states: all $\mathrm{Ti}$ ions preserve
their formal oxidation state $\mathrm{Ti^{4+}}$. 
However, the adsorption of an $\mathrm{Au}$ adatom into a $\mathrm{V_{Ti{5c}}}$ vacancy site 
is strongly exothermic, releasing $-6.38$~eV (or $-6.21$~eV when using 
PBE),
shown in
Table~\ref{tab4}. 
%

This 
particular defective surface 
is found to be extremely reactive. 
%
%
%
%
We have computed the energy required 
to remove
one of the oxygen
atoms $\mathrm{O{\it n}}$ ($n=1$, 2, 3) in the surface layer 
(see Fig. \ref{fig9} for site labeling). 
%
%
%
%
%
The atoms $\mathrm{O1}$ and $\mathrm{O3}$ are
$\mathrm{O_b}$ atoms while the atom $\mathrm{O2}$ is an in--plane oxygen. 
%
Our PBE+U values for the $\mathrm{O1}$, 
$\mathrm{O2}$, 
and
$\mathrm{O3}$ vacancy formation energies are 1.52, 
2.20,
and 1.36~eV
%
%
respectively, which is in 
accord
with previous PBE calculations~\cite{cretien}. 
For the stoichiometric, undoped $\mathrm{TiO_{2}(110)}$
surface,
the corresponding PBE+U value of $E_{\rm{V}}^{\rm{O}}$ is 2.97~eV for 
an oxygen vacancy, $\rm V_O$, in the bridging 
row,
compared to O1 and O3 here. 
%
%
%
These values suggest that 
substituting
a surface $\mathrm{Ti_{5c}}$ atom 
with
an
$\mathrm{Au}$ atom greatly weakens the binding of surface $\mathrm{O}$ 
atoms,
as observed
in the case of $\mathrm{CeO_{2}}$ surfaces~\cite{Ce,Ce2}.

In 
Fig.~\ref{fig5}~{\bf (E)},
the total DOS and PDOS of
the doped $\mathrm{Au@V_{Ti{5c}}}$ surface is depicted in the
presence of an oxygen vacancy.
%
Here, the missing oxygen is the bridging $\mathrm{O3}$~atom,
$\mathrm{O_{3}}$~atom (see Fig.~\ref{fig9}),
the vacancy formation energy 
of which 
is found to be
lowest (1.36~eV)
when Au substitutes a Ti atom at a $\mathrm{Ti_{5c}}$ site, which we call 
``$\mathrm{Au@V_{Ti{5c}},V_{O3}}$''. 
%
%
%
%
Concerning the electronic 
structure,
the main difference between the two
scenarios, i.e. one obtained by substituting a $\mathrm{Ti_{5c}}$ 
with
an $\mathrm{Au}$
atom $\mathrm{Au@V_{Ti{5c}}}$ and the other 
generated by 
substituting a $\mathrm{Ti_{5c}}$ with 
an
$\mathrm{Au}$
atom and by removing a surface $\mathrm{O_b}$ atom
$\mathrm{Au@V_{Ti{5c}},V_{O3}}$, is related to the reduction of
the oxide substrate. 
%
As we can extract
from Fig.~\ref{fig5} {\bf (E)},
in presence
of a vacancy located in a neighboring bridging oxygen row, 
$\mathrm{O3}$, 
the excess electron resulting from this vacancy transfers to the substrate 
and a filled
state 
appears in the band gap.
This gap state stems from a $\mathrm{Ti}$--3{\it d} orbital, thus reducing 
one 
second--layer
$\mathrm{Ti}$ ion to $\mathrm{Ti^{3+}}$.

\subsection{{\em Ab initio} thermodynamics of 
defective $\mathbf{Au}$/$\mathbf{TiO_{2}(110)}$ surfaces}

\begin{figure}[h]
\begin{center}
\begin{tabular}{cc}
\subfigure{\includegraphics[angle=0,scale=0.45]{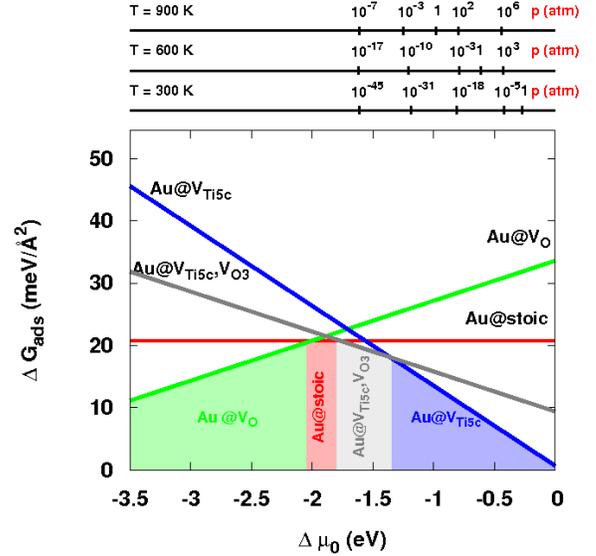}}
\end{tabular}
\caption{%
Free energy $\Delta G_{\rm ads} (T,p)$ for $\mathrm{Au}$ adsorption on
  stoichiometric and defective $\mathrm{TiO_{2}}$(110) surfaces as a function
  of the oxygen chemical potential 
$\Delta \mu_{\rm O}$;
see text for the four different phases.
Conversion to oxygen partial pressures $p$ (upper axes) has been carried out
at constant temperatures of $T=300$, 
600,
and 900~K (see text).
%
%
%
%
%
%
\label{fig10}}
\end{center}
\end{figure}

The 
effects
of temperature and pressure on the relative stability of
$\mathrm{Au}$/$\mathrm{TiO_{2}}$ metal/oxide surfaces
have
been taken into account by
employing the formalism of approximate {\it ab initio} 
thermodynamics,
as sketched in Sec.~\ref{sec:methods}. 
To this 
end,
we compute the free energies of $\mathrm{Au}$ adsorption, 
$\Delta G_{\rm ads} (T,p)$ as given by Eq.~\eqref{eq5}, 
%
%
and report them in Fig.~\ref{fig10} as a function of the
$\mathrm{O}$~chemical potential
including a conversion to oxygen partial pressures at several relevant temperatures.
These free energies are measured relative to the
stoichiometric surface and therefore include the 
free
energy cost of creating whatever vacancy the gold atom may be associated with.

As highlighted by 
the colors in
Fig.~\ref{fig10}, it is possible to identify four 
thermodynamically stable phases. 
%
The first 
phase,
which holds for values of $\mu_{\rm{O}}>-1.35$~eV, 
corresponds to the 
scenario 
where a surface $\mathrm{Ti_{5c}}$ atom has been
replaced by
an
$\mathrm{Au}$ 
adatom, denoted ``$\mathrm{Au@V_{Ti{5c}}}$''. 
%
In oxidative environments, and at the reference conditions
that
are traditionally 
used in most 
computational studies, this structure becomes the 
thermodynamically most stable one. 
The second 
most stable structure thermodynamically
is the one obtained by
removing a surface O~atom from a bridging row based on the 
$\mathrm{Au@V_{Ti{5c}}}$ structure described above, 
which is called ``$\mathrm{Au@V_{Ti{5c}},V_{O3}}$'' since
the missing $\mathrm{O_b}$ atom is the 
$\mathrm{O3}$ atom 
(refer to Fig.~\ref{fig9})
We have shown that the PBE+U energy required for removing the
$\mathrm{O3}$ oxygen atom, i.e. the vacancy formation energy $E_{\rm V}^{\rm O3}$, 
from the surface of the $\mathrm{Au@V_{Ti{5c}}}$ 
structure
amounts to 1.36~eV.
This is significantly lower 
than the required energy of 2.97~eV
to remove an $\mathrm{O_b}$ atom from the ideal, 
stoichiometric $\mathrm{TiO_{2}}$ surface. 
Therefore the $\mathrm{TiO_{2}}$(110) oxide surface becomes a better
oxidant when doped with gold. 
%

As demonstrated 
in
Fig.~\ref{fig10},
the $\mathrm{Au@V_{Ti{5c}},V_{O3}}$ 
surface structure becomes thermodynamically stable for
$-1.79$~eV $<\mu_{\rm{O}} < -1.35$~eV. 
It turns out that these two 
defective surface structures, 
$\mathrm{Au@V_{Ti{5c}}}$ and
$\mathrm{Au@V_{Ti{5c}},V_{O3}}$,  are thermodynamically stable in a wide
range of temperatures $T$ and pressures $p$ that are relevant for applications
in the realm of catalysis. 
%
In contrast,
the adsorption of $\mathrm{Au}$ adatoms on the stoichiometric
$\mathrm{TiO_{2}}$(110)  surface is thermodynamically
stable only in a quite narrow range of values of the $\mathrm{O}$ 
chemical potential of $-2.05$~eV $< \mu_{\rm{O}} <  -1.79$~eV. 
%
Finally,
the adsorption of $\mathrm{Au}$ adatoms on $\mathrm{O}$ vacancies becomes
thermodynamically stable for values of $\mu_{\rm{O}}<-2.05$~eV. 
We therefore conclude that 
under $\mathrm{O}$--rich 
conditions,
the thermodynamically most stable structure is the 
defective surface structure 
$\mathrm{Au@V_{Ti{5c}}}$ obtained by substituting a surface
$\mathrm{Ti_{5c}}$ atom with 
an
$\mathrm{Au}$ adatom, while under $\mathrm{Ti}$--rich 
conditions,
the $\mathrm{Au}$ adatoms are preferentially adsorbed at 
$\mathrm{O}$ vacancies.

\section{Gold--promoted titania: Interactions with CO probe molecules}

\subsection{CO molecular adsorption on $\mathbf{Au}$/$\mathbf{TiO_{2}(110)}$ surfaces}

We have shown that our calculations predict the
two most stable adsorption sites of an
$\mathrm{Au}$ adatom adsorbed on the stoichiometric $\mathrm{TiO_{2}}$(110) surface,
differing 
energetically
by only 
$\sim$0.1~eV.
%
The lowest energy configuration is that
with the $\mathrm{Au}$ adatom adsorbed onto
a bridge site between a surface $\mathrm{O_b}$ atom and a 
first--layer
$\mathrm{Ti_{5c}}$ atom. 
A detailed analysis of the electronic structure of
this configuration shows that all
$\mathrm{Ti}$ ions belonging to the substrate preserve their formal
oxidation state $\mathrm{Ti^{4+}}$. 
The other stable structure
consists of an
$\mathrm{Au}$ adatom sitting on top of 
an
$\mathrm{O_b}$
atom of the (110) surface. 
The metal adsorption on 
an
$\mathrm{O_b}$ top site induces a strong charge
rearrangement at the metal/oxide contact and entails the reduction of a 
second--layer
$\mathrm{Ti}$ ion which becomes formally $\mathrm{Ti^{3+}}$.

\begin{figure*}
\begin{center}
\begin{tabular}{cc}
\subfigure{\includegraphics[angle=0,scale=0.5]{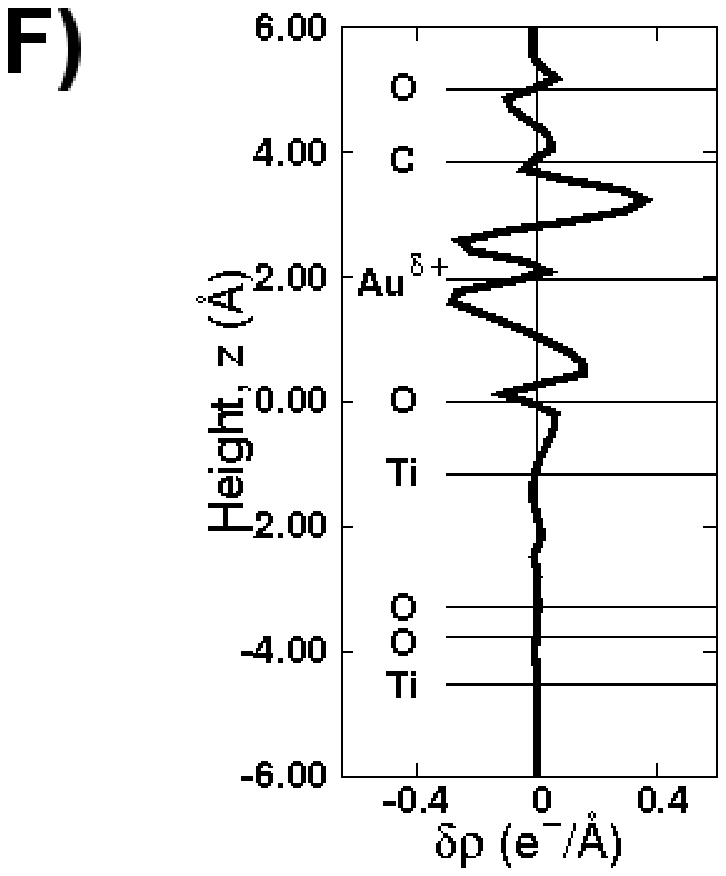}}\hspace{0.50cm}
\subfigure{\includegraphics[angle=0,scale=0.2]{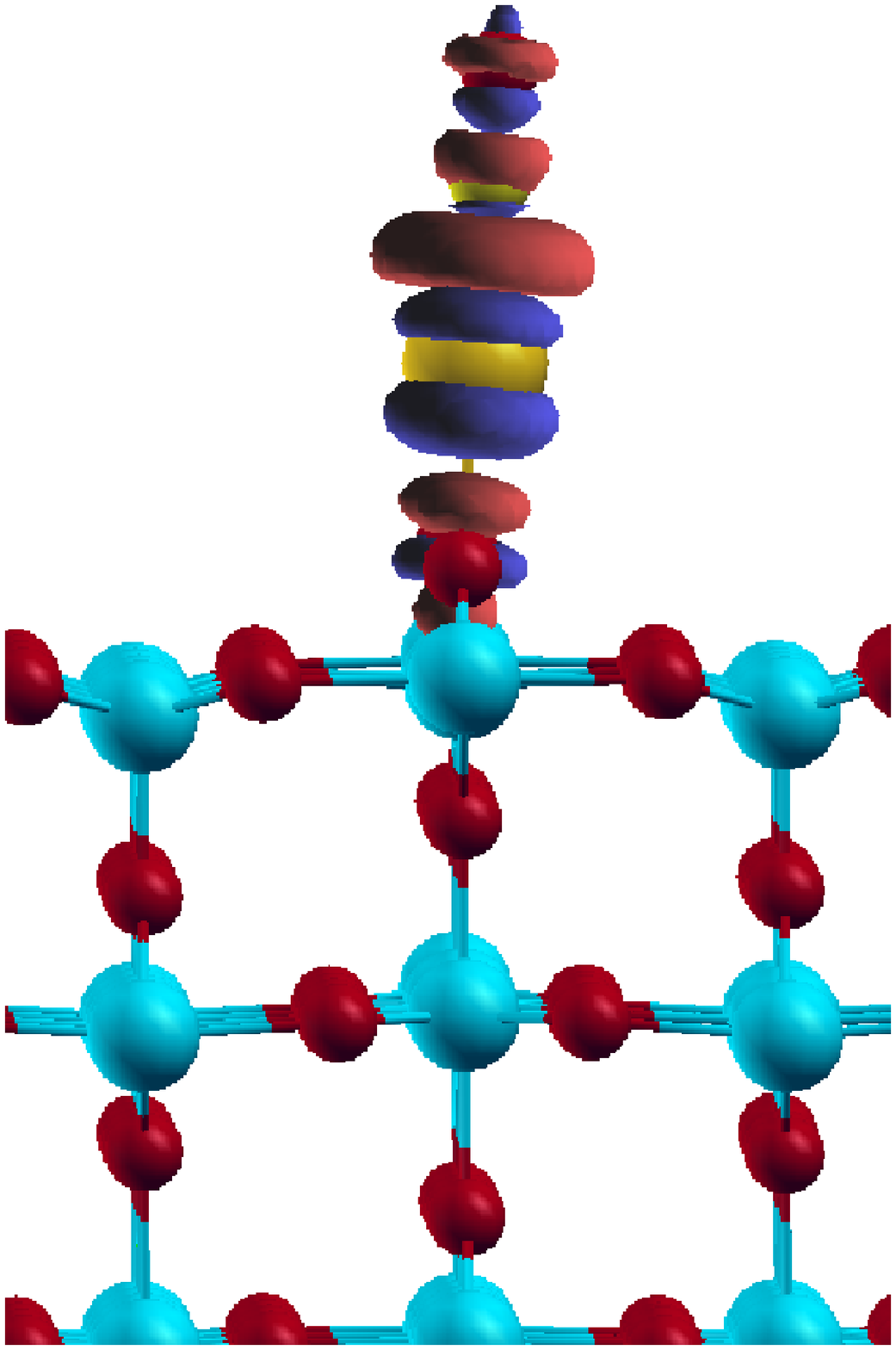}}\hspace{1.00cm}
\subfigure{\includegraphics[angle=0,scale=0.35]{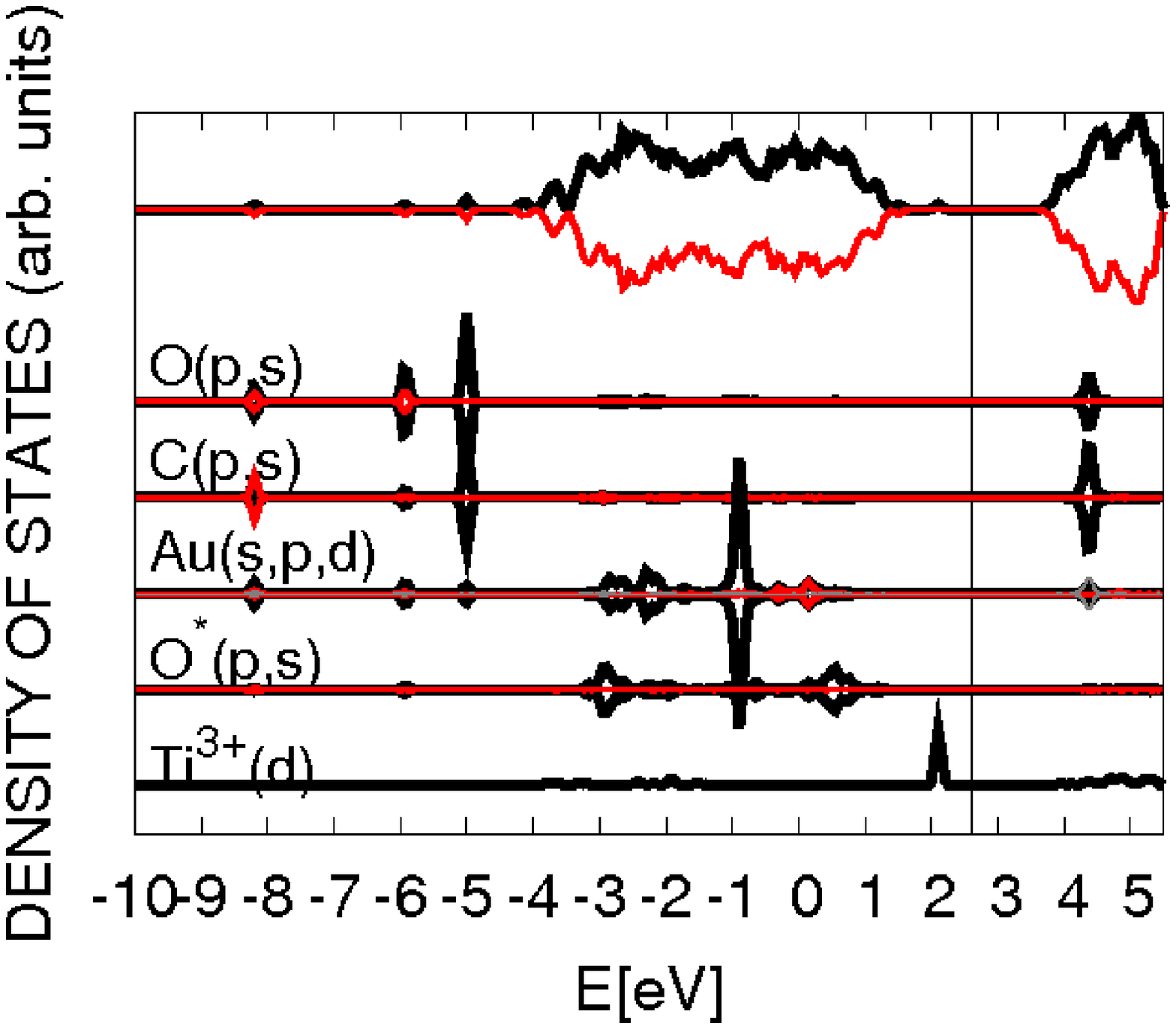}}\\
\subfigure{\includegraphics[angle=0,scale=0.5]{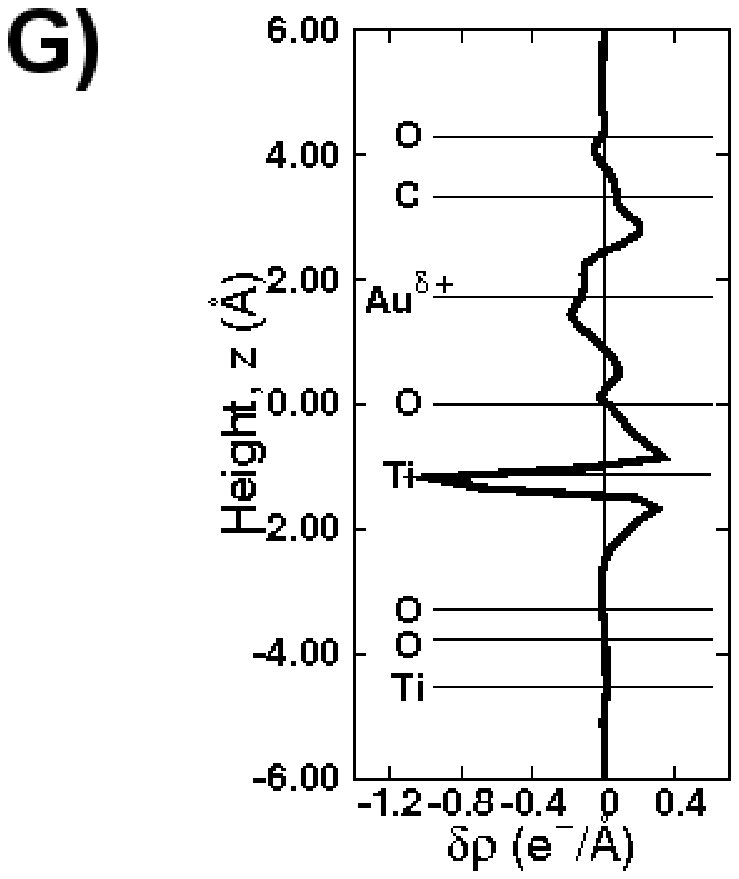}}\hspace{0.25cm}
\subfigure{\includegraphics[angle=0,scale=0.20]{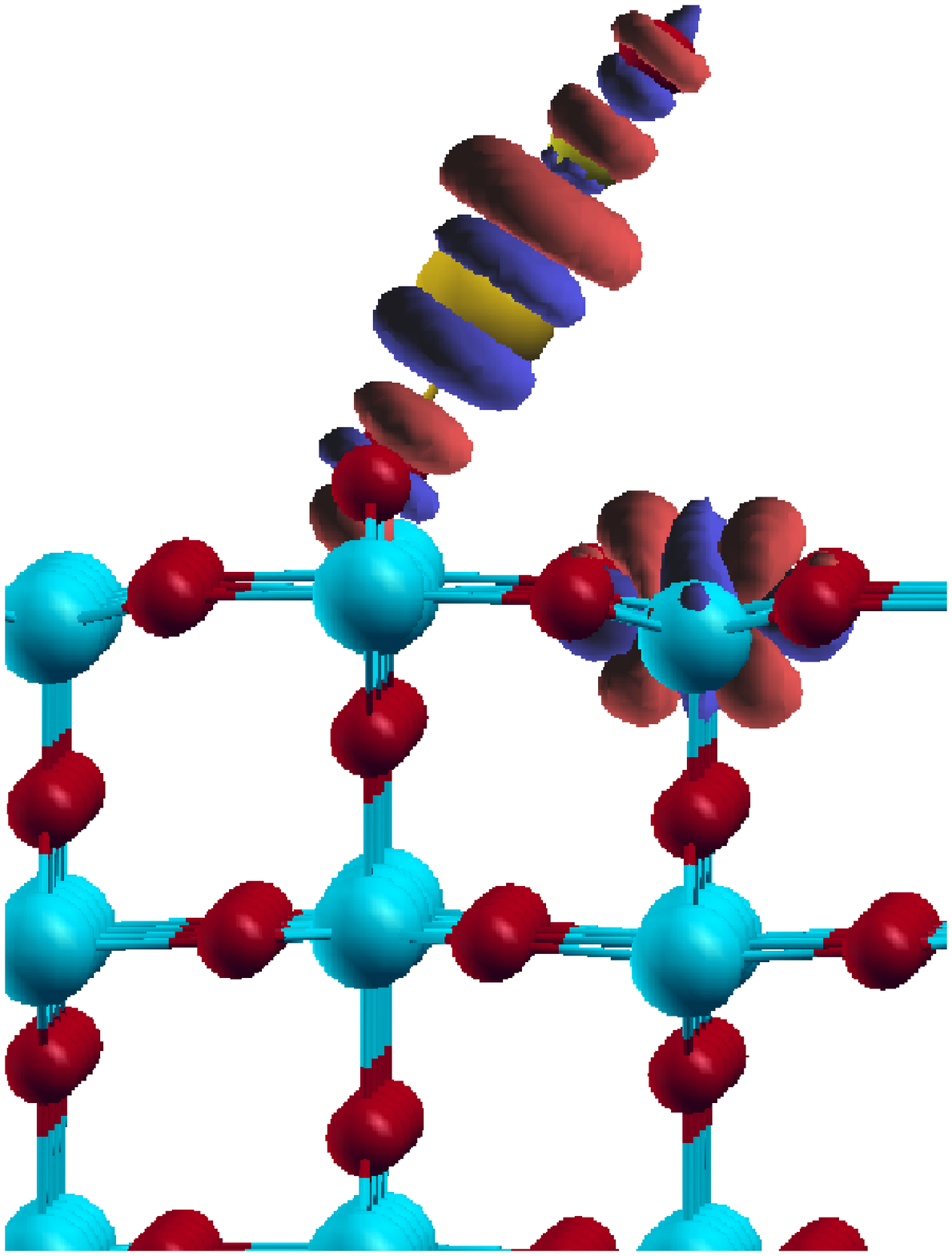}}\hspace{1.00cm}
\subfigure{\includegraphics[angle=0,scale=0.35]{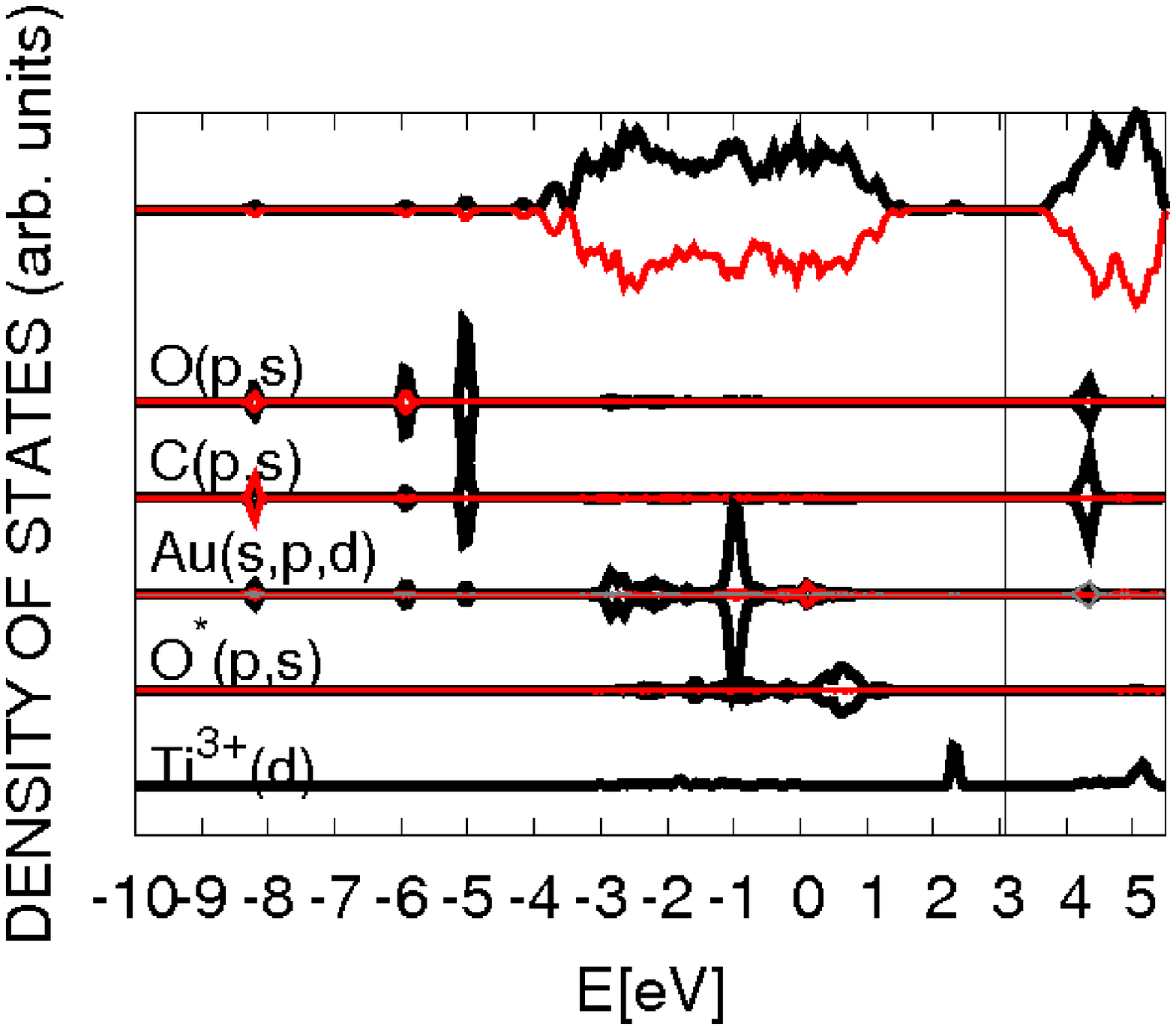}}\\
\subfigure{\includegraphics[angle=0,scale=0.5]{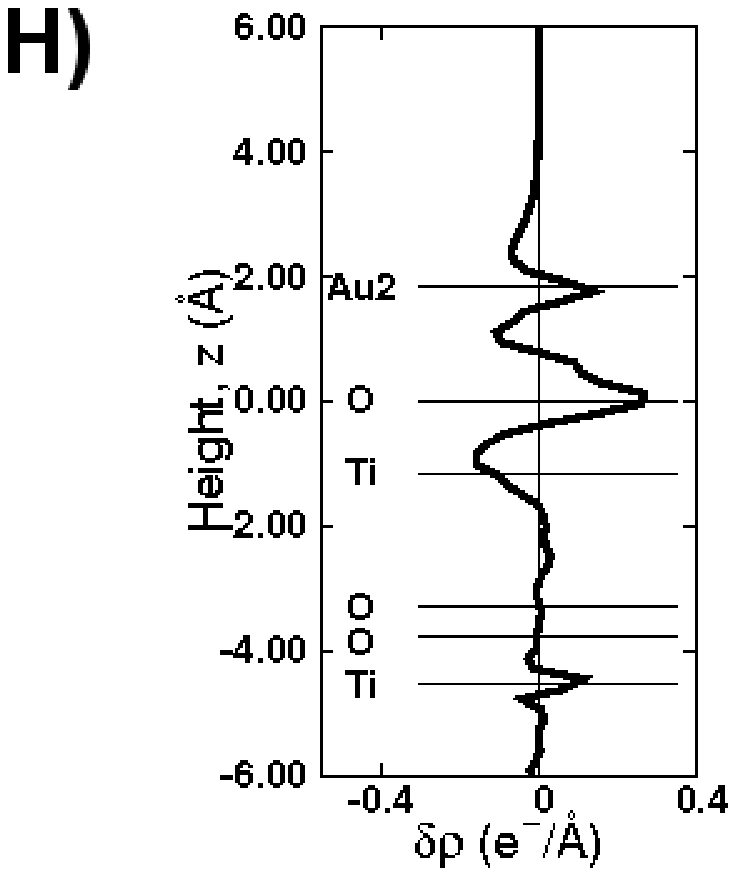}}\hspace{0.50cm}
\subfigure{\includegraphics[angle=0,scale=0.20]{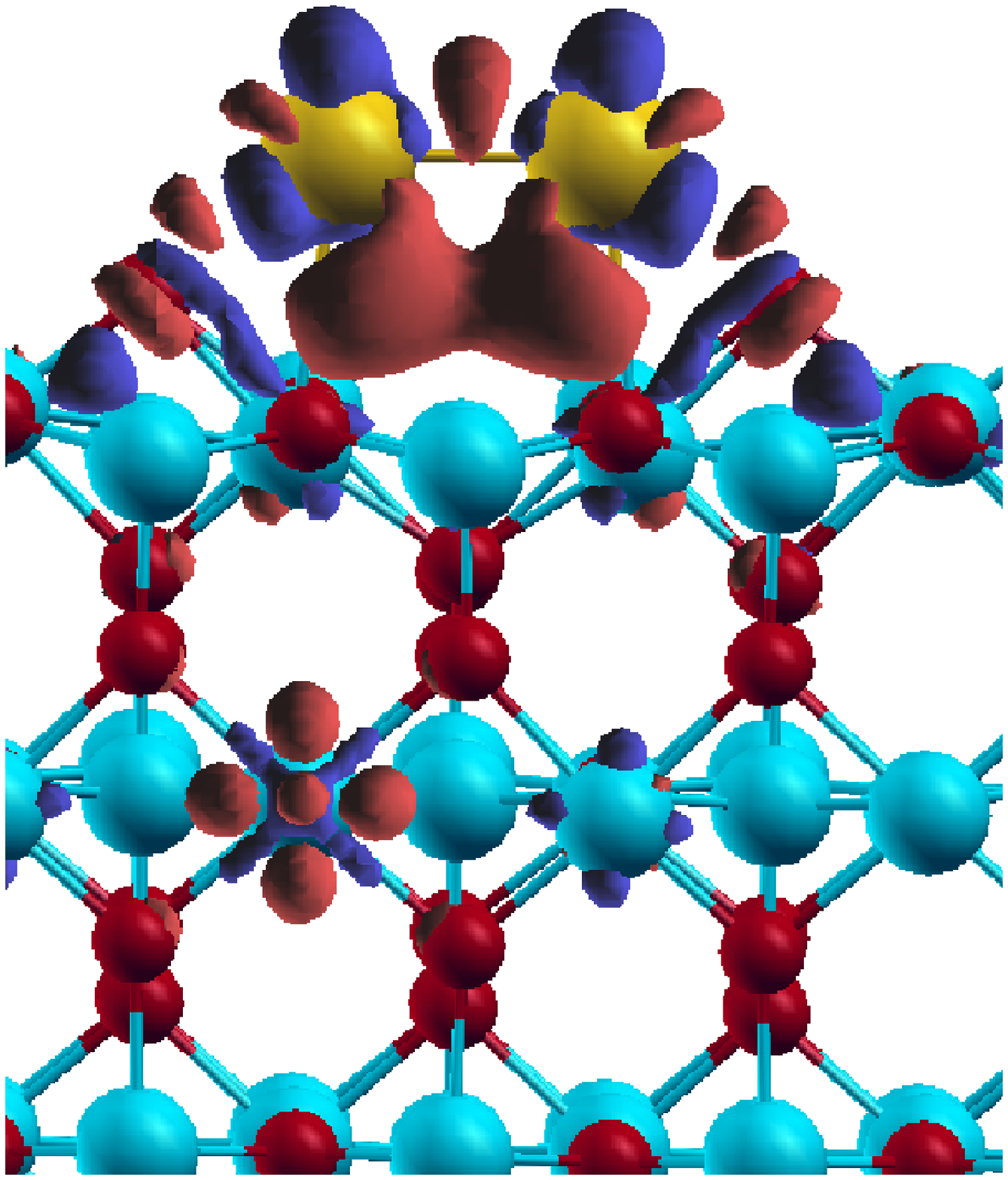}}\hspace{1.00cm}
\subfigure{\includegraphics[angle=0,scale=0.35]{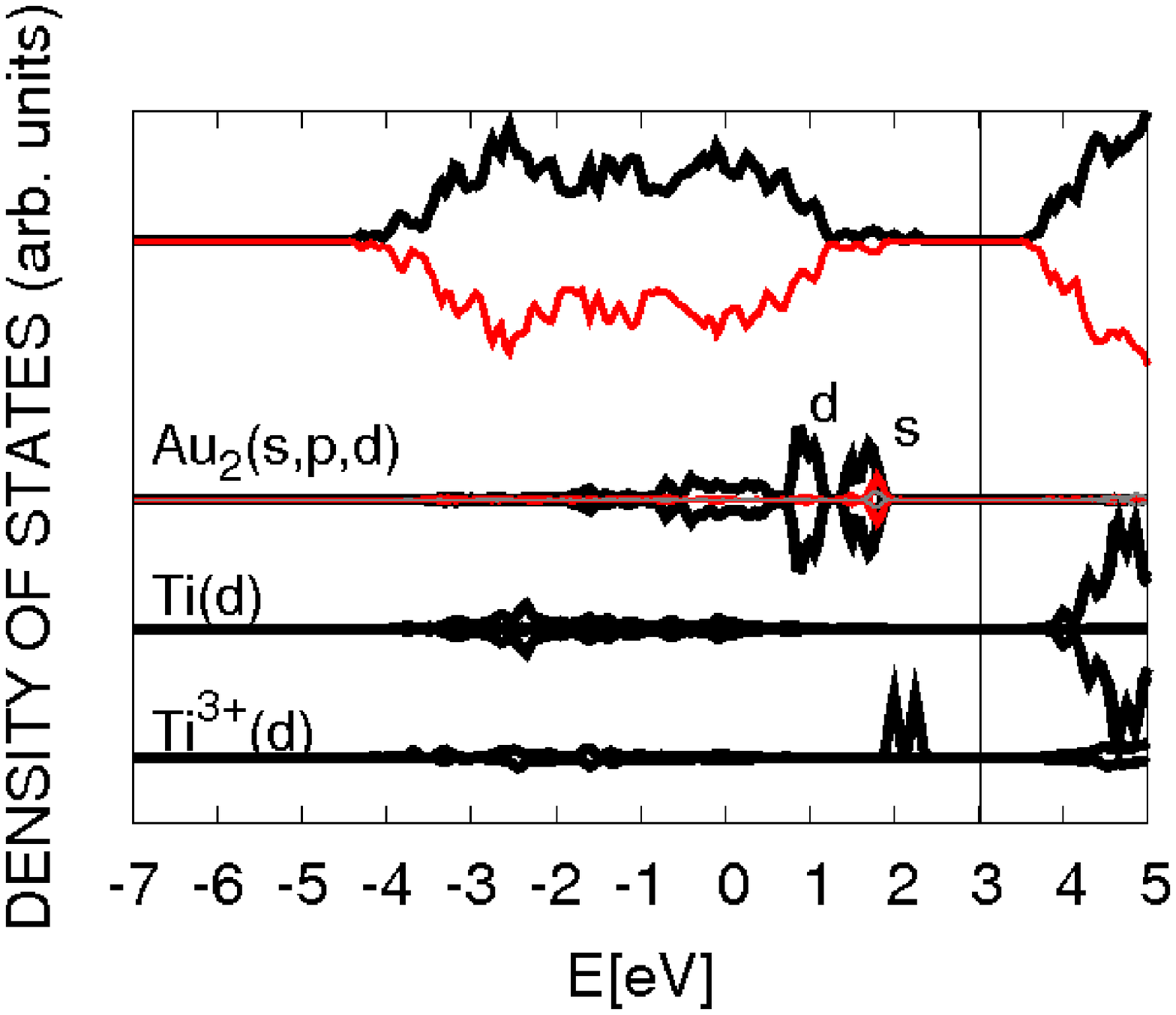}}\\
\subfigure{\includegraphics[angle=0,scale=0.5]{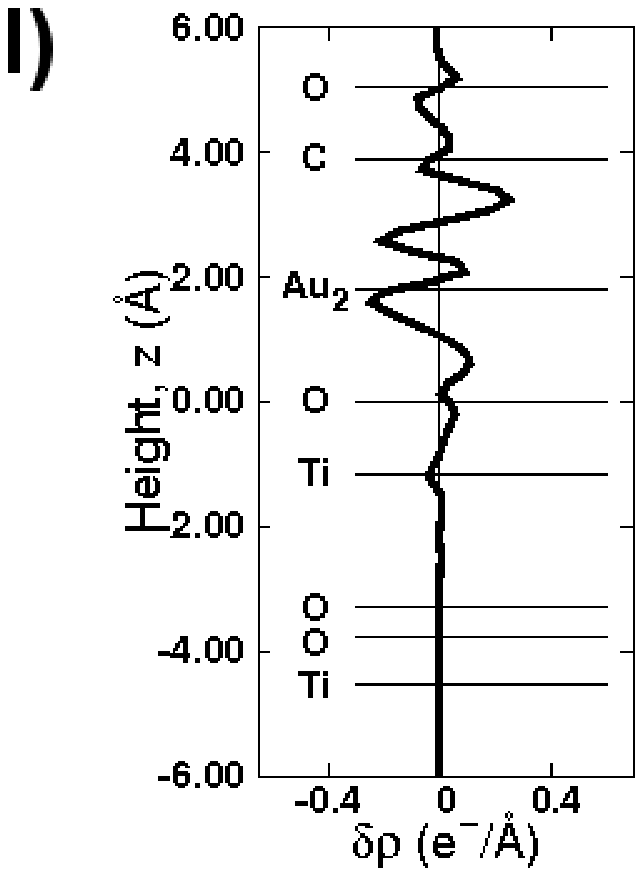}}\hspace{0.50cm}
\subfigure{\includegraphics[angle=0,scale=0.20]{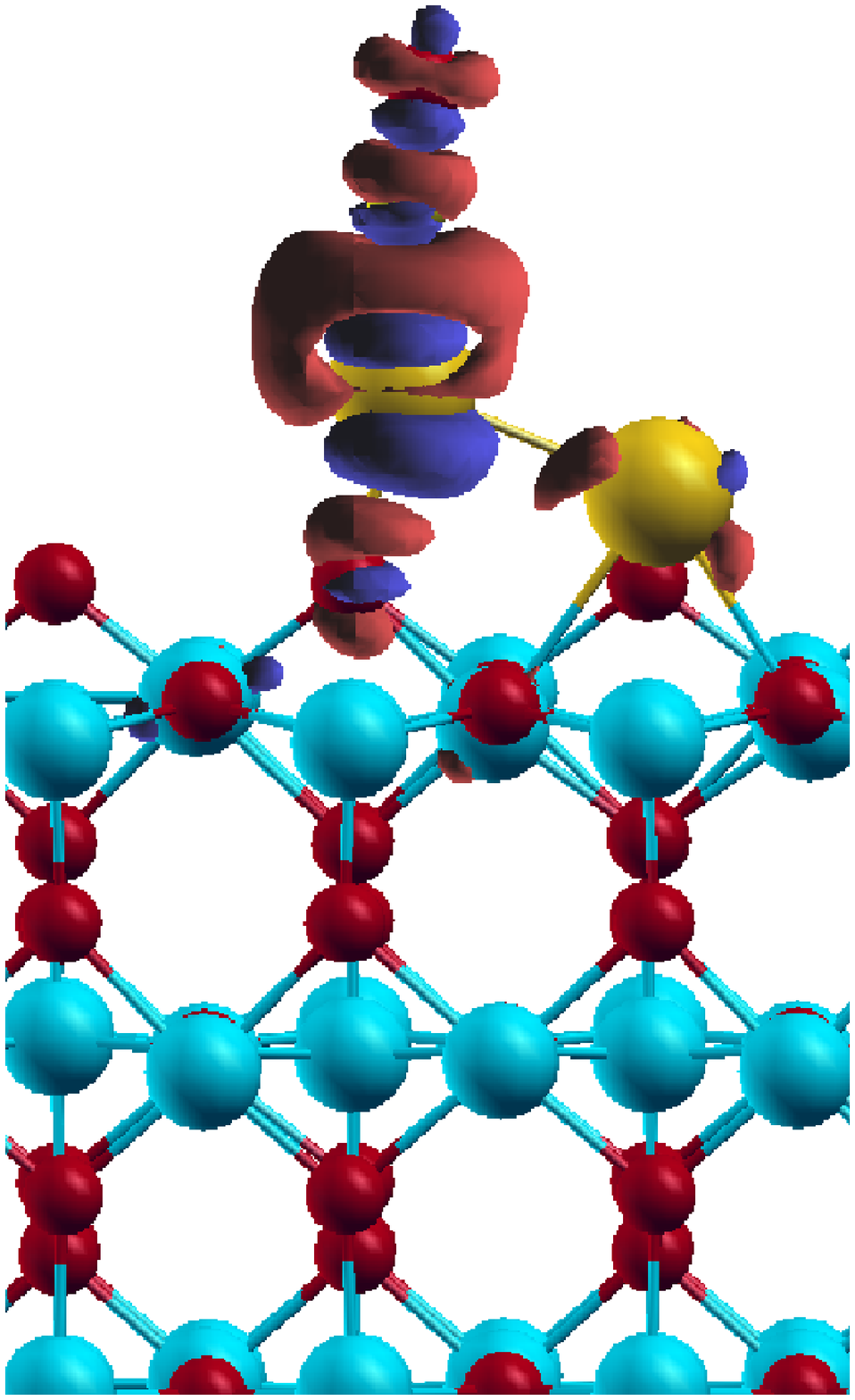}}\hspace{1.00cm}
\subfigure{\includegraphics[angle=0,scale=0.35]{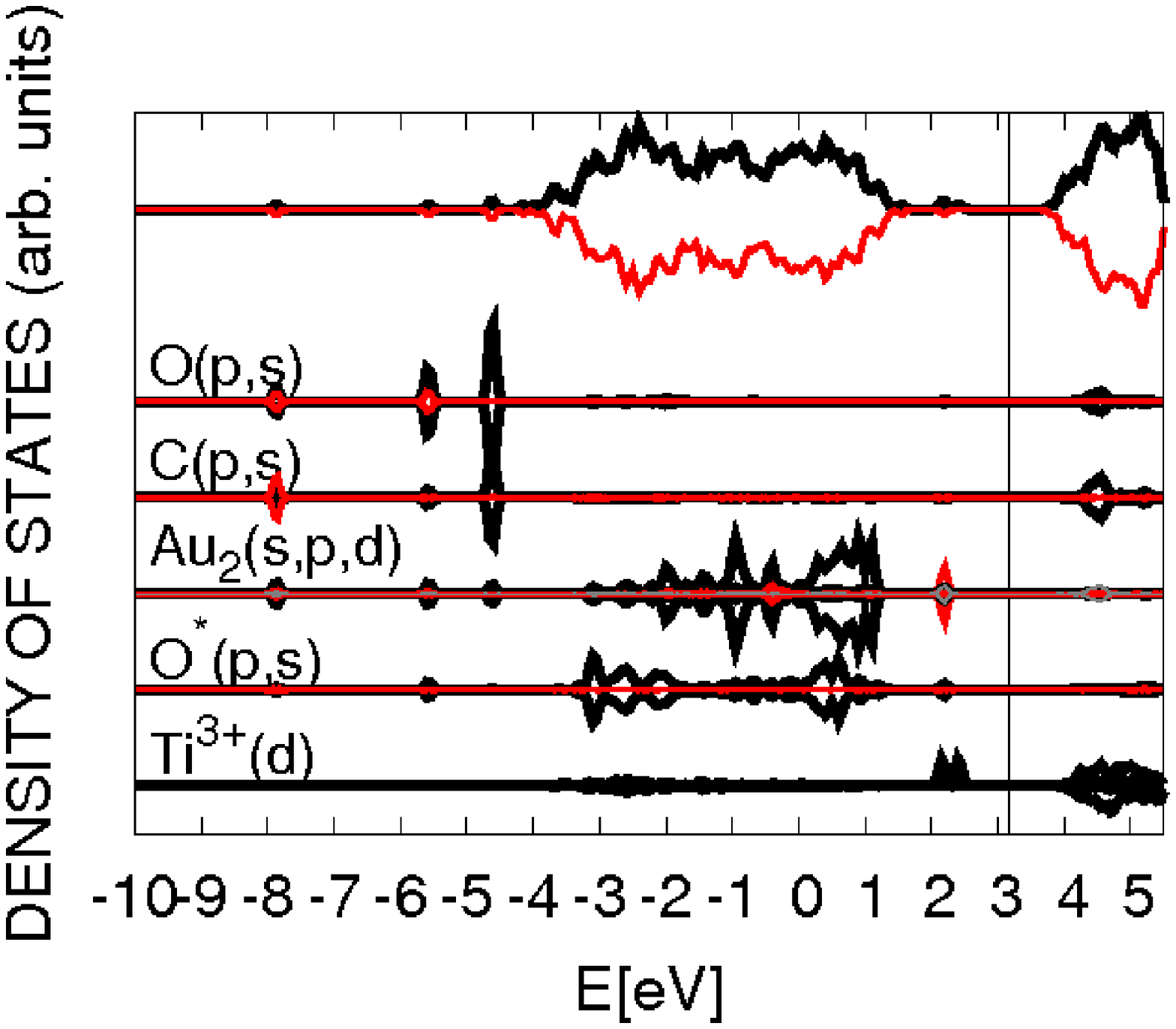}}

\end{tabular}
\caption{%
Electronic structure analyses (based on the PBE+U approach) of
  (F) a $\mathrm{CO}$ molecule
  adsorbed on a supported $\mathrm{Au}$ adatom (originally the $\mathrm{Au}$
  atom was adsorbed
  at
  a bridging $\mathrm{O}$ atom) on the $\mathrm{TiO_{2}}$(110) surface, 
%
  (G) a $\mathrm{CO}$ molecule
  adsorbed on a supported $\mathrm{Au}$ adatom (originally the $\mathrm{Au}$
  atom was adsorbed
  between a bridging $\mathrm{O}$ atom and a $\mathrm{Ti_{5c}}$
  atom)
  on the $\mathrm{TiO_{2}}$(110) surface, 
  (H) of the $\mathrm{Au_2}$ dimer
  adsorbed 
  onto
  surface and 
  (I) a $\mathrm{CO}$ molecule
  bonded to an $\mathrm{Au_{2}}$ dimer adsorbed 
  onto
a surface $\mathrm{O}$
  vacancy on the $\mathrm{TiO_{2}}$(110)
  surface.
The left panel  
represents the bonding charge $\delta\rho (z)$  
integrated over planes
perpendicular to the surface and plotted as a function of the height from
the surface.
The central panel 
displays the bonding charge $\Delta\rho (\vec{r})$ at an isovalue of
$\pm 0.06$~$|e|$/\AA$^{3}$
where electron accumulation and depletion are
represented by red and blue areas, respectively.
The right panel shows the total DOS and atom resolved projected DOS 
(PDOS), as indicated, where
energies are with respect to the Fermi level, which is marked by a solid vertical line.
\label{fig11}}
\end{center}
\end{figure*}

Using these preferred structures,
we now study their interaction with
a $\mathrm{CO}$ admolecule. 
A $\mathrm{CO}$ molecule is placed 
end--on
at 2.5~\AA\ above the $\mathrm{Au}$ adatoms.
%
%
%
%
The adsorption of the $\mathrm{CO}$ molecule on an $\mathrm{Au}$ adatom adsorbed
at
an
$\mathrm{O_b}$ site is found to be strongly exothermic, 
releasing $-2.66$~eV ($-2.33$~eV using 
PBE;
see Table~\ref{tab4}). This is in accord with earlier calculations~\cite{anke}.
%
In this 
configuration,
the $\mathrm{CO}$ molecule is aligned with the
$\mathrm{Au}$ adatom and 
oriented normal to the surface (see Fig.~\ref{fig11}~{\bf (F)}). 
The computed value of the $\mathrm{C}$--$\mathrm{O}$ bond length is 1.15~\AA\
%
%
%
compared to a value of 1.14~\AA\ of an isolated CO molecule 
using the same approach. 
The PBE+U (PBE) 
value
of the distance between the $\mathrm{Au}$ atom 
and the substrate $\mathrm{O_b}$ atom to
which it is bonded is 1.98~\AA\ (1.97~\AA ), while the $\mathrm{Au}$ to $\mathrm{C}$
distance is 1.88~\AA\ (1.88~\AA ).
Analysis of the PDOS (see Fig.~\ref{fig11}~{\bf (F)}) 
reveals that
at this $\mathrm{Au}$/$\mathrm{TiO_{2}}$ 
contact,
only one $\mathrm{Ti^{3+}}$ ion
is present before and after adsorption, 
and that the excess charge populating
the substrate occupies the same second--layer 
$\mathrm{Ti}$--3{\it d} orbital. 
%
Therefore,
the charge redistribution
due to adsorbing a CO molecule 
does not 
further reduce
the oxide support. 
%
However,
the bonding charge analysis 
(also displayed in Fig.~\ref{fig11} {\bf (F)}) 
shows that the $\mathrm{Au}$ adatom is involved in charge depletion (blue areas), 
while both the $\mathrm{C}$ and $\mathrm{O}$ atoms 
accumulate the resulting excess charge
(red areas).

%
%
%
We now consider the interaction between a $\mathrm{CO}$ molecule and the $\mathrm{Au}$ atom
adsorbed 
onto
the bridge site. 
Again, the mechanism is exothermic by $-2.27$~eV (or $-2.17$~eV when using PBE instead of PBE+U). 
%
In this 
configuration,
the $\mathrm{CO}$ molecule is bonded to the $\mathrm{Au}$ atom 
and the $\mathrm{C}$--$\mathrm{O}$ bond length is 
1.15~\AA\, 
as in the previous case. 
Here, however, 
the CO--Au structure is found to be 
tilted with respect to the surface 
at 
an angle of about $60^{\circ}$ (see Fig.~\ref{fig11} {\bf (G)}). 
The PBE+U (PBE) values of the distances between the $\mathrm{Au}$ adatom and the $\mathrm{Ti_{5c}}$ atom
before and after the adsorption of molecular $\mathrm{CO}$ are
2.79 (2.88) and 3.90~\AA\ (3.88~\AA ), 
respectively, while the distances between the metal and the $\mathrm{O}$
bridging atom before and after $\mathrm{CO}$ adsorption are 
2.30 (2.39) and 1.99~\AA\ (2.01~\AA ), respectively. 
%
%
The analysis of the PDOS as depicted in Fig.~\ref{fig11}~{\bf (G)} 
shows, in this case, that at the $\mathrm{Au}$/$\mathrm{TiO_{2}}$ 
contact,
the charge redistribution reduces the oxide 
support, 
contrary to the previous scenario. 
%
%
We recall that when $\mathrm{Au}$ is adsorbed 
onto
a bridge 
site,
all the
substrate $\mathrm{Ti}$ ions preserve their formal $\mathrm{Ti^{4+}}$
oxidation state. 
Upon $\mathrm{CO}$ adsorption 
onto
the $\mathrm{Au}$ atom
occupying a bridge 
position,
one $\mathrm{Ti}$ ion is reduced.
The additionally reduced $\mathrm{Ti^{3+}}$ ion, 
which belongs to the first $\mathrm{TiO_{2}}$ layer, 
is precisely the ion that forms
a bond with the $\mathrm{Au}$ in the bridge configuration before
$\mathrm{CO}$ adsorption.

The configuration in Fig. \ref{fig11}~{\bf (G)}, where the
CO--Au structure is 
tilted by $\sim 60^{\circ}$ with respect to the 
substrate, is used as an
initial condition for a short
AIMD run at room temperature in order to probe dynamical instabilities. 
%
%
In less than 
0.4~ps,
the CO--Au complex bends and sits 
perpendicular to the surface, therefore recovering the structure
obtained upon molecular $\mathrm{CO}$ adsorption 
on gold sitting on top of 
an
$\mathrm{O_b}$ atom.
The difference in the binding energies for $\mathrm{CO}$ molecule adsorption
on the top and bridge configurations 
described above
can be attributed in part 
to the tilting of the $\mathrm{Au}$--$\mathrm{CO}$ complex 
with respect to the surface plane.
More important, however, is 
the position of the reduced $\mathrm{Ti^{3+}}$ site. 
Previous studies~\cite{PMD,deskins2} on charge localization 
induced by $\mathrm{O_b}$ vacancies on $\mathrm{TiO_{2}}$ surfaces have shown 
that excess electrons, 
which are trapped
at specific $\mathrm{Ti}$ sites, 
migrate easily by phonon--assisted hopping to other $\mathrm{Ti}$ atoms, 
thus exploring different electronic structure topologies~\cite{PMD}.
%
In particular,
it was found 
that the most stable sites for charge localization
belong to the 
second subsurface layer under $\mathrm{Ti_{5c}}$ rows. 
The topologies 
where
the excess charge is shared
between surface $\mathrm{Ti_{5c}}$ atoms and second or third subsurface layer
sites below $\mathrm{Ti_{5c}}$ rows were found to be about 0.2--0.3 and
0.3--0.4~eV higher in energy, respectively. 
%
It is therefore
conceivable  
that we would observe
a similar 
dynamics 
for
excess charge de-- and re--localization.

\subsection{CO adsorption on $\mathbf{Au_{1}}$/$\mathbf{TiO_{(2-x)}(110)}$ 
and $\mathbf{Au_{2}}$/$\mathbf{TiO_{(2-x)}(110)}$ surfaces}
\label{sec:co-au-tio2} 

The interaction between a $\mathrm{CO}$ molecule and a single $\mathrm{Au}$
adatom adsorbed at a bridging site or on top of 
an
$\mathrm{O_b}$ atom on the 
$\mathrm{TiO_{2}}$(110) surface has been investigated above. 
We now
concentrate on the interaction between $\mathrm{CO}$ molecule and
$\mathrm{Au}$ adatom adsorbed at a bridging $\mathrm{O}$ vacancy.
It has been shown that the
metal adatom strongly interacts with this reduced 
oxide support.
%
The resulting $\mathrm{Au^{\delta-}}$ adatom is 
highly stable, considering its
binding energy of $-1.54$~eV. 
Our calculations predict a fairly weak interaction between molecular
$\mathrm{CO}$ and the $\mathrm{Au^{\delta-}}$ adatom. 
The $\mathrm{CO}$ molecule initially located 
end--on
at 2.5~\AA\ above
the $\mathrm{Au^{\delta-}}$ species is found to bind with about $-0.41$~eV. 
%
This binding energy is close to an order of magnitude lower than the binding energy of
the positively charged $\mathrm{Au}$ adatom adsorbed on 
the stoichiometric $\mathrm{TiO_{2}(110)}$ surface, which binds with $-2.66$~eV.

%
%
%

Our calculations show that while positively charged $\mathrm{Au^{\delta +}}$ ions 
supported on the $\mathrm{TiO_{2}}$(110) surface are shown to
activate molecular $\mathrm{CO}$, negatively charged $\mathrm{Au^{\delta -}}$ adspecies, 
which are incorporated into surface $\mathrm{O}$ vacancies, interact
only weakly with $\mathrm{CO}$ adsorbates.
This is in line with the 
scenario
of $\mathrm{Au}$ adsorbed on the
$\mathrm{CeO_{2}}$ substrate~\cite{Ce,Ce1},
where it has been shown that
the higher stability of $\mathrm{Au}$ adatoms adsorbed
onto
surface is the basis of the deactivation mechanism during $\mathrm{CO}$~oxidation. 
The oxidation mechanism proposed~\cite{Ce} involves three steps:
the spillover of the $\mathrm{CO}$ molecule, the actual oxidation via a
lattice oxygen atom leading to $\mathrm{CO_{2}}$ desorption, and the diffusion
of the $\mathrm{Au}$ adatom into the newly formed $\mathrm{O}$ vacancy, which
results in
negatively charged $\mathrm{Au^{\delta -}}$ adspecies that 
prevents
the adsorption of molecular $\mathrm{CO}$. 
Since the present AIMD simulations have shown that $\mathrm{Au}$ adatoms
diffuse rather easily on the stoichiometric $\mathrm{TiO_{2}}$(110) surface,
we would expect a deactivation process with titania similar to what was observed on the 
ceria surface.                 

\begin{figure}[h]
\begin{center}
\begin{tabular}{cc}
\subfigure[]{\includegraphics[angle=0,scale=0.21]{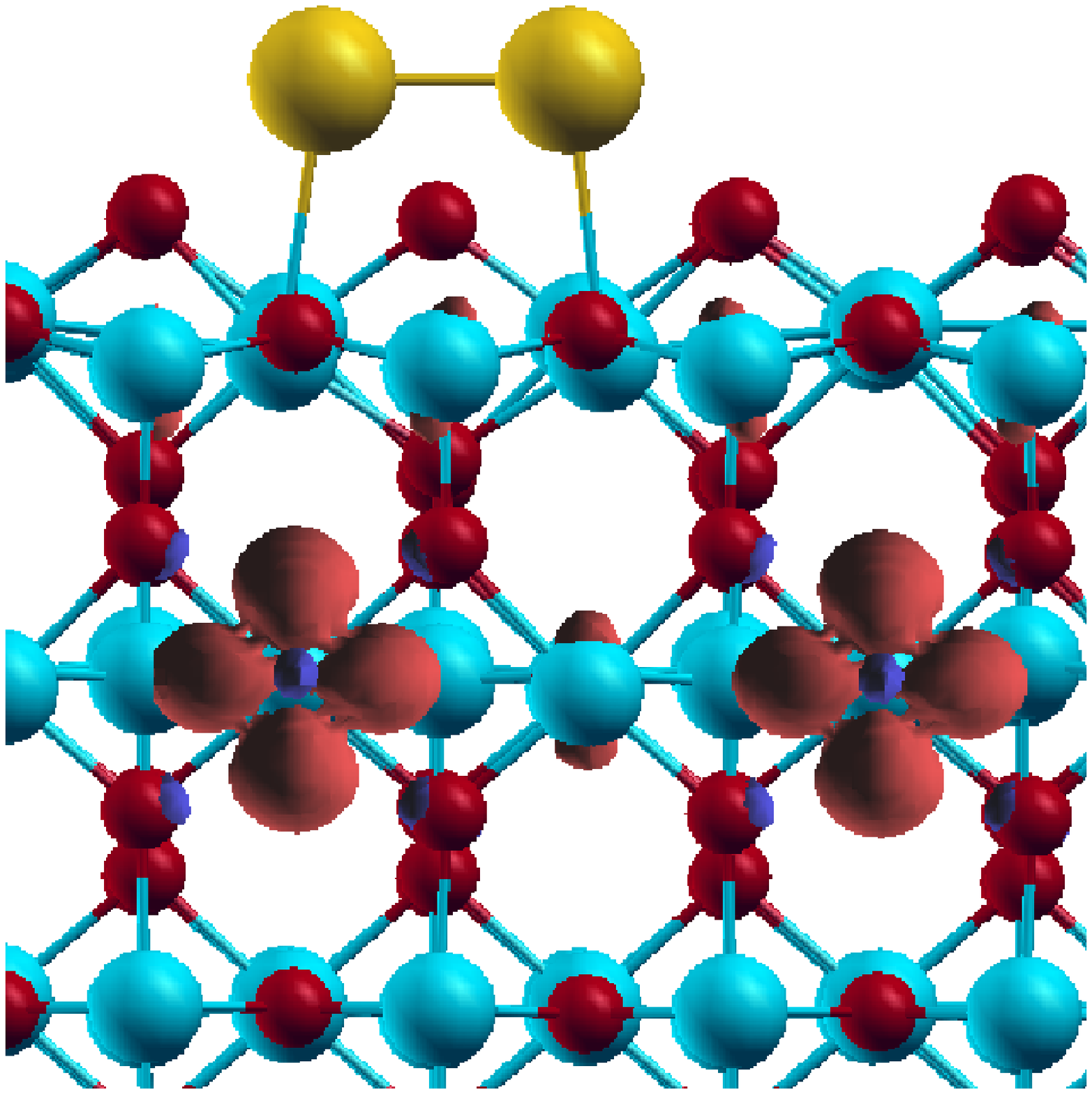}}\hspace{1.00cm}
\subfigure[]{\includegraphics[angle=0,scale=0.21]{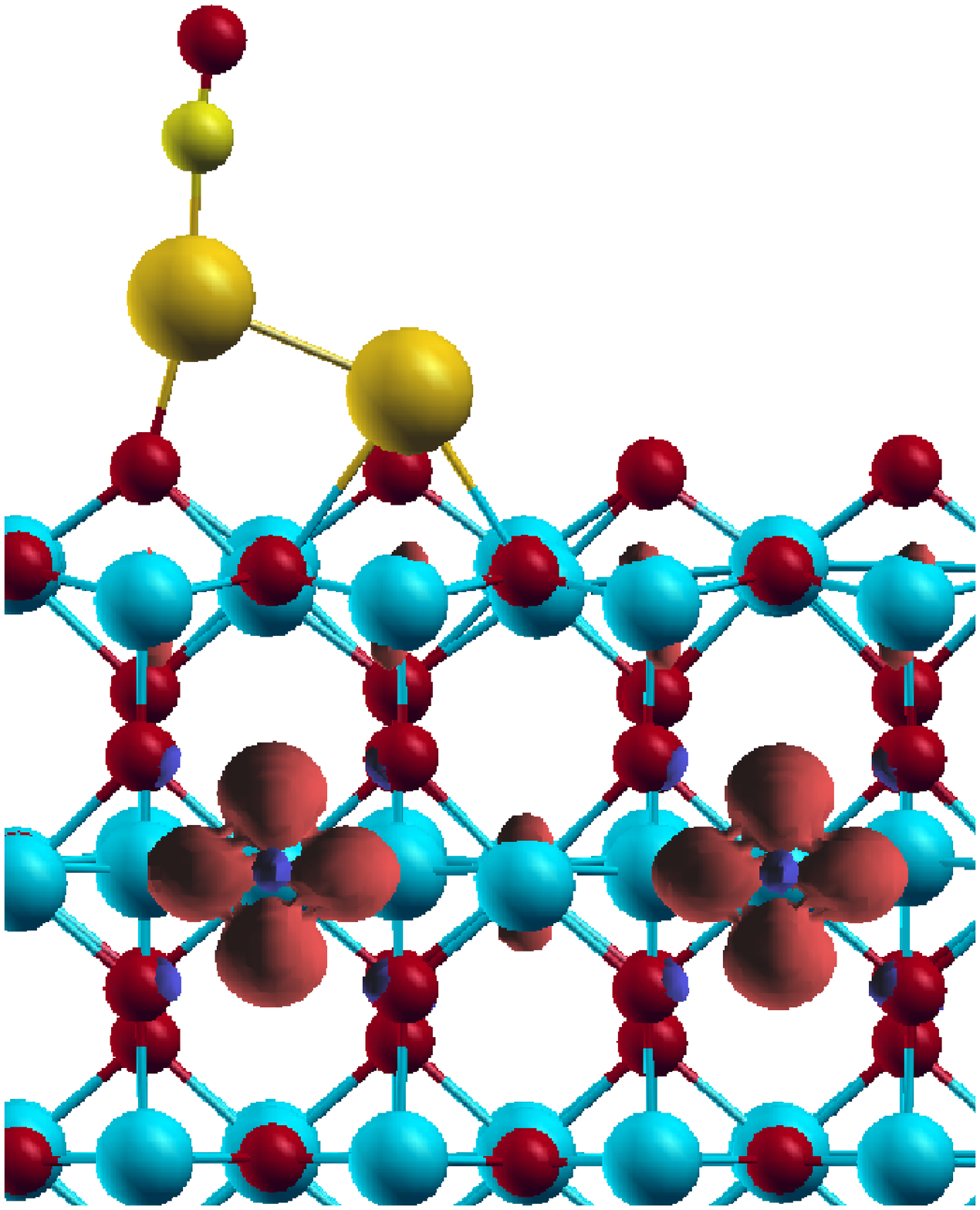}}
\end{tabular}
\caption{%
%
%
%
Spin density at $\pm 0.005$~$|e|$/\AA$^{3}$
of the $\mathrm{Au2}$ dimer adsorbed 
onto
a surface $\mathrm{O}$ vacancy 
on the $\mathrm{TiO_{2}}$(110) surface 
(a),
and the same system after coadsorbing a CO molecule 
(b);
see text.
%
%
\label{fig12}}
\end{center}
\end{figure}

Despite the fact that $\mathrm{CO}$ only weakly interacts with negatively charged
$\mathrm{Au^{\delta -}}$ adspecies, several theoretical and experimental studies suggest that 
$\rm V_O$ vacancies in bridging oxygen rows, $\mathrm{O_b}$, 
are active nucleation sites for $\mathrm{Au}_n$ clusters on the
$\mathrm{TiO_{2}}$(110) surface~\cite{matthey,ti1,ti2}. 
It is therefore expected that nucleation and growth of $\mathrm{Au}_n$ clusters on the
rutile surface is intimately related to the presence of surface oxygen
vacancies.
%
To get a glimpse,
we next investigated
the limiting case of a gold dimer, $\mathrm{Au_{2}}$,
adsorbed 
at
an $\mathrm{O_b}$ vacancy. 
The optimized structure of an $\mathrm{Au_{2}}$ dimer adsorbed 
onto
a 
surface bridging $\mathrm{O}$ vacancy is depicted in 
Fig.~\ref{fig12}~(a). 
%
The computed value of the adsorption energy of $\mathrm{Au_{2}}$ on
the reduced $\mathrm{TiO_{2}(110)}$ surface, calculated with respect to 
the isolated $\mathrm{Au_{2}}$ molecule, is $-1.17$~eV ($-1.19$~eV with PBE). 
The adsorption mechanism induces reduction of an additional $\mathrm{Ti}$~ion, leading to 
a gold dimer into an $\mathrm{O}$ vacancy 
and two 
second--layer
$\mathrm{Ti^{3+}}$ ions 
(see Fig.~\ref{fig11} {\bf (H)}). 
The two gold atoms are located 1.20~\AA\ above the $\mathrm{O}$ vacancy site,
whereas the distances between the $\mathrm{Au}$ atoms and the nearest--neighbor 
$\mathrm{Ti}$ atoms are 
2.80~AA and the
$\mathrm{Au}$--$\mathrm{Au}$ distance amounts to 2.51~\AA\ 
compared to 2.53~\AA\ of the isolated gold dimer using the same method.

Next, a $\mathrm{CO}$ molecule is positioned above the $\mathrm{Au_{2}}$
dimer adsorbed 
onto
the surface $\mathrm{O}$ vacancy. 
%
After relaxation,
the $\mathrm{CO}$ molecule reaches the most stable
configuration of 
Fig.~\ref{fig12}~(b).
%
The computed binding energy is $-1.22$~eV 
($-1.00$~eV using 
PBE), in good agreement
with previous studies~\cite{anke}. 
This value is 
about half of that
obtained for adsorption of $\mathrm{CO}$ on a
single $\mathrm{Au}$ adatom adsorbed on the stoichiometric surface ($-2.66$~eV) 
but distinctly lower 
than the energy
calculated on the
$\mathrm{Au}$/$\mathrm{TiO_{(2-x)}}$ surface ($-0.40$~eV). 
We therefore find that a cluster as small as $\mathrm{Au_{2}}$ nucleated 
at 
an
$\mathrm{O}$ vacancy on the reduced $\mathrm{TiO_{2}}$(110) surface 
favors the formation of stable $\mathrm{CO}$ adsorbates. 
%
Specifically,
the $\mathrm{CO}$ molecule binds to the
$\mathrm{Au}$ atom 
that is farthest
from the $\mathrm{O}$ vacancy
and the Au--Au distance elongates from 2.51 to 2.78~\AA . 
The interaction between $\mathrm{CO}$ and the metal oxide does not induce further
reduction of the substrate; 
before and after adsorption, two reduced $\mathrm{Ti^{3+}}$ ions are present
(see Fig.~\ref{fig11} {\bf (I)}).
In Fig.~\ref{fig12}, 
we plot the spin density of the
$\mathrm{Au_{2}}$ dimer adsorbed on the reduced
$\mathrm{TiO_{2}}$ surface~(a)
and the spin density after adsorbing a 
$\mathrm{CO}$ molecule to this $\mathrm{Au_{2}}$/$\mathrm{TiO_{(2-x)}}$ 
metal/oxide substrate~(b).
In each panel, we can see that both $\mathrm{Ti^{3+}}$ ions belong
to subsurface layer sites below $\mathrm{Ti_{5c}}$ rows.

It is clear from these calculations
that a gold ``cluster''
as small as the $\mathrm{Au_{2}}$ dimer adsorbed 
onto
a bridging $\mathrm{O}$ vacancy 
leads to a promoted support that strongly binds to molecular $\mathrm{CO}$.
%
Furthermore,
the step from one to two gold atoms greatly
changes 
the electronic properties,
thus lending support to the general view
that the reactivity of gold nanoparticles nucleated at
$\mathrm{O}$ vacancies is strongly size dependent.

\section{Conclusions and outlook}
\label{conclusions}

We have performed periodic density functional based calculations that account
for the on--site Coulomb interaction via a Hubbard correction (``GGA+U'')  
on stoichiometric, 
reduced,
and gold-promoted rutile $\mathrm{TiO_{2}}$(110) surfaces. 
Structure optimizations, {\em ab initio} thermodynamics calculations, and
{\em ab initio} molecular dynamics simulations have been carried out 
both using PBE+U and plain PBE in order to provide a broad picture on the 
interactions of these surfaces with several catalytically important molecules.

In agreement with plain PBE calculations and experimental 
results,
we find
that $\mathrm{H}$ atoms preferentially adsorb 
on
surface
$\mathrm{O_b}$ atoms of the stoichiometric surface.
Adsorption of hydrogen results in its reduction and transfer of
close to one electron per $\mathrm{H}$ atom to the substrate.
Using PBE+U, however, the electron transferred by a single $\mathrm{H}$
atom into the substrate localizes preferentially at 
second--layer $\mathrm{Ti}$ sites
(whereas plain PBE yields a delocalized state).
Both PBE+U and PBE calculations show a decrease in the adsorption energy as a
function of coverage 
and a maximum coverage of about 
60--70\%.
Oxygen vacancies are the preferred adsorption sites for $\mathrm{H_{2}O}$ 
dissociation and similar values for the adsorption
energies of water are obtained with and without the Hubbard correction. 
Our PBE+U calculations predict the presence of two reduced $\mathrm{Ti^{3+}}$ ions 
before and after the adsorption of water, indicating that the adsorption of water 
does not further reduce the support
(whereas plain PBE again yields delocalized excess electrons).

Akin to plain PBE calculations, PBE+U predicts a weak interaction 
between $\mathrm{CO}$ molecules and the reduced $\mathrm{TiO_{2}}$ substrate. 
The data indicate that $\mathrm{CO}$ adsorbs at both 
oxygen vacancies in the bridging surface rows, $\mathrm{O_b}$, 
and fivefold coordinated titania sites in the first layer,
$\mathrm{Ti_{5c}}$. 
While PBE calculations give
similar energy values for $\mathrm{CO}$ 
at oxygen vacancies
and at $\mathrm{Ti}_{5c}$, PBE+U calculations show that the adsorption of
$\mathrm{CO}$ at those $\mathrm{Ti}_{5c}$ sites 
which are 
nearest neighbors of the oxygen vacancies
is energetically disfavored.
Upon $\mathrm{CO}$ adsorption on the reduced oxide,
the resulting charge redistribution does not reduce further the titania substrate.
%

Addressing next the interaction of titania with gold, 
both PBE+U and PBE calculations predict two most stable
adsorption sites for
$\mathrm{Au}$ adatom adsorption on the stoichiometric $\mathrm{TiO_{2}}$
surface. 
On the one hand, once an $\mathrm{Au}$ adatom is adsorbed 
at a bridge site
between a
surface $\mathrm{O_b}$ atom and 
a first--layer $\mathrm{Ti_{5c}}$ atom,
PBE+U suggests
that all the $\mathrm{Ti}$ ions belonging to the substrate
preserve their formal oxidation state 
$\mathrm{Ti^{4+}}$,
indicating a very
weak oxidation 
of $\mathrm{Au}$. 
On the other hand, if the $\mathrm{Au}$
adatom is adsorbed on top of 
an
$\mathrm{O_b}$ atom, a net charge transfer from the metal to surface, leading to a positively charged
$\mathrm{Au^{\delta +}}$, is observed.
%
In this case,
the metal adsorption
entails the reduction of a second--layer $\mathrm{Ti}$ ion, which formally
becomes $\mathrm{Ti^{3+}}$.
{\em Ab initio} molecular dynamics reveals that gold adatoms 
on stoichiometric $\mathrm{TiO_{2}}$(110) are very mobile.
%
They are found to migrate easily 
along the [001]~direction,
either along the top of bridging oxygen rows or around the area between such rows.
%
In the former case, we observe an interesting subsurface charge 
delocalization and relocalization dynamics of the excess charge.

Promotion of the (110) rutile surface via 
adsorption of gold atoms or substitution of Ti by Au results in a system with 
greatly changed electronic properties and thus modified reactivities 
in the realm of heterogeneous catalysis. 
%
This is traced back to the oxidation state of the gold adatoms, which is
determined by the site at which the $\mathrm{Au}$ atom is adsorbed as well as 
by the stoichiometry of the substrate.
Isolated $\mathrm{Au}$ atoms supported by the stoichiometric $\mathrm{TiO_{2}}$ surfaces 
are shown to induce a significant charge redistribution at the metal/oxide contact. 
The calculations, both PBE+U and PBE, show positively
charged $\mathrm{Au^{\delta +}}$ adspecies supported on the stoichiometric
surface which activate $\mathrm{CO}$ admolecules.
In stark contrast, the negatively charged $\mathrm{Au^{\delta -}}$ adspecies,
which are incorporated into surface $\mathrm{O_b}$ vacancies,
interact 
only weakly with the same $\mathrm{CO}$ molecules.
We have shown that structures obtained by substituting a 
first--layer $\mathrm{Ti_{5c}}$ ion with an
$\mathrm{Au}$ atom weakens the bond of 
surface $\mathrm{O}$ atoms, akin to recent observations with ceria.
{\it Ab initio} thermodynamics mappings of the surface phase diagram
predict that the 
``$\mathrm{Au@V_{Ti{5c}}}$'' and 
``$\mathrm{Au@V_{Ti{5c}},V_{O3}}$''
of temperatures and pressures 
relevant for catalytic applications. 
%
Finally,
we have shown that although a single $\mathrm{Au}$ adatom
bound to a surface $\mathrm{O}$ vacancy weakly binds $\mathrm{CO}$, 
a second gold atom adsorbed simultaneously into the $\mathrm{O}$ vacancy 
leads to the formation of a gold dimer, $\mathrm{Au_{2}}$, 
featuring
distinctly different electronic and binding properties.

We expect that these insights will be of great help in understanding 
the catalytic activity of gold-promoted titania interfaces with liquid water
as used in selective oxidation reactions of more complex molecules.

\begin{acknowledgments}
%
We are grateful to Bernd Meyer and Martin Muhler for fruitful discussions.
This work has been supported by the German Research Foundation (DFG) via
the Collaborative Research Center
SFB~558 ``Metal--Substrate Interactions in Heterogeneous 
Catalysis,''
by Research Department ``Interfacial Systems Chemistry'' (RD~IFSC), and by 
Fonds der Chemischen Industrie (FCI).
Computational resources were provided by 
NIC (J\"ulich), {\sc Bovilab@RUB} (Bochum), and by RV--NRW.
\end{acknowledgments}


\end{document}